\documentclass{article}
\usepackage[preprint]{neurips_2023}
\usepackage[utf8]{inputenc} 
\usepackage[T1]{fontenc}    
\usepackage{hyperref}       
\usepackage{url}            
\usepackage{booktabs}       
\usepackage{amsfonts}       
\usepackage{amsmath}
\usepackage{makecell}
\usepackage{nicefrac}       
\usepackage{microtype}      
\usepackage{verbatim}
\usepackage{courier}
\usepackage{float}
\usepackage{graphicx}
\usepackage{subcaption}
\usepackage[table,xcdraw]{xcolor}

\makeatletter
\renewcommand{\@notice}{}
\makeatother

\title{Insights from Benchmarking Frontier Language Models on Web App Code Generation}
\author{%
	Yi Cui \\
	ONEKQ Lab\\
	{yi@onekq.ai} \\
}

\begin{document}	
\maketitle

\begin{abstract}
This paper presents insights from evaluating 16 frontier large language models (LLMs) on the WebApp1K benchmark, a test suite designed to assess the ability of LLMs to generate web application code. The results reveal that while all models possess similar underlying knowledge, their performance is differentiated by the frequency of mistakes they make. By analyzing lines of code (LOC) and failure distributions, we find that writing correct code is more complex than generating incorrect code. Furthermore, prompt engineering shows limited efficacy in reducing errors beyond specific cases. These findings suggest that further advancements in coding LLM should emphasize on model reliability and mistake minimization.
\end{abstract}
	
\section{Introduction}
In \citep{webapp1k-paper}, we introduced WebApp1K, a benchmark to evaluate web app code generation performance of frontier LLMs. The performance results of these 16 models are summarized in Tab.~\ref{tab:leaderboard}.
\begin{table}[h!]
\centering
\begin{tabular}{|l|c|c|c|}
\hline
\textbf{Model} & \textbf{pass@1} & \textbf{pass@5} & \textbf{pass@10} \\ \hline
gpt-4o-2024-08-06 & 0.885 & 0.9047 & 0.909 \\ \hline
claude-3.5-sonnet & 0.8808 & 0.8845 & 0.886 \\ \hline
gpt-4o-2024-05-13 & 0.8702 & 0.9013 & 0.909 \\ \hline
gpt-4o-mini & 0.8271 & 0.8534 & 0.858 \\ \hline
mistral-large-2 & 0.7804 & 0.8191 & 0.831 \\ \hline
deepseek-coder-v2-instruct & 0.7002 & 0.8009 & 0.827 \\ \hline
gemini-1.5-pro & 0.6813 & 0.7678 & 0.795 \\ \hline
gemini-1.5-flash & 0.57 & 0.6427 & 0.663 \\ \hline
deepseek-coder-v2-lite-instruct & 0.4606 & 0.6144 & 0.653 \\ \hline
mixtral-8x22b-instruct & 0.3074 & 0.4821 & 0.533 \\ \hline
llama-v3-70b-instruct & 0.3323 & 0.4462 & 0.489 \\ \hline
llama-v3p1-405b-instruct & 0.302 & 0.4053 & 0.437 \\ \hline
llama-v3p1-8b-instruct & 0.2512 & 0.3941 & 0.432 \\ \hline
llama-v3p1-70b-instruct & 0.1027 & 0.1848 & 0.246 \\ \hline
mixtral-8x7b-instruct & 0.1269 & 0.196 & 0.218 \\ \hline
llama-v3-8b-instruct & 0.0679 & 0.1183 & 0.139 \\ \hline
\end{tabular}
\caption{pass@k results for frontier LLMs}
\label{tab:leaderboard}
\end{table}

In this report, we share insights gained from the code written by these 16 models To prevent benchmark contamination, we do not reveal the actual code, but their outcome aggregated by certain measures. The artifacts are on GitHub and Huggingface: the dataset containing all 1000 problems of WebApp1K\citep{webapp1k-dataset}, the script\citep{webapp1k-script} to run WebApp1K, and the leaderboard\citep{webapp1k-leaderboard}. 

The rest of this report is organized as follows. Sec.~\ref{sec:benchmark} reveals the difficulty of WebApp1K to each model in terms of test failures. Sec.~\ref{sec:loc} analyzes the LOC (lines of code) distributions. Sec.~\ref{sec:errors} provides a deep dive to errors made by models. Sec.~\ref{sec:related} presents related works. Sec.~\ref{sec:conclude} concludes our findings and discusses future directions.
\section{Benchmark Difficulty}\label{sec:benchmark}
\subsection{Failure Distributions}
\begin{figure}[h]
    \centering
    \includegraphics[width=0.7\textwidth]{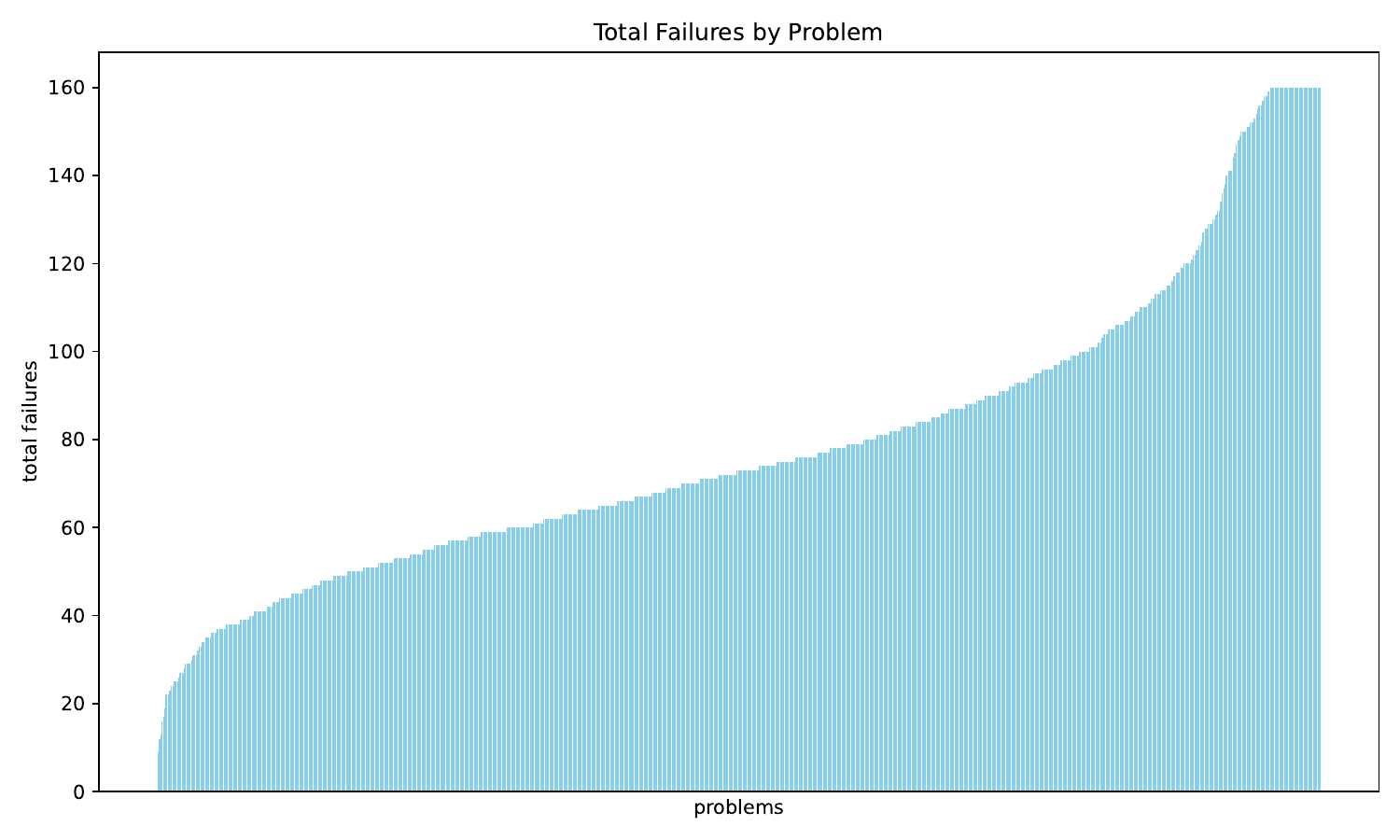}
    \caption{Failures per Problem}
    \label{fig:failures}
\end{figure}

We begin by examining the difficulty of the benchmark. We made each model solve each coding problem for 10 times, which gives us 160 solutions per problem. If a solution passes the tests, it is considered a succes, otherwise a failure. Fig.~\ref{fig:failures} shows number of failures per problem. The more failures a problem collects, the more difficult it is.

As the figure shows, the majority of the problems are situated on the left side of the graph, characterized by low failure rates, indicating that these problems were relatively easy, especially for top-ranked models. Conversely, a small cluster of problems on the far right exhibit extremely high failure rates. The most difficult 38 problems have never been solved by any model. This pattern still holds after we break down Fig.~\ref{fig:failures} by applications. More details are available in Appendix \ref{sec:difficulty_apps}.
\subsection{Does the Code Build?}
Of the total 160,000 solutions produced (10 runs by each of the 16 models for 1000 problems), only 172 have syntax errors, i.e. the build failure rate is 0.1\%. In particular, the solutions by Claude 3.5 and Mistral Large 2 have no syntax errors.

This means all models are able to write high-quality and compilable code. Yet some of them did not manage to meet test expectations, some explicit and others implicit, therefore causing the failures.
\section{Line-of-Code (LOC) Analysis}\label{sec:loc}
Thanks to the modularized design of the React framework, the solutions output by all models universally follow the template outlined in Tab.~\ref{tab:solution}, with no need for any explicit prompting. As such, we seek to another proxy signal, LOC (line-of-code), to gain insights.
\begin{table}[!t]
	\centering
	\begin{tabular}{|l|}
		\hline
		\begin{minipage}{\dimexpr\textwidth-2\fboxsep-2\fboxrule}
			\vspace{2mm}
			\begingroup
			\renewcommand{\ttdefault}{pcr} 
			\scriptsize
			\begin{verbatim}
				// Import Statements
				import React from 'react';

				// Component Declaration
				const componentName = (...) -> {
				  // function body

				  // JSX-based UI layout
				  return (
				    <div> ... </div>
				  );
				};

				// Export Statements
				export default componentName;
			\end{verbatim}
			\endgroup
		\end{minipage} \\
		\hline
	\end{tabular}
	\caption{A React Solution to a WebApp1K Problem}
	\label{tab:solution}
\end{table}
\subsection{LOC Distribution by Models}
\begin{table}[h!]
    \centering
    \begin{tabular}{lll}
        \toprule
        \textbf{Model} & \textbf{Median LOC} & \textbf{pass@1} \\
        \midrule
        mixtral-8x7b-instruct          & 35 & 0.1269 \\
        llama-v3-8b-instruct           & 39 & 0.0679 \\
        gpt-4o-2024-05-13              & 39 & 0.8702 \\
        llama-v3p1-405b-instruct       & 40 & 0.3020 \\
        gpt-4o-2024-08-06              & 40 & 0.8850 \\
        deepseek-coder-v2-instruct     & 40 & 0.7002 \\
        gpt-4o-mini                    & 40 & 0.8271 \\
        mistral-large-2                & 41 & 0.7804 \\
        gemini-1.5-flash               & 41 & 0.5700 \\
        llama-v3p1-8b-instruct         & 42 & 0.2512 \\
        mixtral-8x22b-instruct         & 43 & 0.3074 \\
        claude-3.5-sonnet              & 43 & 0.8808 \\
        llama-v3-70b-instruct          & 43 & 0.3323 \\
        deepseek-coder-v2-lite-instruct & 43 & 0.4606 \\
        gemini-1.5-pro                 & 45 & 0.6813 \\
        llama-v3p1-70b-instruct        & 46 & 0.1027 \\
        \bottomrule
    \end{tabular}
    \caption{Models Ranked by Median LOC with $pass@1$}
    \label{tab:loc_by_models}
\end{table}

In Tab.~\ref{tab:loc_by_models}, we rank models by their median LOC alongside their respective $pass@1$ scores. Picking one $pass@k$ is sufficient because all scores produced basically the same model rankings as shown in Tab.~\ref{tab:leaderboard}.

We observe that the median LOCs across all models stay close, ranging from 35 to 46. We believe this narrow range is largely enforced by the conciseness and expressiveness of the React framework itself. Also there is no strong correlation between the conciseness (median LOC) and correctness ($pass@1$). For example, mixtral-8x7b-instruct, which has the shortest median LOC, ranks quite low on $pass@1$ (0.1269). Conversely, stronger models like claude-3.5-sonnet and gpt-4o-2024-08-06, generate longer code. Other models, e.g. deepseek-coder-v2-instruct and gemini-1.5-pro, strike a balance between median.

\begin{figure}[h!]
	\centering
	\begin{subfigure}{0.32\textwidth}
		\includegraphics[width=\textwidth]{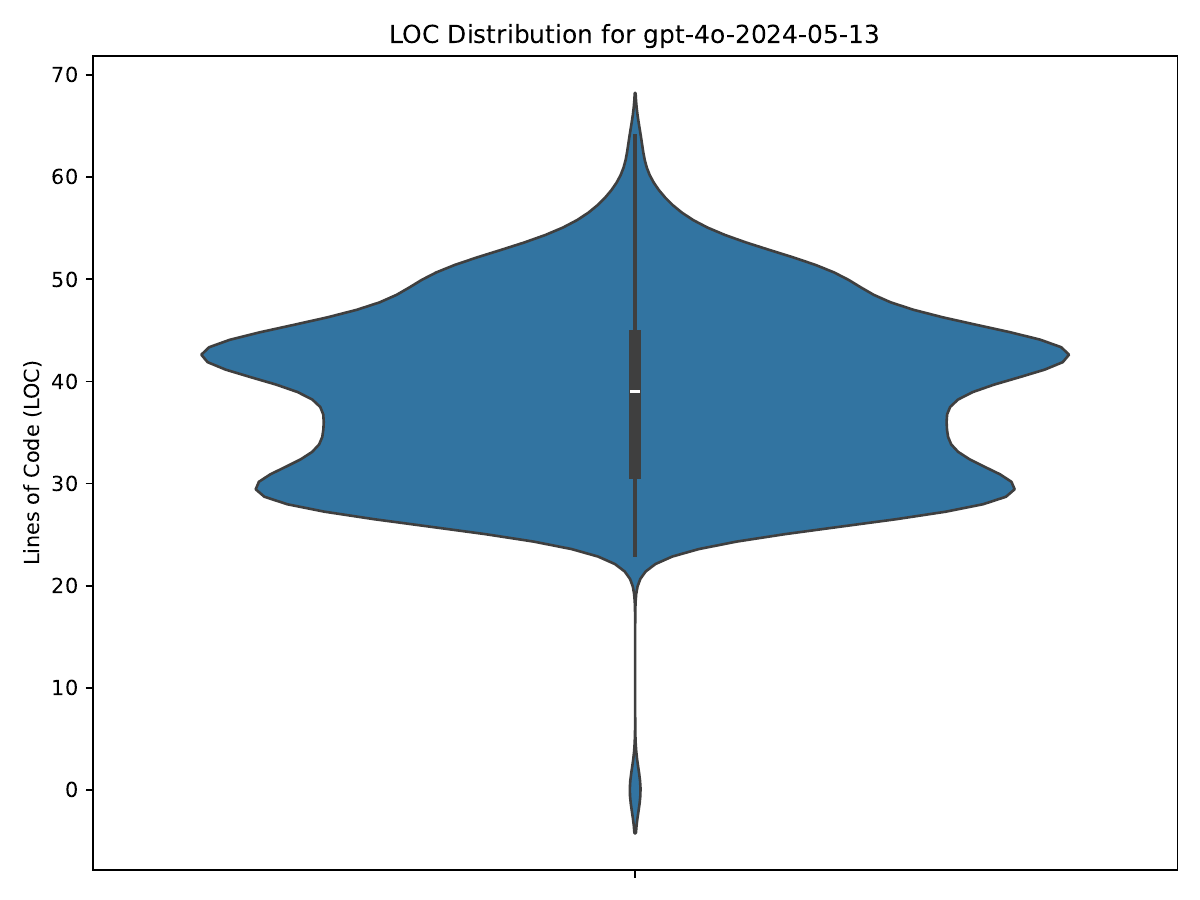}
		\caption{gpt-4o-2024-05-13}
	\end{subfigure}
	\begin{subfigure}{0.32\textwidth}
		\includegraphics[width=\textwidth]{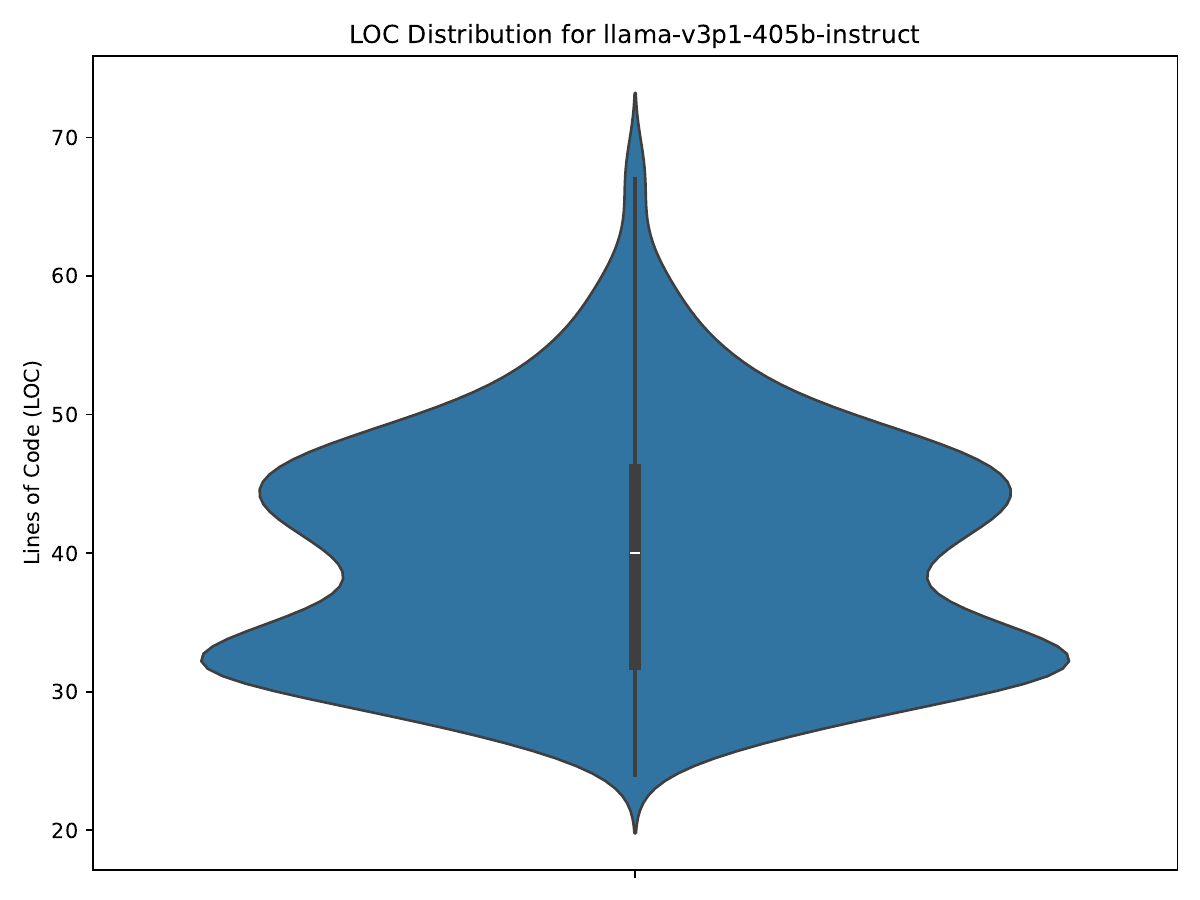}
		\caption{llama-v3p1-405b-instruct}
	\end{subfigure}
	\begin{subfigure}{0.32\textwidth}
		\includegraphics[width=\textwidth]{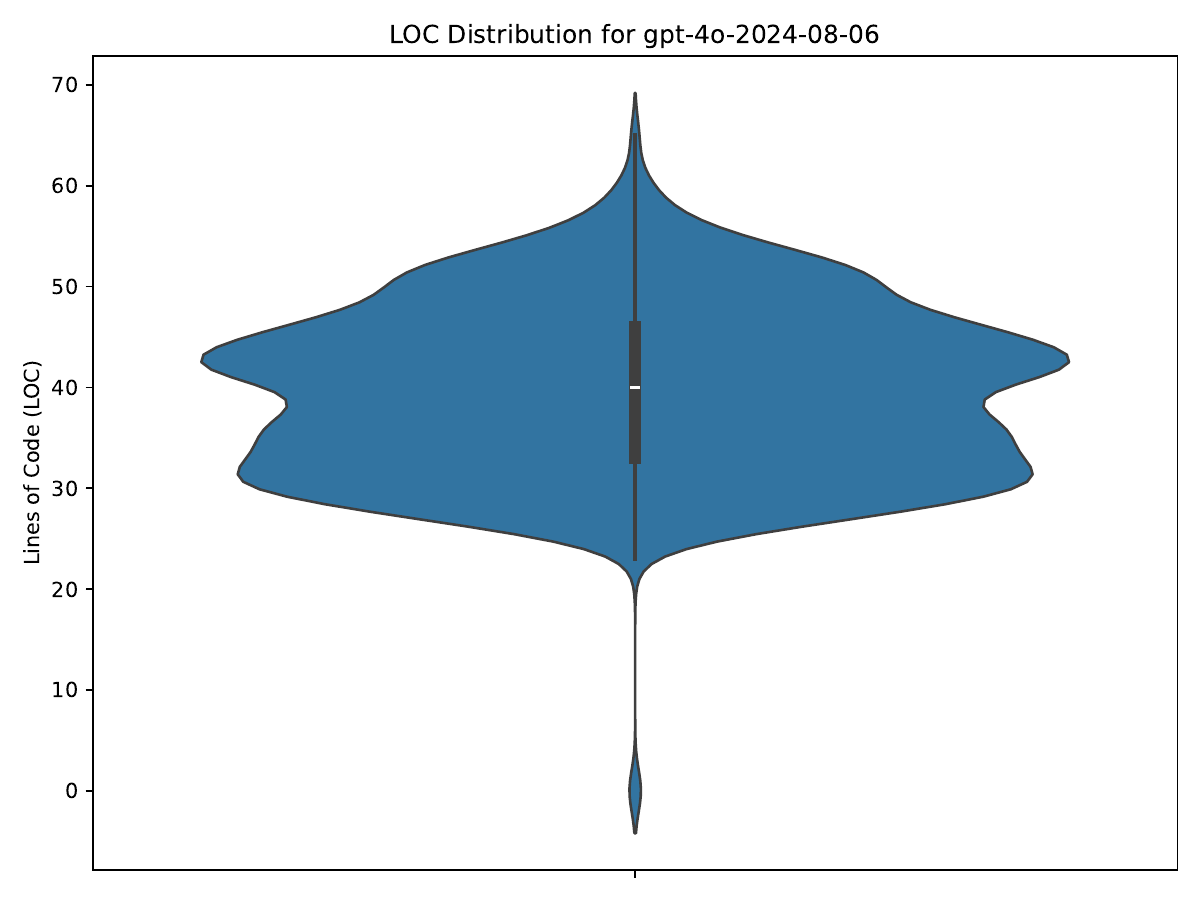}
		\caption{gpt-4o-2024-08-06}
	\end{subfigure}
	
	\begin{subfigure}{0.32\textwidth}
		\includegraphics[width=\textwidth]{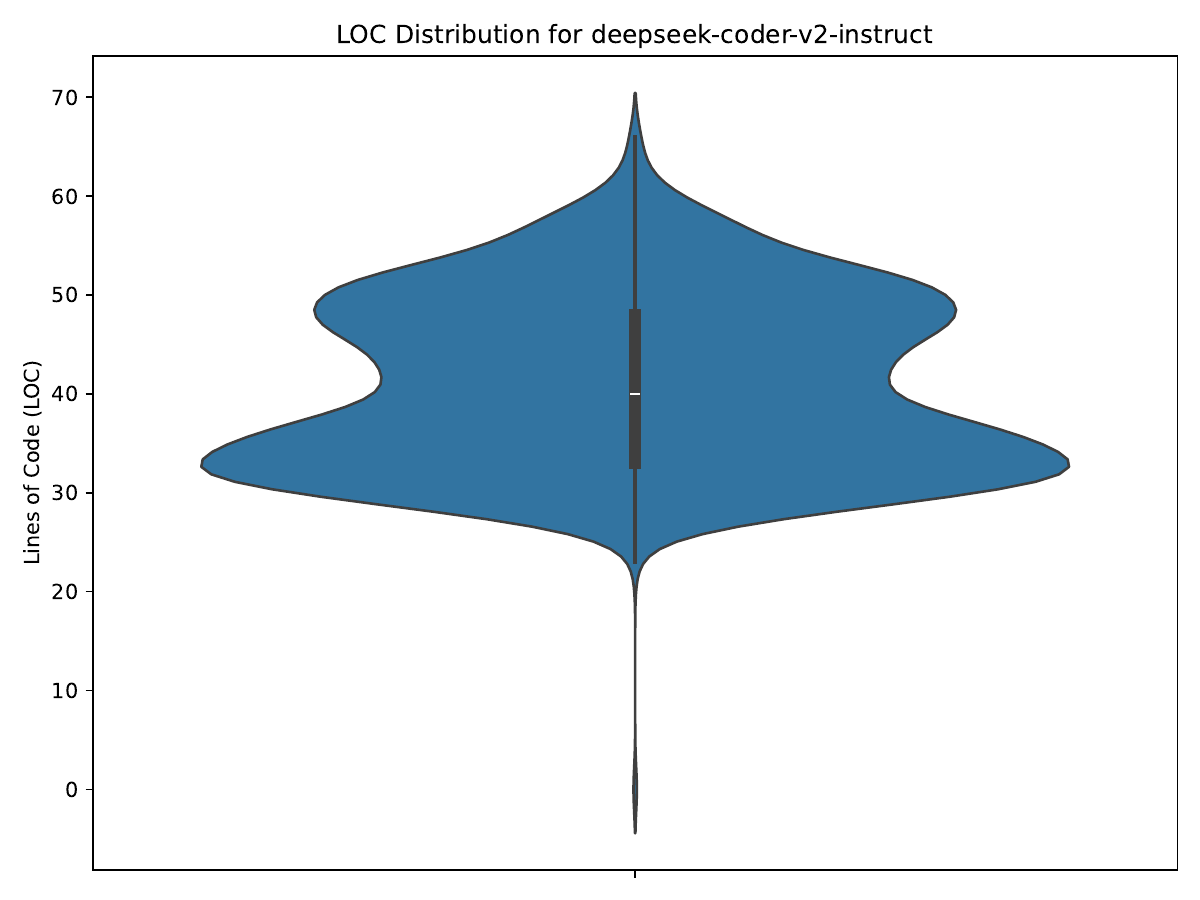}
		\caption{deepseek-coder-v2-instruct}
	\end{subfigure}
	\begin{subfigure}{0.32\textwidth}
		\includegraphics[width=\textwidth]{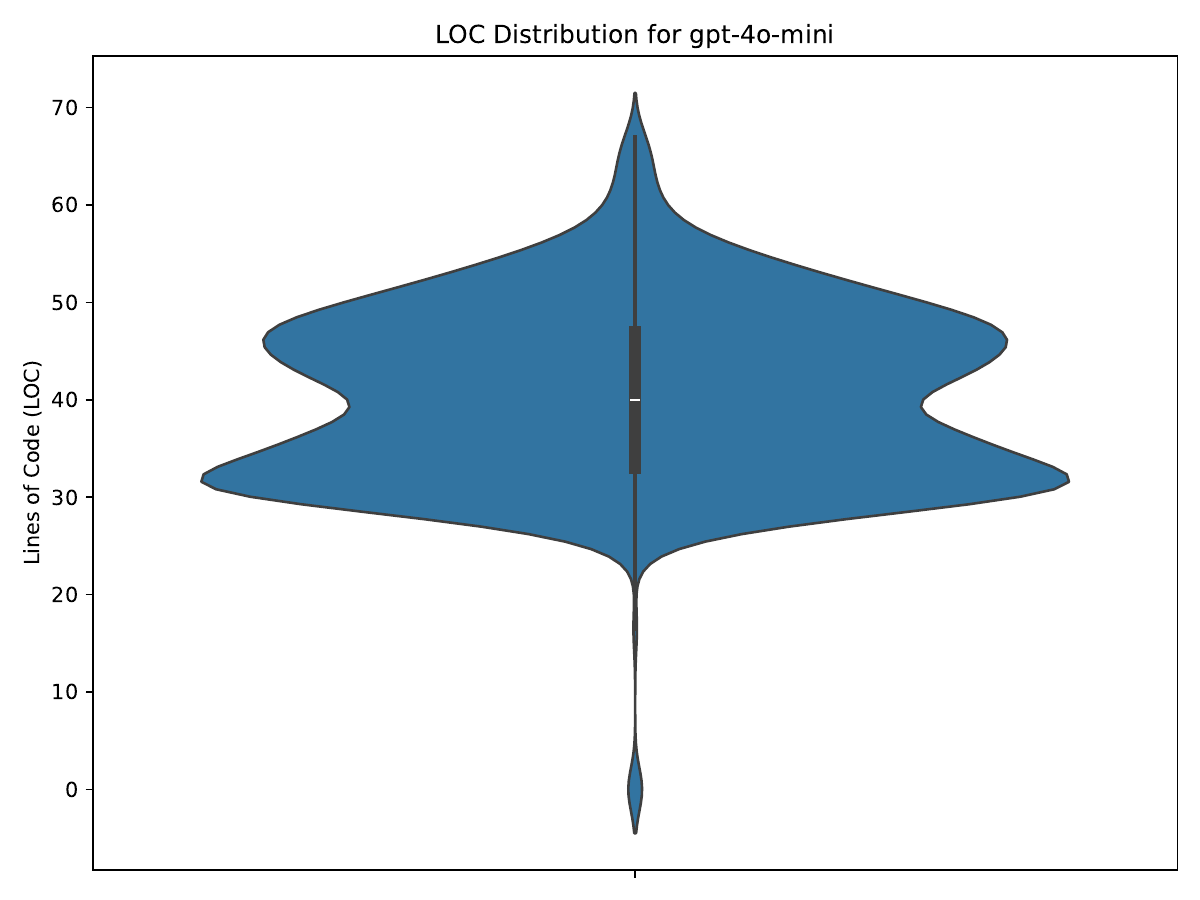}
		\caption{gpt-4o-mini}
	\end{subfigure}
	\begin{subfigure}{0.32\textwidth}
		\includegraphics[width=\textwidth]{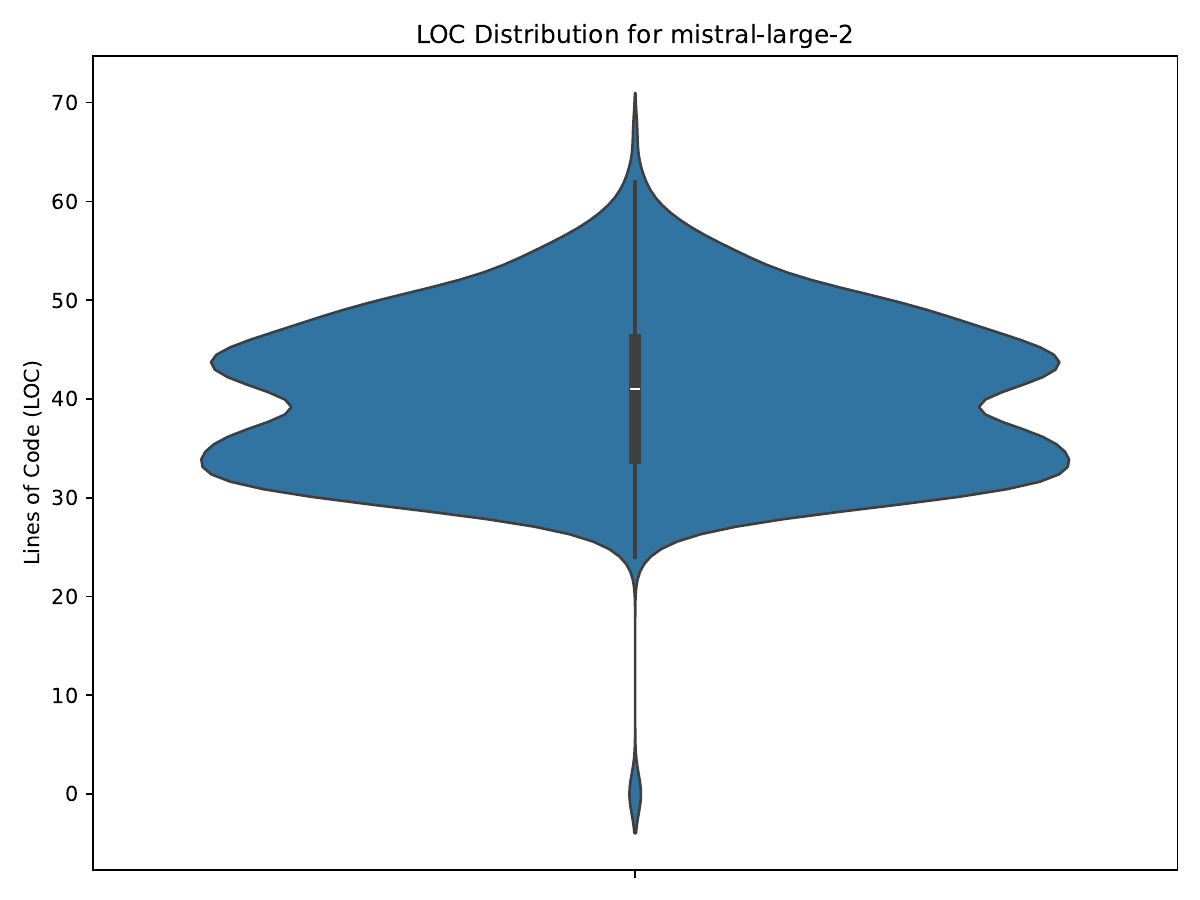}
		\caption{mistral-large-2}
	\end{subfigure}
	
	\begin{subfigure}{0.32\textwidth}
		\includegraphics[width=\textwidth]{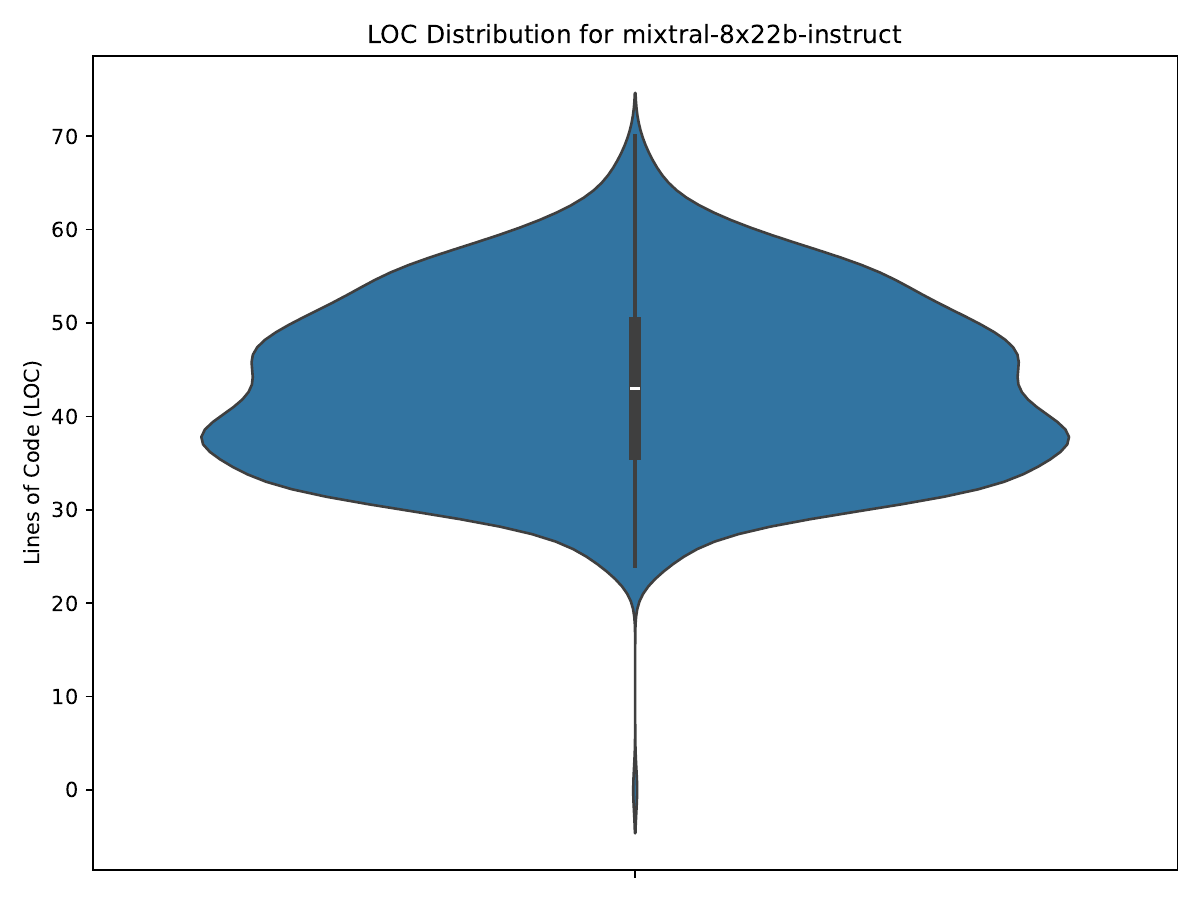}
		\caption{mixtral-8x22b-instruct}
	\end{subfigure}
	\begin{subfigure}{0.32\textwidth}
		\includegraphics[width=\textwidth]{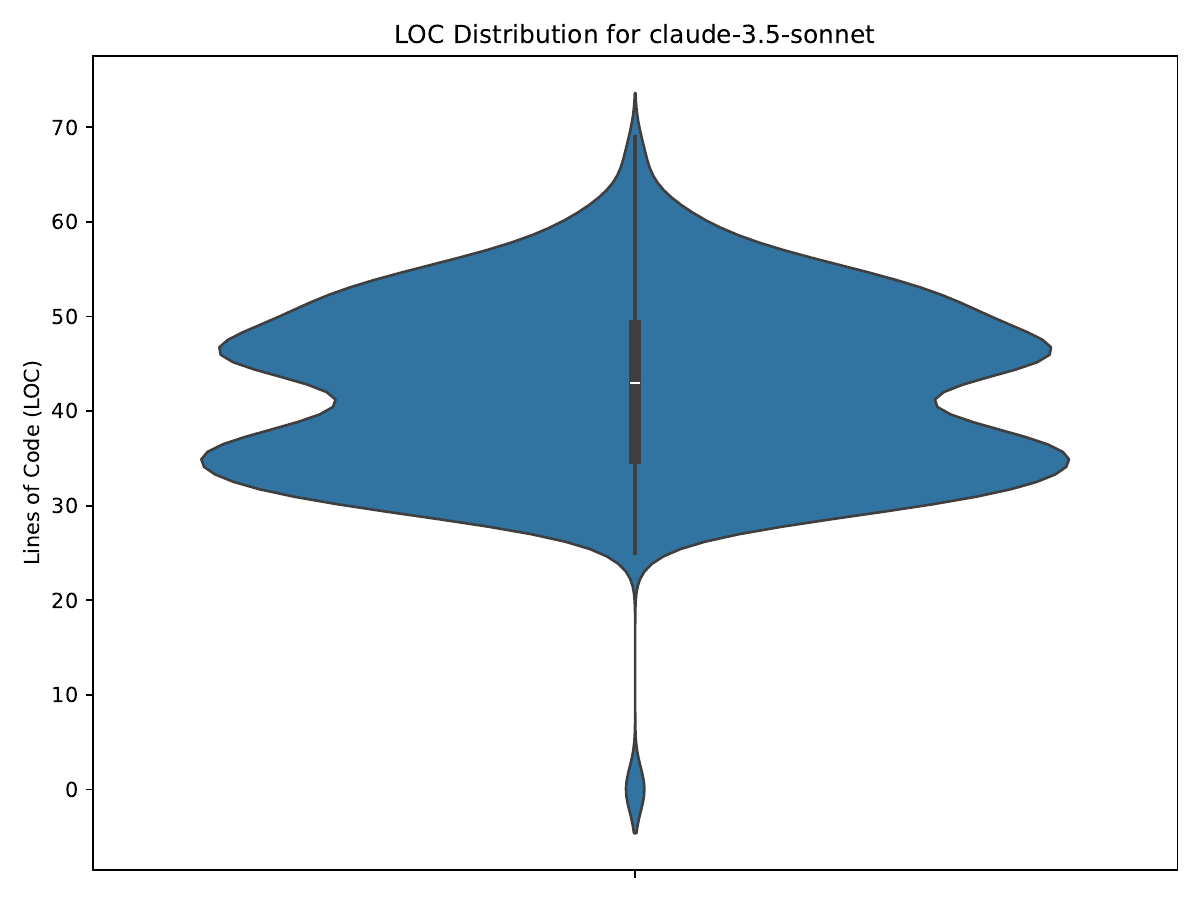}
		\caption{claude-3.5-sonnet}
	\end{subfigure}
	\begin{subfigure}{0.32\textwidth}
		\includegraphics[width=\textwidth]{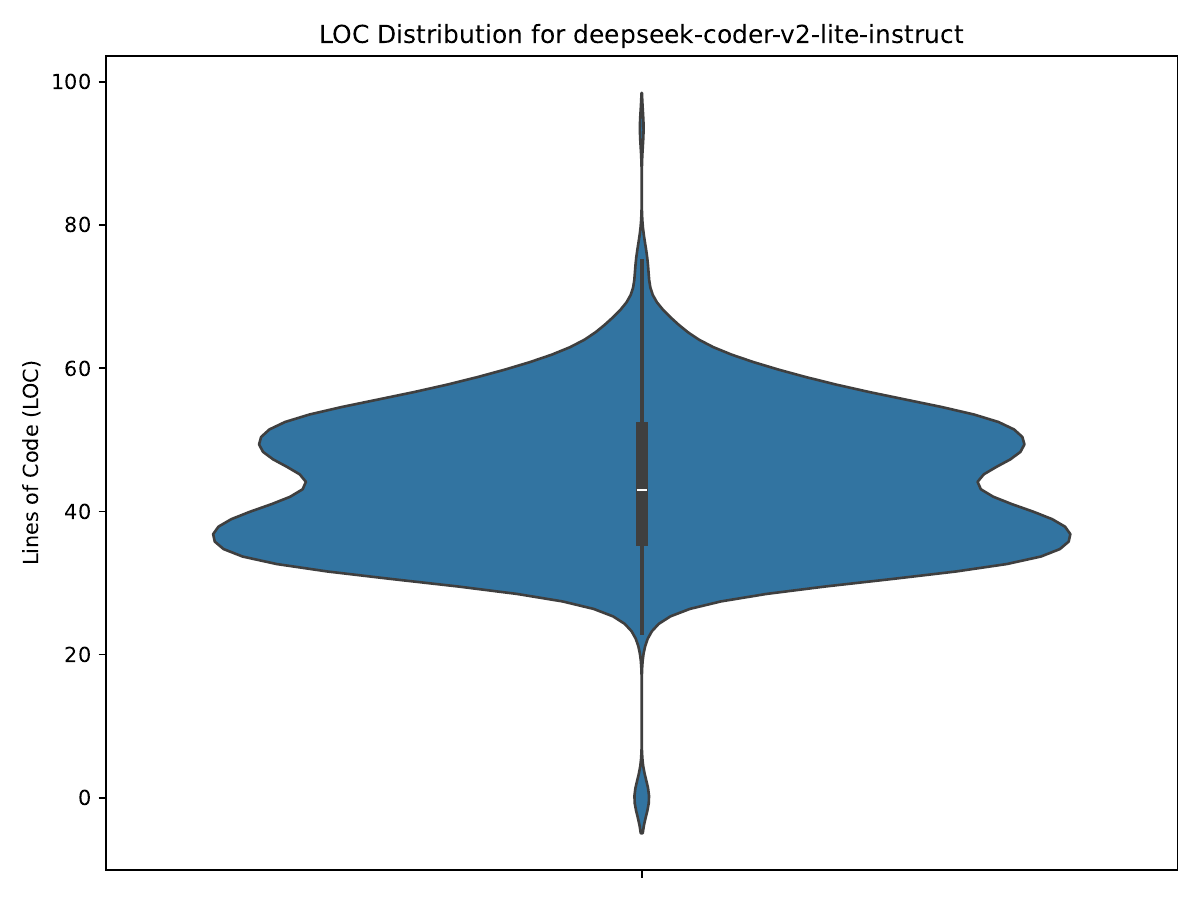}
		\caption{deepseek-coder-v2-lite-instruct}
	\end{subfigure}
	
	\begin{subfigure}{0.32\textwidth}
		\includegraphics[width=\textwidth]{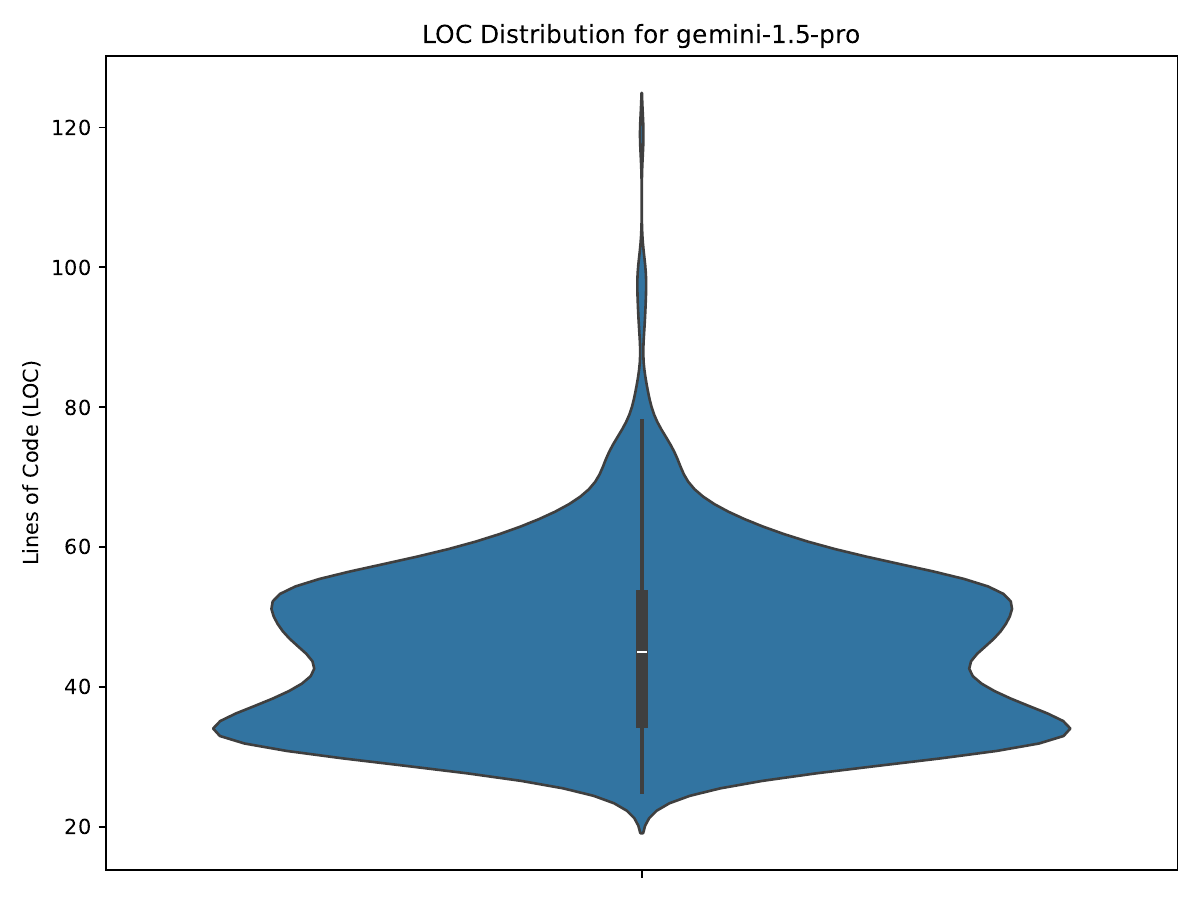}
		\caption{gemini-1.5-pro}
	\end{subfigure}
	
	\caption{LOC Distribution by Model (BiModal)}
	\label{fig:loc_distribution_models_bimodal}
\end{figure}

\begin{figure}[h!]
	\centering
	\begin{subfigure}{0.32\textwidth}
		\includegraphics[width=\textwidth]{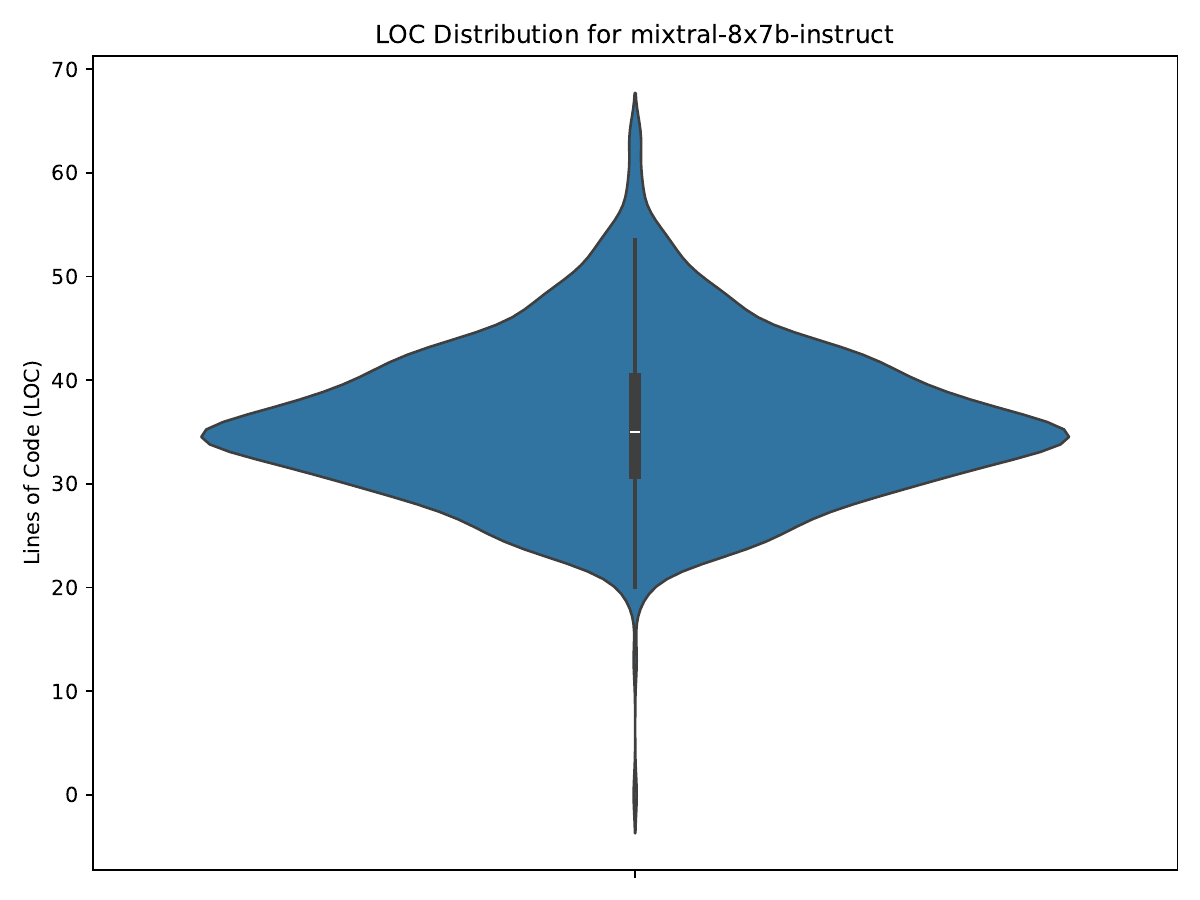}
		\caption{mixtral-8x7b-instruct}
	\end{subfigure}
	\begin{subfigure}{0.32\textwidth}
		\includegraphics[width=\textwidth]{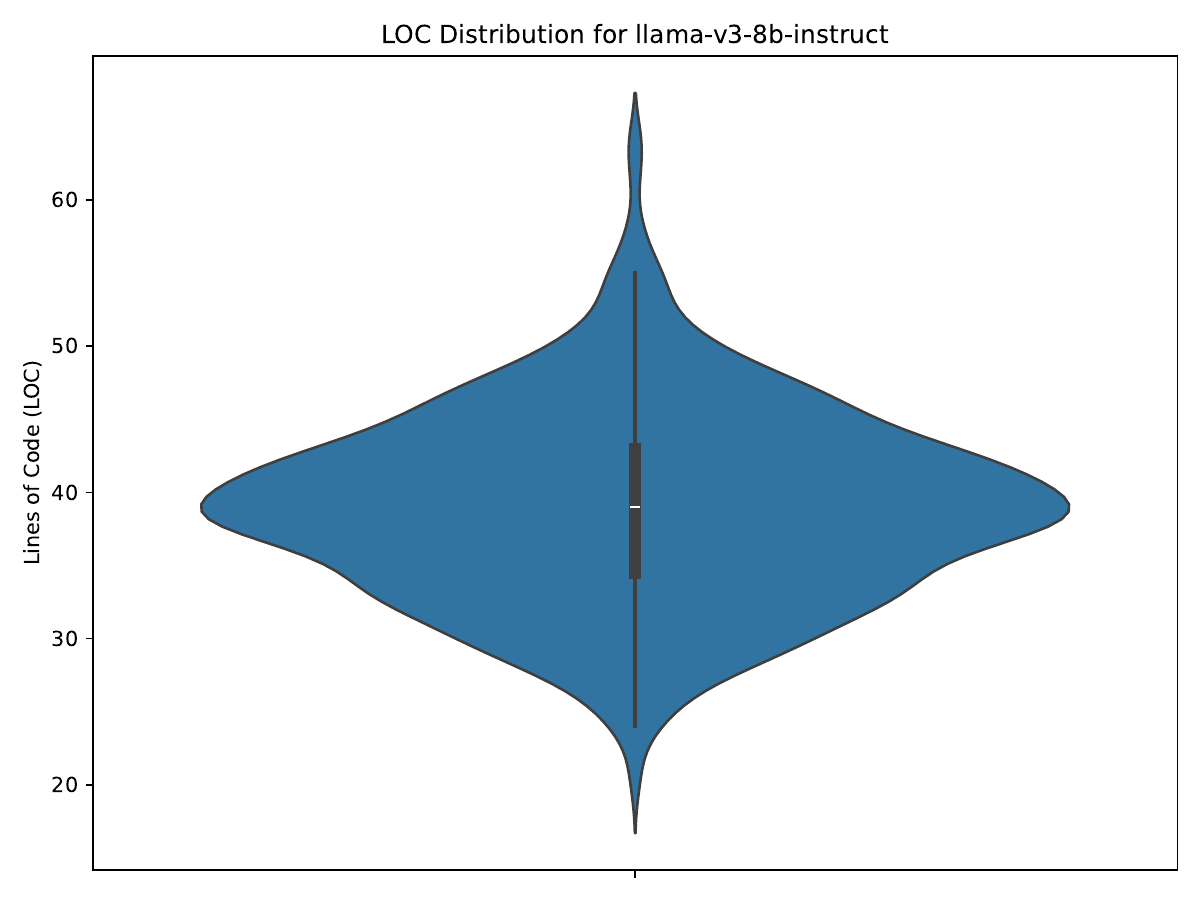}
		\caption{llama-v3-8b-instruct}
	\end{subfigure}
	\begin{subfigure}{0.32\textwidth}
		\includegraphics[width=\textwidth]{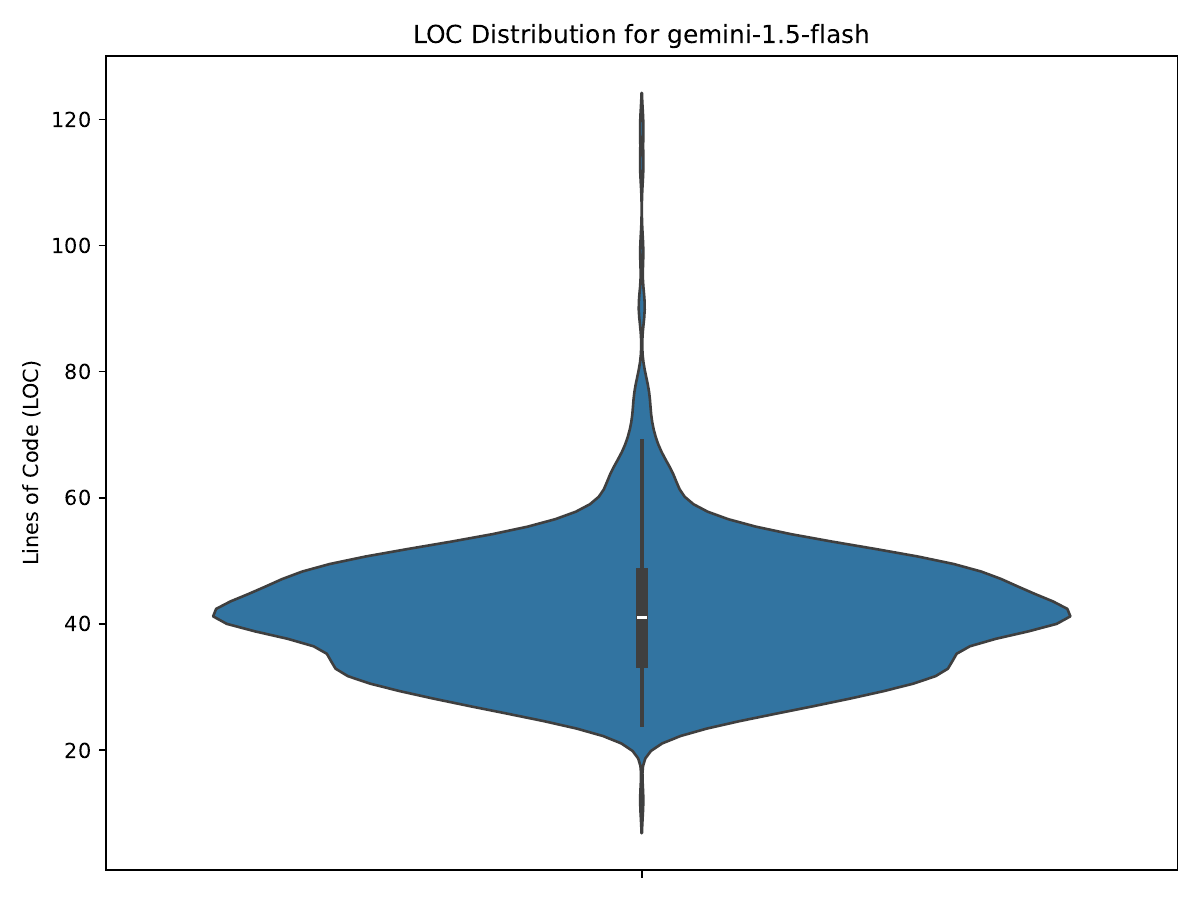}
		\caption{gemini-1.5-flash}
	\end{subfigure}
	
	\begin{subfigure}{0.32\textwidth}
		\includegraphics[width=\textwidth]{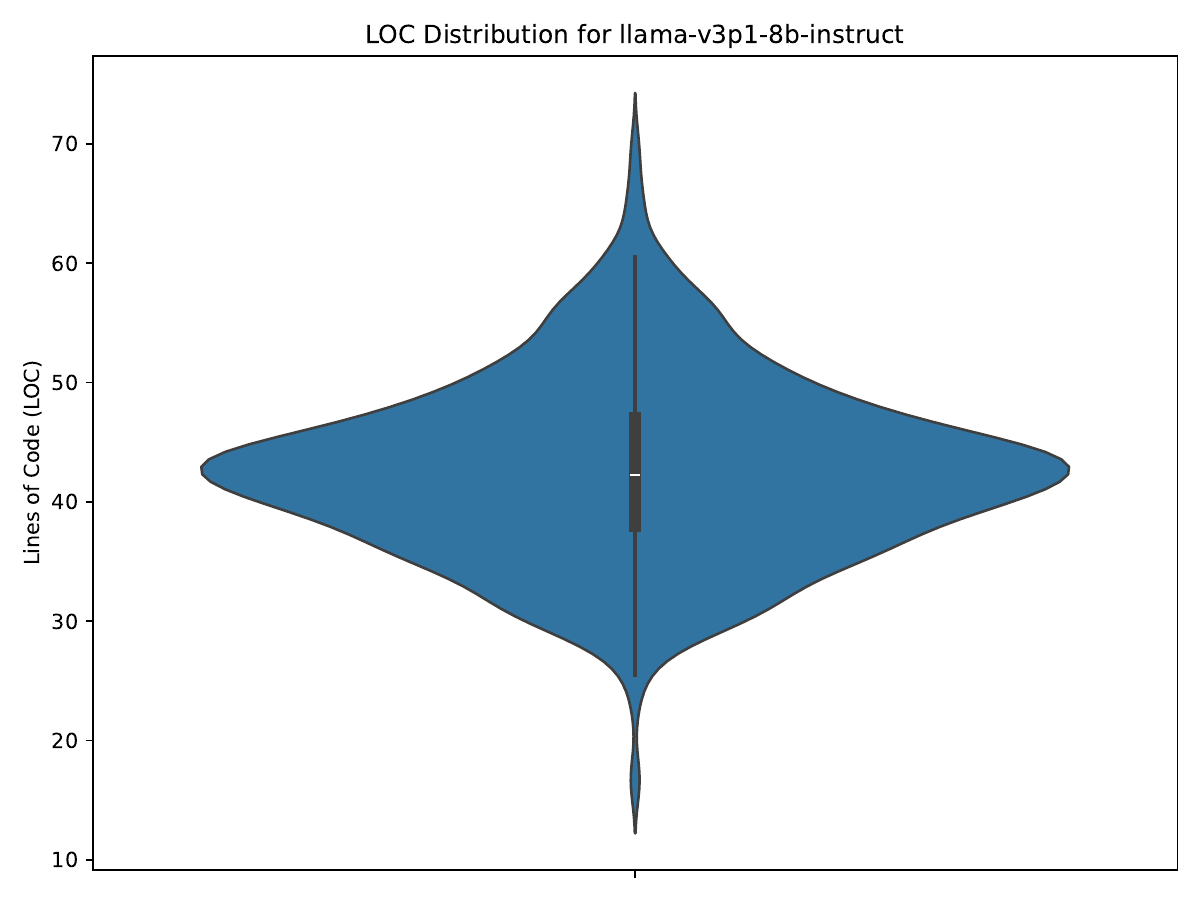}
		\caption{llama-v3p1-8b-instruct}
	\end{subfigure}
	\begin{subfigure}{0.32\textwidth}
		\includegraphics[width=\textwidth]{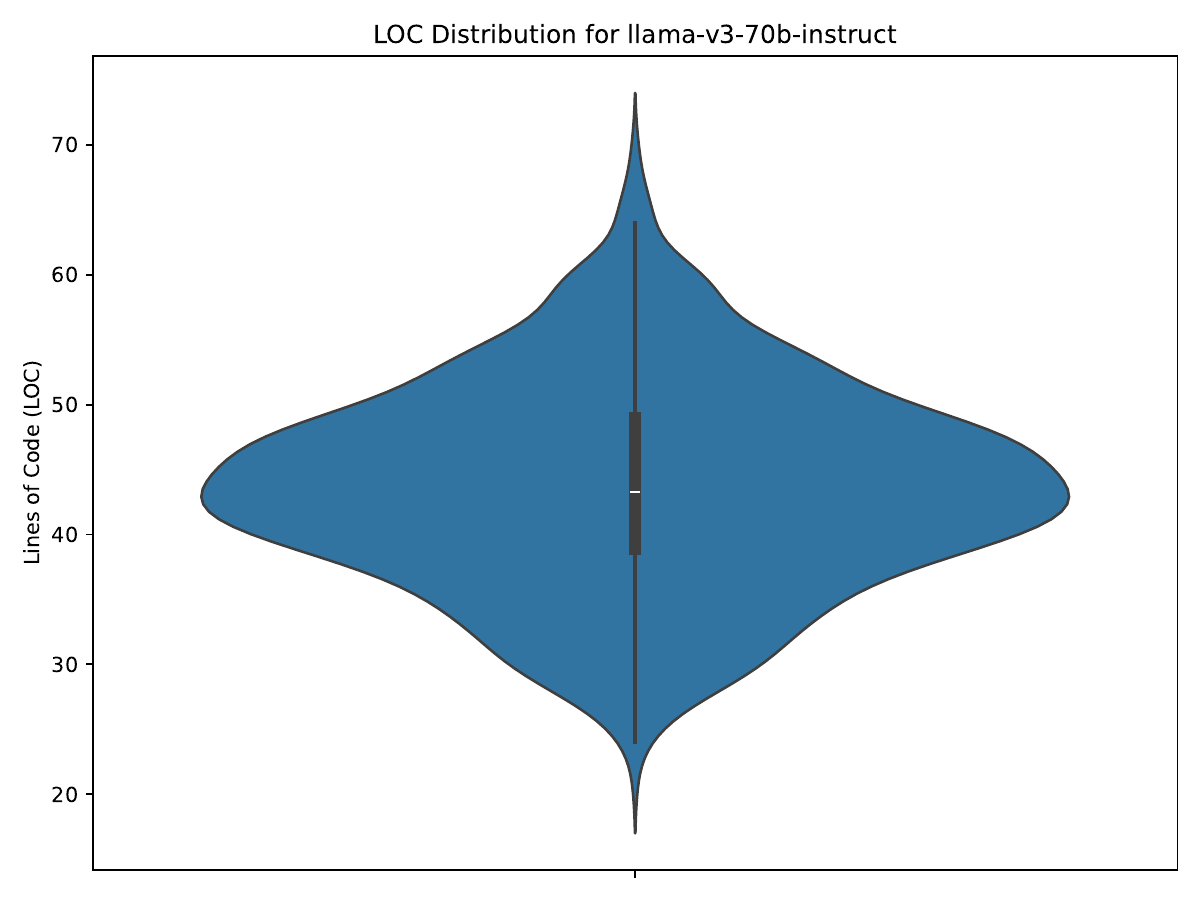}
		\caption{llama-v3-70b-instruct}
	\end{subfigure}
	\begin{subfigure}{0.32\textwidth}
		\includegraphics[width=\textwidth]{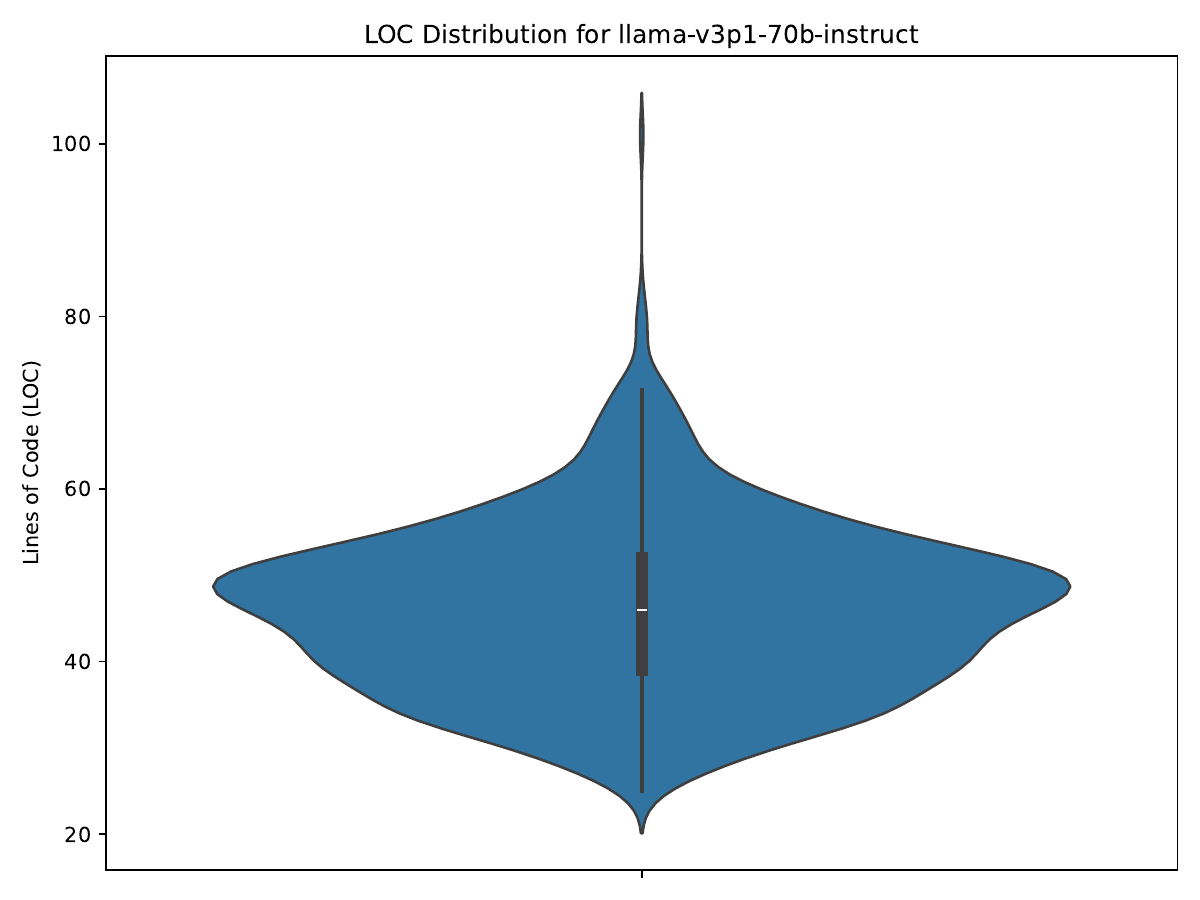}
		\caption{llama-v3p1-70b-instruct}
	\end{subfigure}
	
	\caption{LOC Distribution by Model (UniModal)}
	\label{fig:loc_distribution_models_unimodal}
\end{figure}

Next, we use violin charts to visualize LOC distribution of each model. The distributions are either bimodal or unimodal, and they are collected in Fig.~\ref{fig:loc_distribution_models_bimodal} and Fig.~\ref{fig:loc_distribution_models_unimodal} respectively.

Notably, all high-performing models with high $pass@1$ scores are located in Fig.~\ref{fig:loc_distribution_models_bimodal}. These models, such as the gpt-4o variants and deepseek-coder series, demonstrate higher variability in their LOC distributions, i.e. bimodal. The two distinct peaks in these models' distributions suggests that they generate both shorter and longer code lengths, depending on the task. Importantly, the median LOC values for these bimodal models consistently fall between the two peaks, highlighting a balance in their code generation. Also the higher of the two peaks often corresponds to smaller LOC. This suggests that while these models can produce longer code when necessary, they tend to generate shorter, more optimized code in most cases. 

In contrast, Fig.~\ref{fig:loc_distribution_models_unimodal} contains smaller models. Some exhibit near-perfect normal distributions, e.g. mixtral-8x7b-instruct and llama-v3-8b-instruct. These models generate LOC distributions that are tightly centered around their medians, indicating more consistent and predictable behavior. The lack of bimodal characteristics in these distributions reflects a more stable output across tasks, but with lower complexity compared to the larger models in Fig.~\ref{fig:loc_distribution_models_bimodal}.

We also study LOC distribution sharded by applications and have the similar observations. More details can be found in Appendix \ref{sec:loc_distribution_apps}.
\subsection{Impact of Success/Failure}\label{sec:loc_successfail}
\begin{figure}[h!]
    \centering
    \begin{subfigure}{0.49\textwidth}
        \centering
        \includegraphics[width=\linewidth]{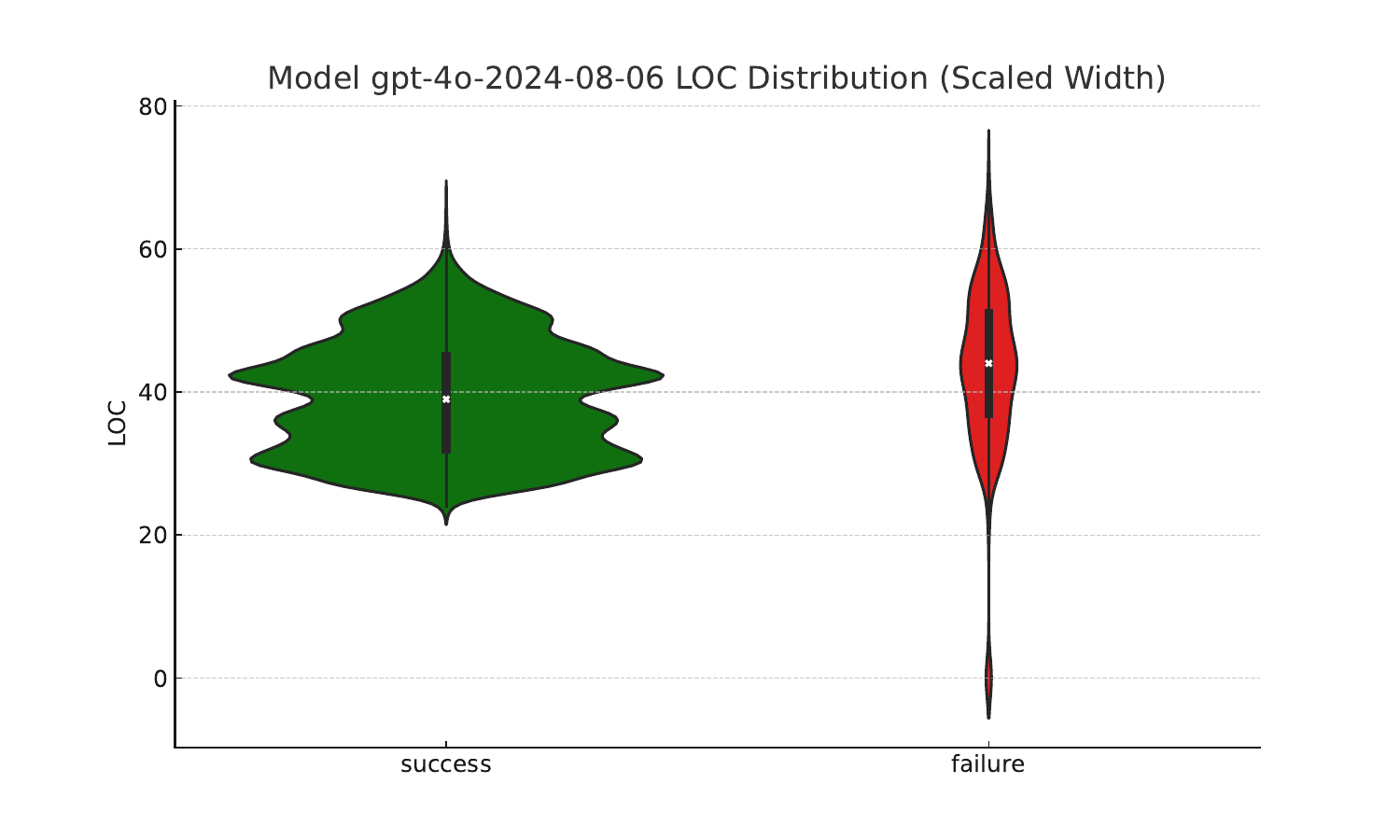}
        \caption{gpt-4o-2024-08-06 ($pass@1$ = 0.885)}
    \end{subfigure}
    \begin{subfigure}{0.49\textwidth}
        \centering
        \includegraphics[width=\linewidth]{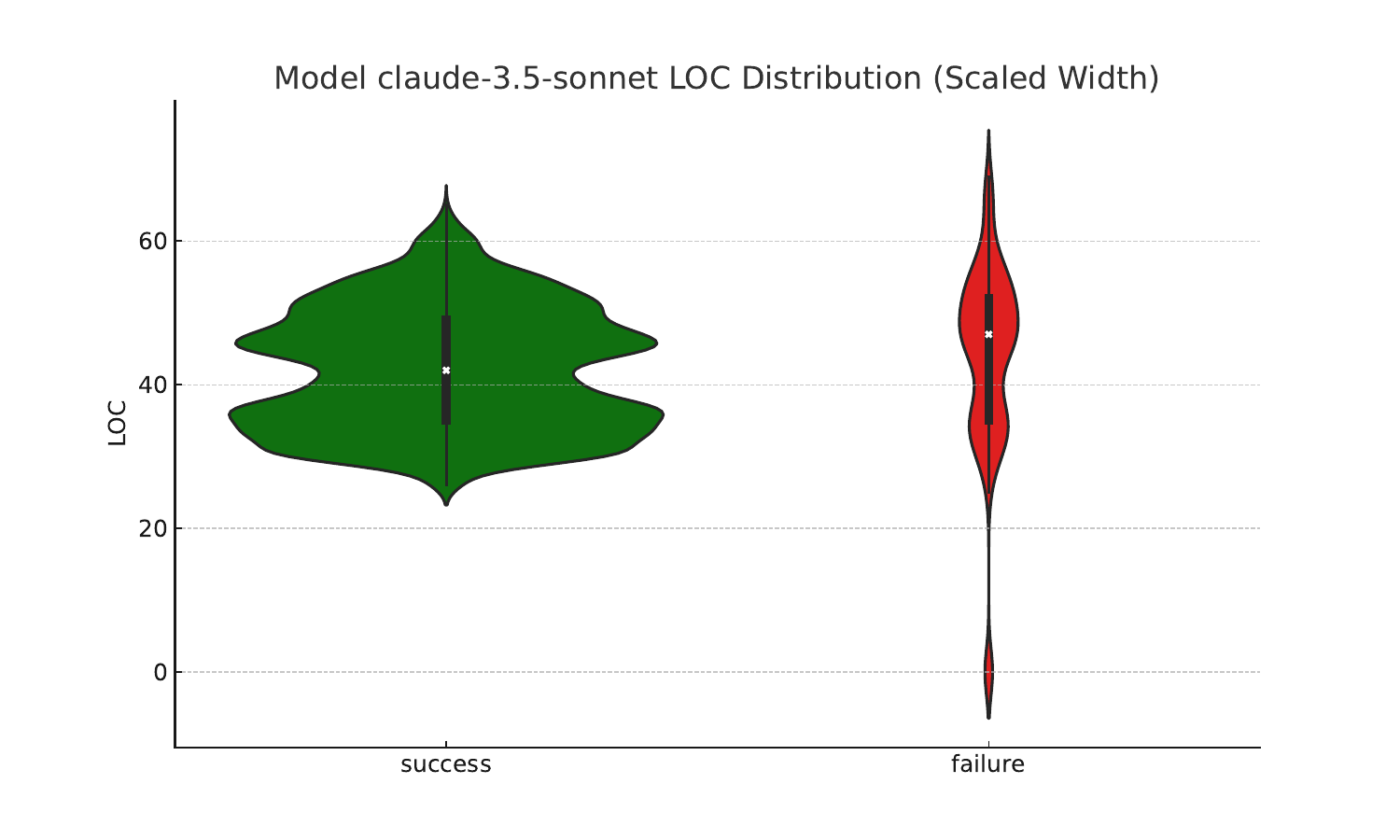}
        \caption{claude-3.5-sonnet ($pass@1$ = 0.8808)}
    \end{subfigure}

    \begin{subfigure}{0.49\textwidth}
        \centering
        \includegraphics[width=\linewidth]{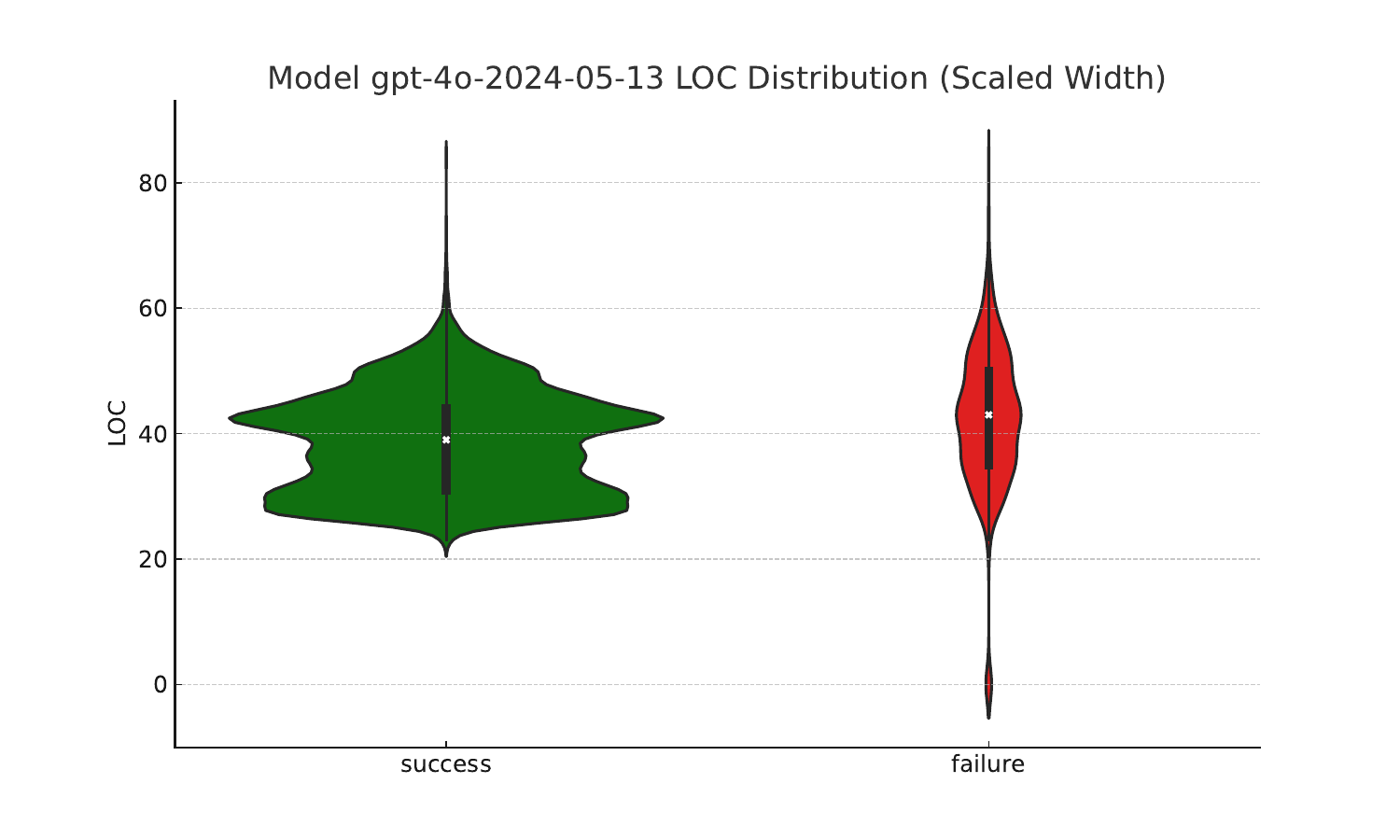}
        \caption{gpt-4o-2024-05-13 ($pass@1$ = 0.8702)}
    \end{subfigure}
    \begin{subfigure}{0.49\textwidth}
        \centering
        \includegraphics[width=\linewidth]{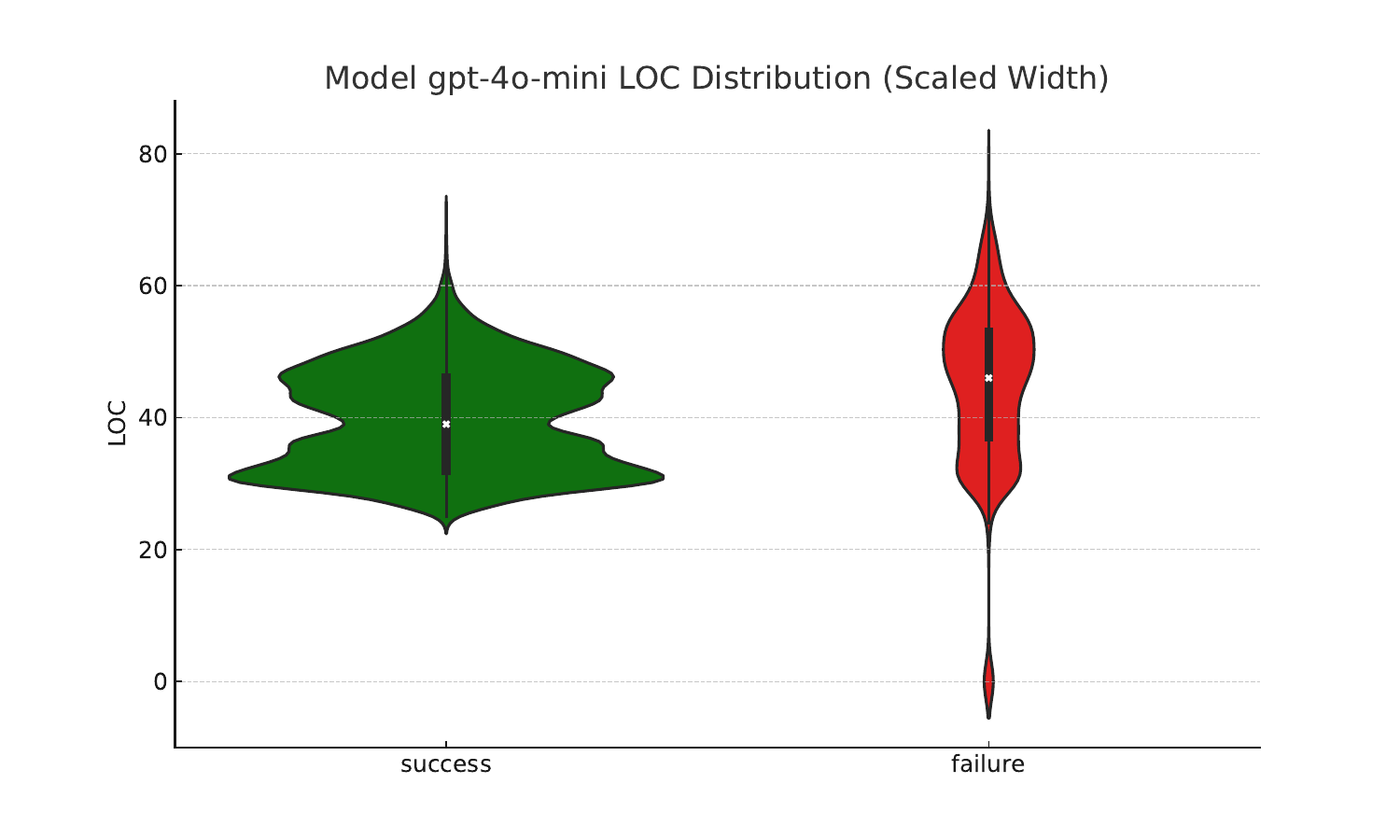}
        \caption{gpt-4o-mini ($pass@1$ = 0.8271)}
    \end{subfigure}
    \caption{LOC Distribution by Model of High $pass@1$: Success vs Failure}
    \label{fig:loc_success_distribution_models}
\end{figure}

\begin{figure}[h!]
    \centering
    \begin{subfigure}{0.49\textwidth}
        \centering
        \includegraphics[width=\linewidth]{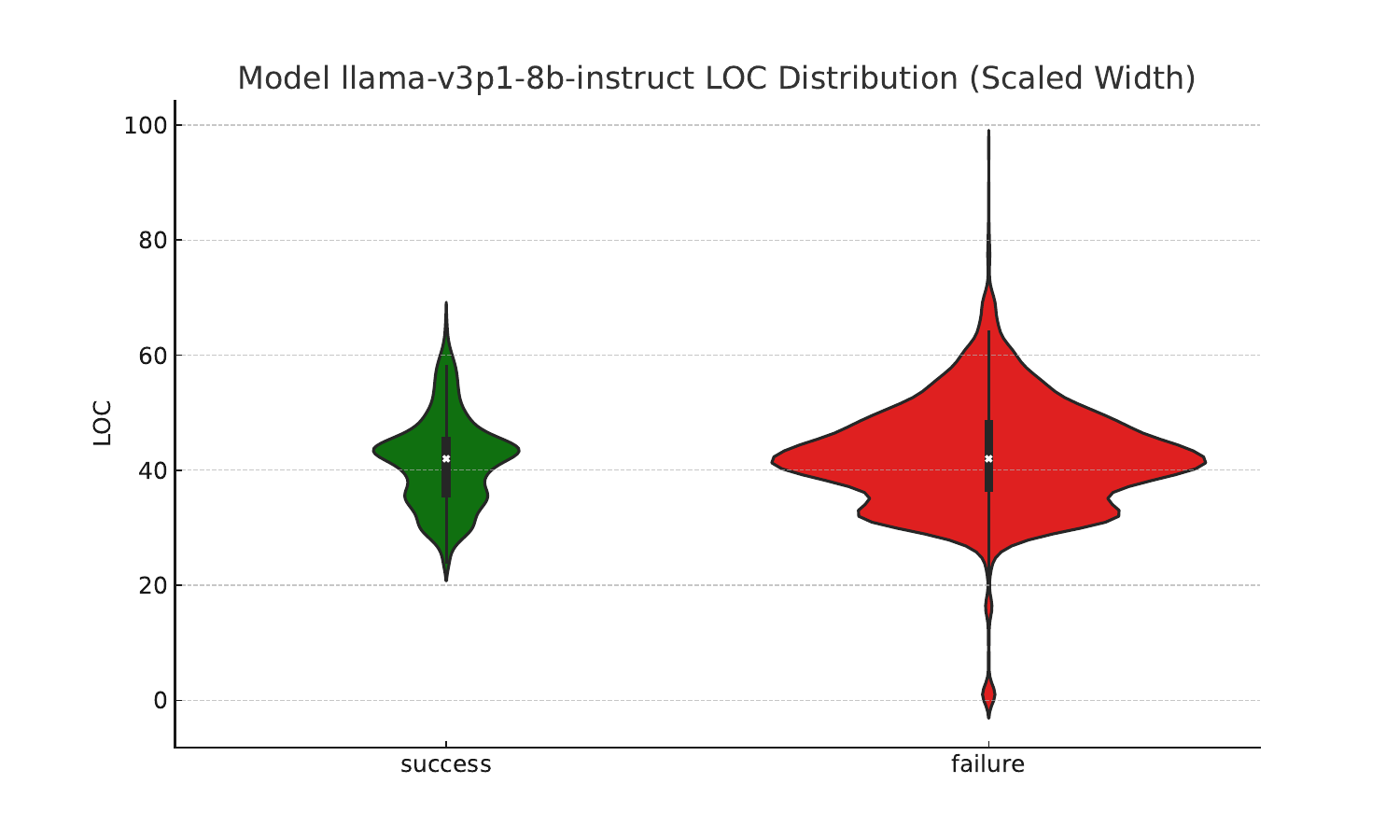}
        \caption{llama-v3p1-8b-instruct (pass@1 = 0.2512)}
    \end{subfigure}
    \begin{subfigure}{0.49\textwidth}
        \centering
        \includegraphics[width=\linewidth]{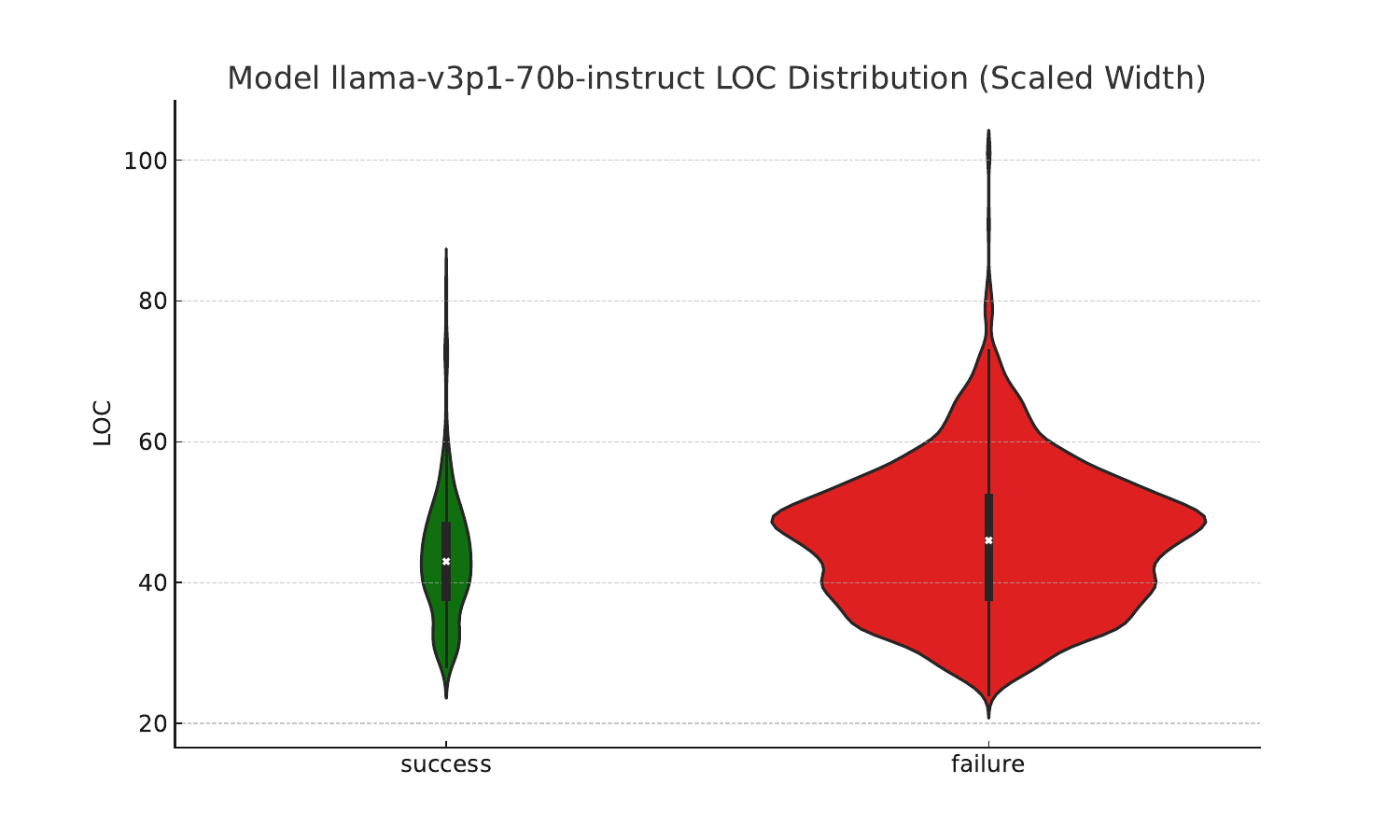}
        \caption{llama-v3p1-70b-instruct (pass@1 = 0.1027)}
    \end{subfigure}

    \begin{subfigure}{0.49\textwidth}
        \centering
        \includegraphics[width=\linewidth]{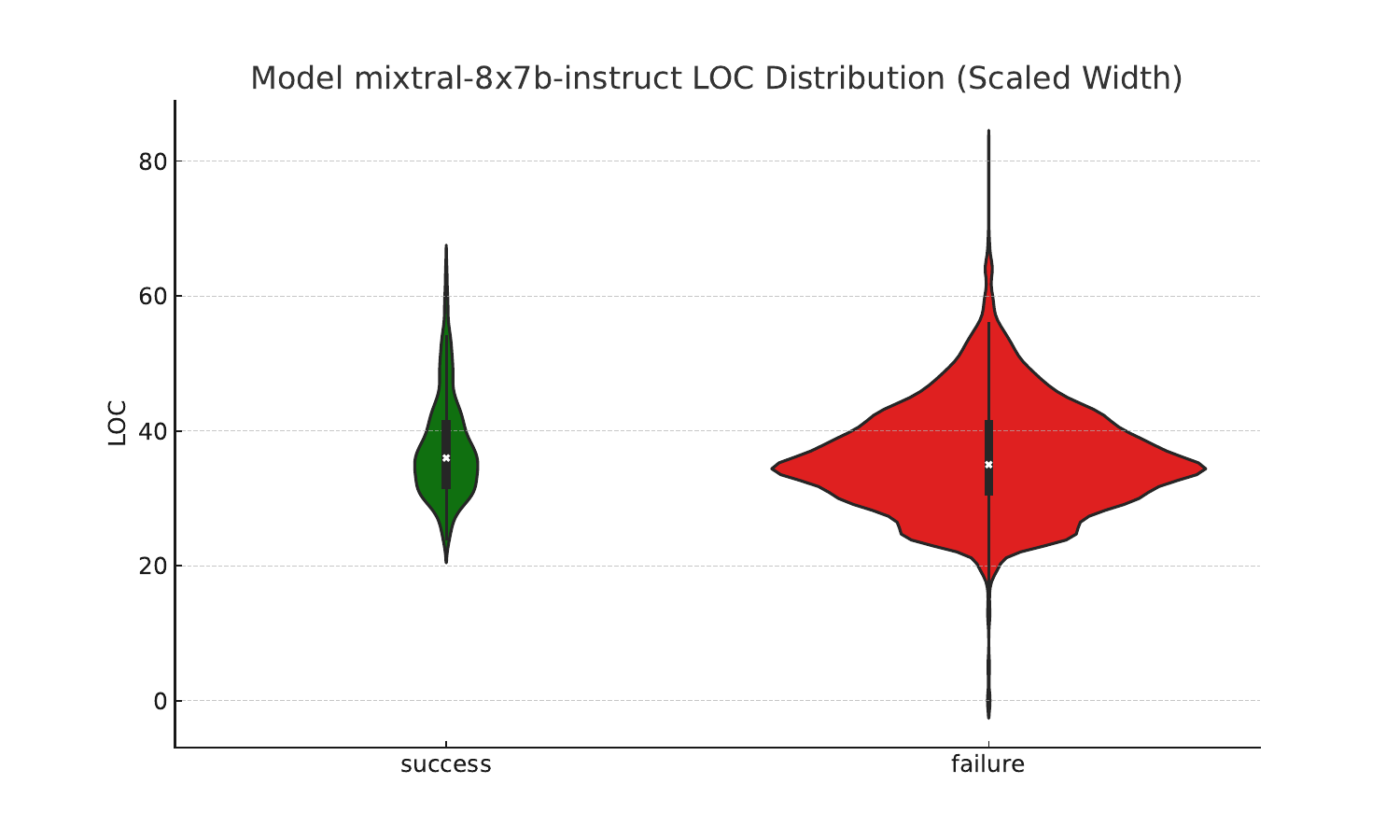}
        \caption{mixtral-8x7b-instruct (pass@1 = 0.1269)}
    \end{subfigure}
    \begin{subfigure}{0.49\textwidth}
        \centering
        \includegraphics[width=\linewidth]{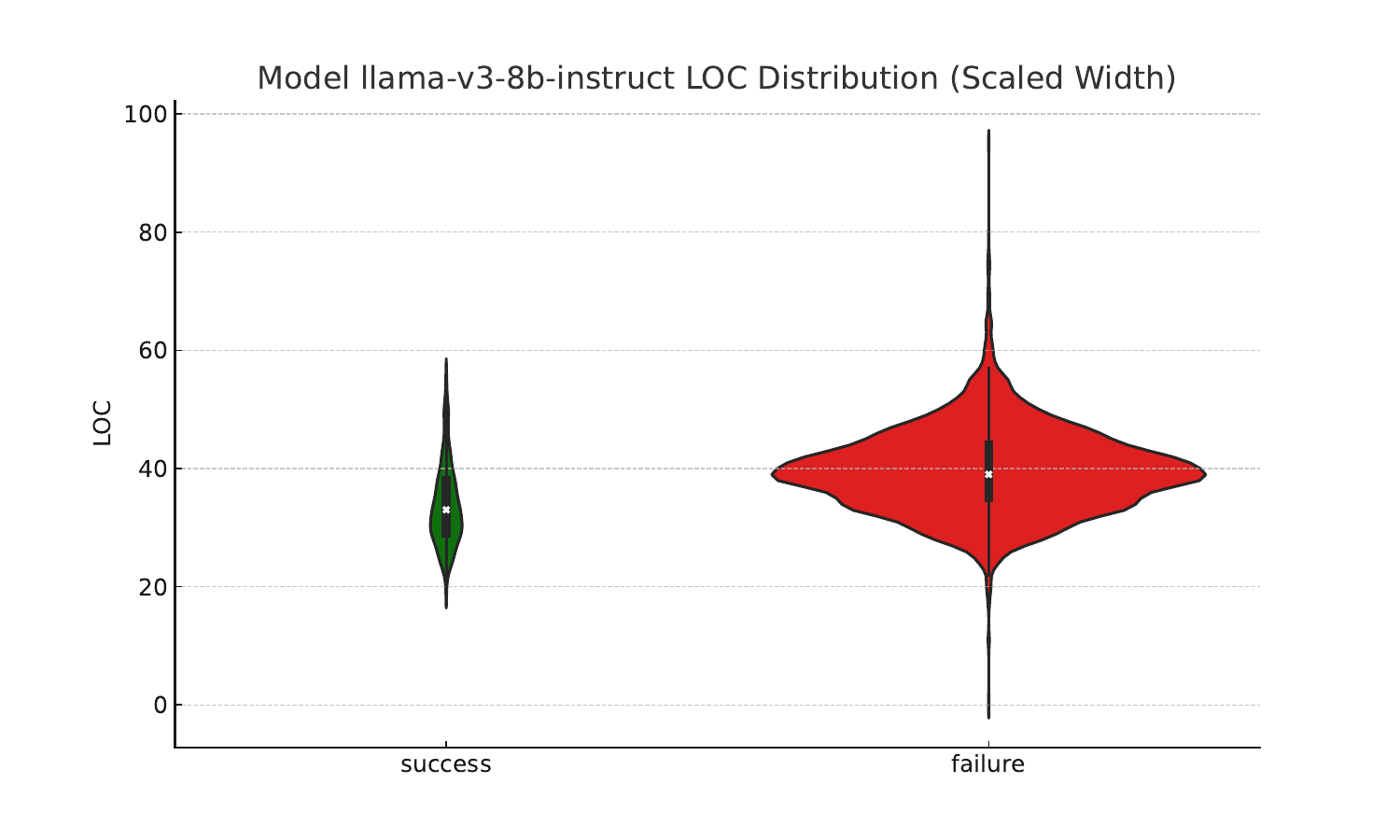}
        \caption{llama-v3-8b-instruct (pass@1 = 0.0679)}
    \end{subfigure}
    \caption{LOC Distribution by Model of Low $pass@1$: Success vs Failure}
    \label{fig:loc_fail_distribution_models}
\end{figure}

To get more insights, we search for statistical distinction between successful model outputs and failed outputs. In Fig.~\ref{fig:loc_success_distribution_models} and \ref{fig:loc_fail_distribution_models}, we visualize the LOC distribution separately for succssful outputs and failed ones, for each model. The graphs are ranked by $pass@1$, where higher $pass@1$ means bigger success sample set and smaller failure sample set. We normalize the width of each violin chart by its sample set size, hence resulting in the thinnest failure graph for the model with the highest $pass@1$. The graph gradually grows wider as the model performance degrades. The opposite pattern is observed for the success violin chart.

An important finding here is that the success distribution is always more complex than its failure counterpart, with more peaks and deviations. Fig.~\ref{fig:loc_fail_distribution_models} groups lower performing models whose failure sample set dominates the success sample set. The failure LOC distributions are unimodal, in contrast with the multimodal distributions of top models in Fig.~\ref{fig:loc_success_distribution_models}. This implies the inherent complexity involved in writing correct code even when the mean LOC is less than 50.

The success/failure LOC distributions of remaining 8 models are collected in Appendix \ref{sec:loc_successfail_distribution_models}. Also Appendix \ref{sec:loc_successfail_distribution_apps} shows LOC distributions sharded by applications.
\section{Error Analysis}\label{sec:errors}
When testing model-generated code, the failed solutions end up with error logs, one log file for each JS file. In this section, we study these logs and share our findings. 
\subsection{Error Types}
There are seven types of errors, which we code them to A through G. They are summarized in Tab.~\ref{tab:errors}.
\begin{table}[h!]
\centering
\begin{tabular}{|c|l|p{3cm}|p{5cm}|}
\hline
\textbf{Error Code} & \textbf{Name} & \textbf{Verbatim Error} & \textbf{Root Cause} \\ \hline
\cellcolor[HTML]{FDB515} A & Version Mismatch & TypeError & Deprecated framework functions are used \\ \hline
\cellcolor[HTML]{F4711E} B & Text Mismatching & TestingLibrary ElementError & Attributes or texts of HTML tags do not match test expectations \\ \hline
\cellcolor[HTML]{C51E3A}C & API Call Mismatch & expect(received) & Mock APIs are called less or more than expected \\ \hline
\cellcolor[HTML]{FF6FFF} D & Uninstalled Module & Cannot find module & Imported module is not installed \\ \hline
\cellcolor[HTML]{09D0EF} E & Invalid API Call & fetch-mock &The call signature does not match the test expectation \\ \hline
\cellcolor[HTML]{20B2AA} F & Scope Violation & ReferenceError & An out-of-scope call is made to a locally-defined function\\ \hline
\cellcolor[HTML]{33B864} G & Missing UI Element & Element type is invalid & No UI element is defined in the code \\ \hline
\end{tabular}\caption{Error Table}
\label{tab:errors}
\end{table}

The verbatim errors are the original error messages or codes captured by the log. Each of them is broadly scoped to contain a wide array of behaviors. However, in the context of our benchmark, we find all verbatim errors are projected to narrowband of behaviors attributed to the same root causes.
\begin{figure}[h!]
    \centering
    \includegraphics[width=0.6\textwidth]{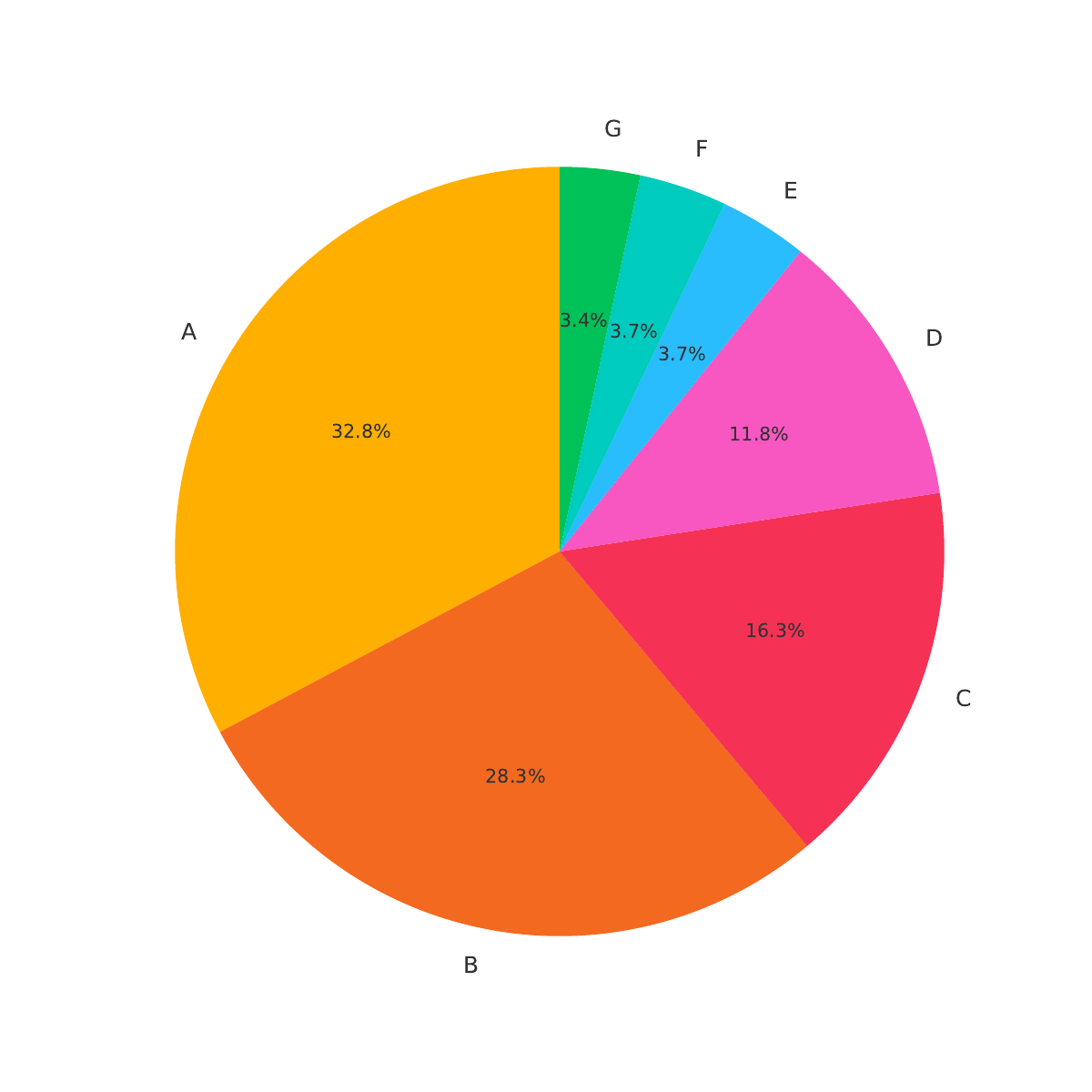}
    \caption{Error Type Distribution}
    \label{fig:errors}
\end{figure}

Fig.~\ref{fig:errors} shows the error type distribution. Note that the same error type can appear in the same log multiple times because for each benchmark challenge, the code needs to pass two unit tests, each containing multiple expectations.
\subsection{Singular and Twin Errors}
\begin{figure}[h!]
    \centering
    \includegraphics[width=0.6\textwidth]{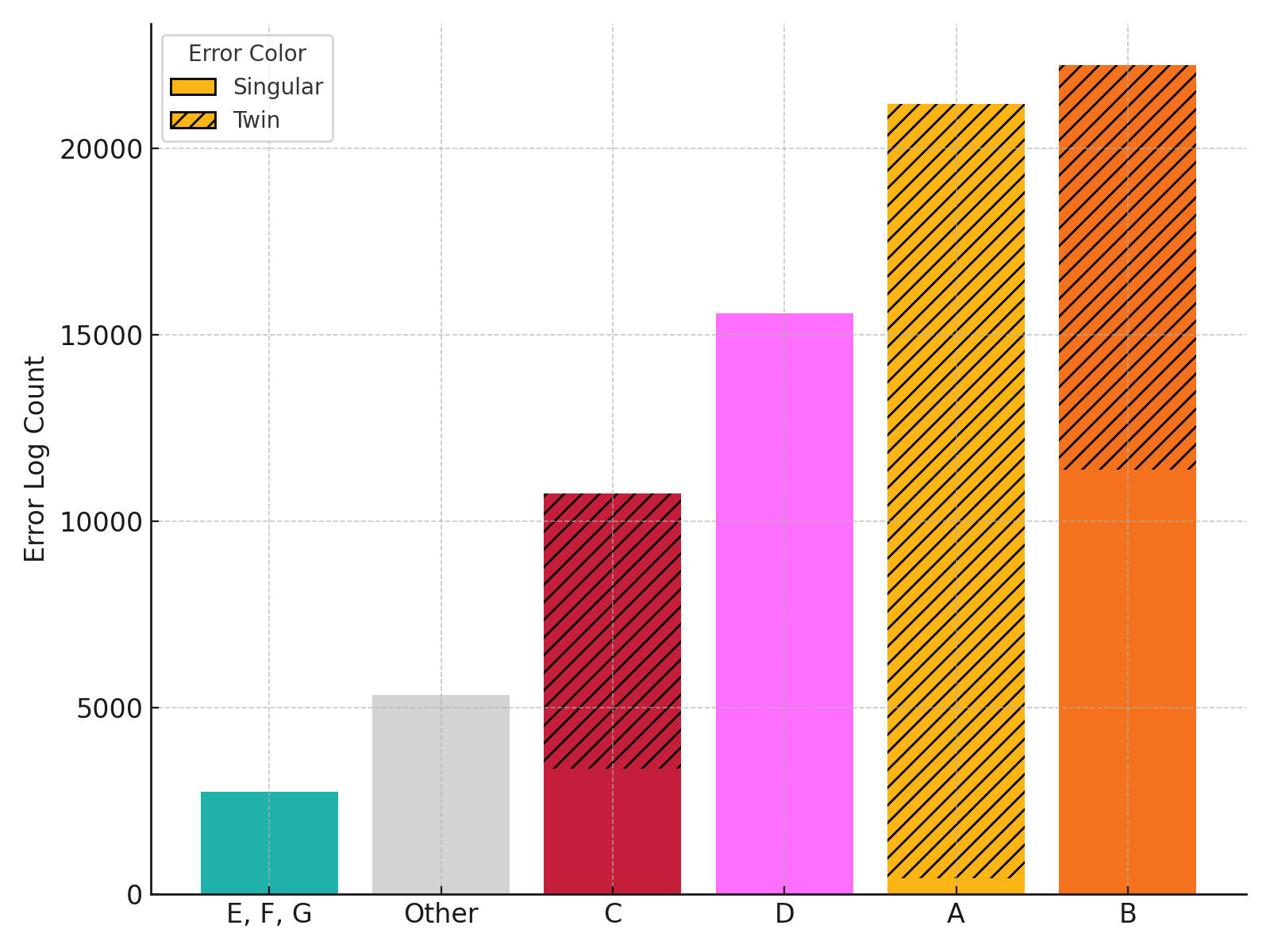}
    \caption{Distribution of Singular and Twin Errors}
    \label{fig:error_logs}
\end{figure}

An error log can contain a combination of many error types, indicating the code is poorly implemented. But this is not the dominant pattern. 93\% of error logs contain either a singular error or twin errors. Fig.~\ref{fig:error_logs} shows the distribution of singular and twin errors.

Singular error means the log contains only one error pointing to a single line. Twin errors are two errors of the same type, preeminently pointing to the same error line. Since the code needs to pass two unit tests, often times the same bug offends both tests. This means that even upon failures, all models produce quality code, but with only one bug.
\subsection{Error Distribution by Models}
\begin{figure}[h!]
    \centering
    \includegraphics[width=0.9\textwidth]{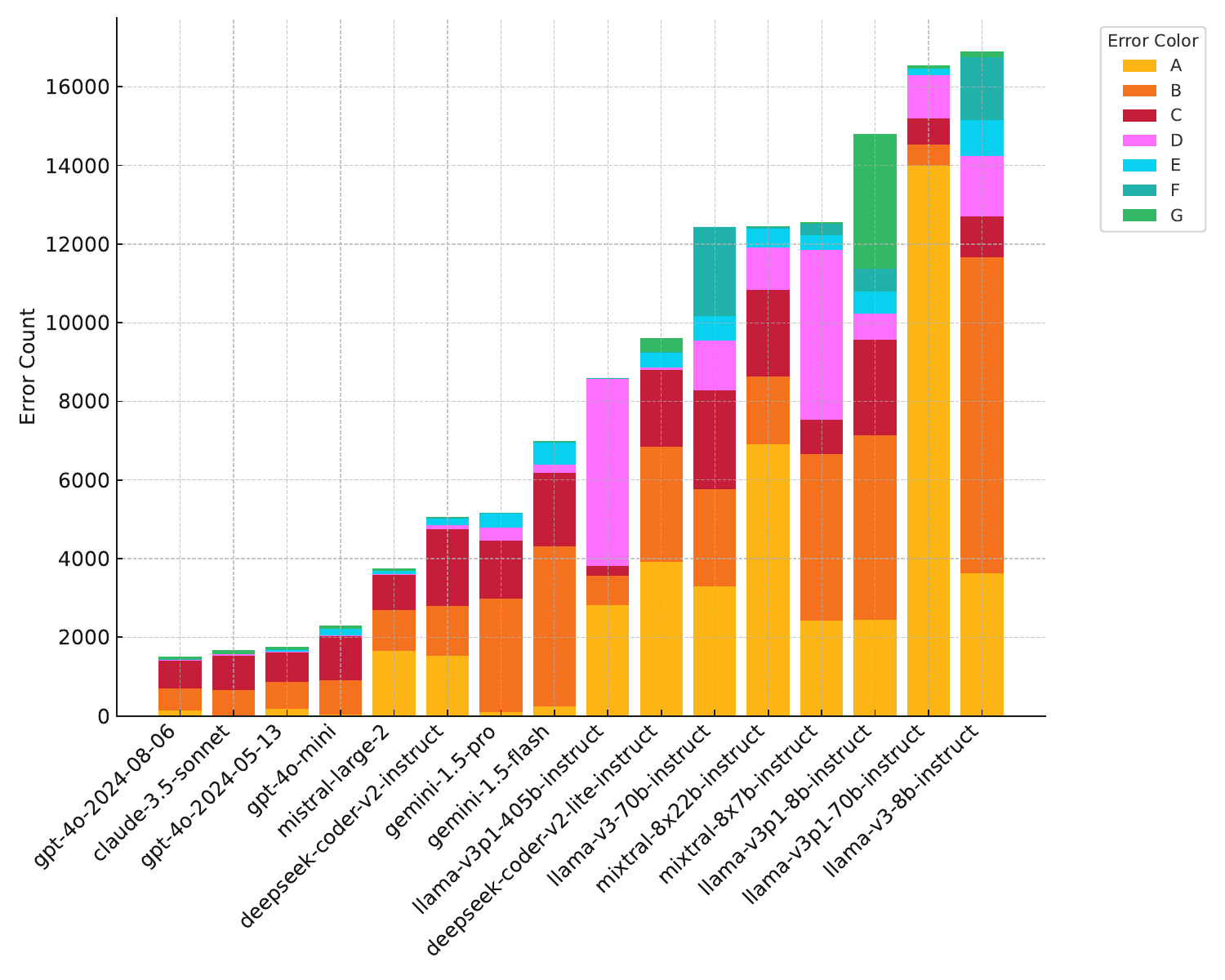}
    \caption{Error Distribution by Models}
    \label{fig:errors_models}
\end{figure}

In Fig.~\ref{fig:errors_models}, we show the error distribution separately for each model. The most important finding here is that no model is immune to any of the seven error types, even when the raw error counts differ by one order of magnitude bewteen two extremes. 

This means that all models possess the same knowledge and capabilities to write high-quality code which meets test expectations, and same inherent vulnerabilities resulting in the same types of errors. But top models distinguish themselves at lower error rates, i.e. ability to make fewer bugs.

We also study error distribution sharded by applications, whose results can be found in Appendix \ref{sec:errors_apps}.
\subsection{Targeted Prompt Optimization}
Prompt engineering is a common practice to improve model outputs by paying additional efforts to refine model inputs. After root causing each error type, we apply targeted prompt optimization to see if it can help weaker models to reduce bugs in their outputs.

Specifically, following the standard prompt below, we add a sentence reminding the model to avoid a specific type of errors.
\begin{align}
	&\text{Generate }\{file\_name\}\text{ to pass the tests below: } \label{eq:prompt} &\\
	&\{success\_test\_code\}\{failure\_test\_code\}.\text{ RETURN CODE ONLY.} \nonumber &
\end{align}

Type A is the only error type we manage to achieve significant improvement. To give more background, $useHistory$ is a commonly used framework function deprecated in React v6, 2021. As such, type A error is triggered each time $useHistory$ appears in the code. 

We choose llama-v3p1-70b-instruct because it makes the most type A errors (Fig.~\ref{fig:errors_models}) among all models. We use different prompts to emphasize to the model not to call $useHistory$. As shown in Tab.~\ref{tab:prompt_type_a}, the more specific the prompt for the model to follow, the better the error reduction.
\begin{table}[h!]
	\centering
	\begin{tabular}{|p{9cm}|p{2cm}|l|}
		\hline
		\textbf{Prompt} & \textbf{Specificity} & \textbf{Error Reduction} \\
		\hline
		USE useNavigate (React Router v6), NOT useHistory (React Router v5). & Functions & 100\% \\
		\hline
		Use React Router v6, not v5. & Packages and versions & 100\% \\
		\hline
		If you call useHistory (React Router v5), REPLACE it with useNavigate (React Router v6). & Functions & 99.93\% \\
		\hline
		Use the LATEST VERSION of React and React Router. & Packages only & 85.87\% \\
		\hline
		Use the LATEST VERSION for all packages. & Too many packages & \textbf{0\%}\\
		\hline
	\end{tabular}
	\caption{Prompts to Reduce Type A Errors (llama-v3p1-70b-instruct)}
	\label{tab:prompt_type_a}
\end{table}

Tab.~\ref{tab:prompt_type_bcd} shows prompts targeted for error types B, C, and D, as well as the model making the most errors. However, none of them can reduce errors by more than 5\% because of inherent challenges to make the prompt more specific.
\begin{table}[h!]
	\centering
	\begin{tabular}{|c|l|p{5cm}|p{3cm}|}
		\hline
		\textbf{Error Type} &\textbf{Model} & \textbf{Prompt} & \textbf{Lack of Specificity} \\
		\hline
		B & llama-v3-8b-instruct & Make sure the text or test-id of UI elements match the test code. & test-id and text are different per test. \\
		\hline
		C & llama-v3p1-8b-instruct & Trigger mocked API EXACTLY ONCE. Avoid no trigger and duplicate trigger which might be caused by useEffect. & API name is different per test. \\
		\hline
		D & llama-v3p1-8b-instruct & ONLY USE modules appeared in test files. & Too many modules to enumerate. \\
		\hline
	\end{tabular}
	\caption{Prompts to Reduce Type B, C, D Errors}
	\label{tab:prompt_type_bcd}
\end{table}

We skip experiments for error types E, F, and G, because of their relatively small sample sizes.
\section{Related Works}\label{sec:related}
The development and evaluation of large language models (LLMs) for code generation have been an area of significant research interest in recent years.
\subsection{Benchmarks}
Benchmarks are essential for evaluating the effectiveness and generalizability of models across various software engineering tasks. CodeSearchNet\citep{codesearchnet} is a benchmark to evaluate semantic code search performance. HumanEvalPack\citep{humanevalpack} assesses the ability of models like Codex to generate correct Python code from natural language prompts. Defects4J\citep{defects4j} is widely used to assess LLMs’ ability to handle bug localization and repair in Java. XLCoST\citep{xlcost} evaluates how well LLMs can work across different programming languages. BugsJS\citep{bugsjs} collects real-world JavaScript bugs and is used to evaluate LLM ability to detect and fix bugs in web applications. ClassEval\citep{classeval} evaluates class-level code generation. 
\subsection{Error Analysis}
Error analysis is a critical area of research that focuses on understanding and improving the LLM weaknesses on software tasks. BugAID\citep{bugaid} is a system to discover JavaScript bug patterns in JavaScript and web applications. DeepFix\citep{deepfix} is a deep learning-based system repairing errors in C programs. TSSB-3M\citep{tssb3m} is a large dataset of single-statement bugs across multiple languages. ManySStuBs4J\citep{manysstubs4j} is a dataset of bug-fix pairs commonly used to train models to detect subtle errors in Java.
\subsection{Prompt Engineering}
Quite a few studies focus on prompt engineering to improve LLM performance on coding tasks. In the era of GPT-3\citep{gpt3} and Codex\citep{codex}, prompt engineering has been used for code translation tasks. Chain-of-Code\citep{chainofcode} expands on Chain-of-Thought by way of pseudocode. DotPrompts\citep{dotprompts} leverages prompts for code summarization. APE (Automatic Prompt Engineer)\citep{ape} automates prompt creation by exploring different configurations to LLM code repair performance.
\subsection{Code Complexity}
Code complexity research focus on understanding how models handles complex code and generates efficient code. CoCoNut\citep{coconut} is a syntax-guided neural machine translation system for automatic program repair.
AST-T5\citep{ast-t5} incorporates Abstract Syntax Trees (AST) into T5 to understand the structure of complex code. InCoder\citep{incoder} is a model designed for code generation and infilling tasks.
\subsection{Error Reduction}
Error reduction strategies focus on minimizing the number of mistakes made by models during code generation, completion, and repair tasks. CYCLE\citep{cycle} is a self-refining model designed to reduce error rates by iterating outputs through code evaluation. CodeRL\citep{coderl} use reinforcement learning to reduces syntax errors via immediate feedback at training time. AlphaRepair\citep{
alpharepair} incorporates static analysis feedback into zero-shot learning. \section{Conclusions and Future Works}\label{sec:conclude}
In this report, we study WebApp1K results on 16 frontier LLMs, particularly failure rates, LOC distributions, and error types. Here are some tentative insights.
\begin{enumerate}
\item A failed solution is often one bug away from a correct one. This suggests that all models possess the necessary knowledge and capabilities, but mistake minimization is the key differentiator between top and weak models.
\item Success code outputs exhibit more complex patterns (LOC distribution) than failed code outputs, implifying more factors influencing the model output.
\item Prompt optimization is only effective when errors can be described (and hence avoided) in an exact and specific way.
\end{enumerate}

We hope these insights are useful to the LLM community, especially model trainers. Below are some future tasks.
\begin{enumerate}
\item We will make the benchmark more challenging, forcing LLMs to write more lines of code to cover more scenarios.
\item We will incorporate more frameworks (e.g. Vue) and languages (e.g. Python) to increase the benchmark coverage.
\item We will continue to explore and evaluate new prompting techniques since they are crucial to LLM practitioners.
\end{enumerate}
\appendix
\section{Appendix: Benchmark Difficulty per Application}\label{sec:difficulty_apps}
\begin{figure}[htbp]
    \centering
    \begin{subfigure}[b]{0.3\textwidth}
        \includegraphics[width=\textwidth]{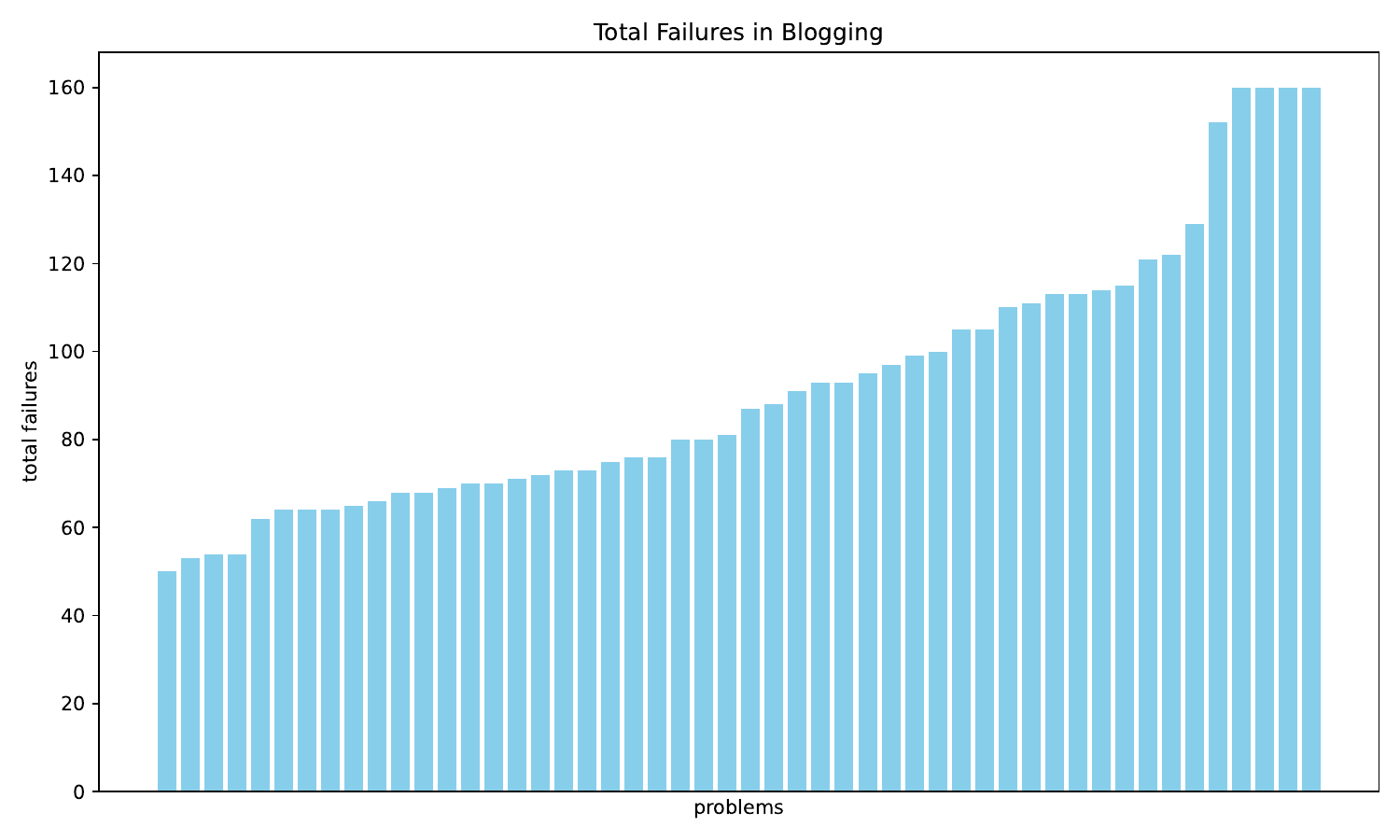}
        \caption{Blogging}
    \end{subfigure}
    \begin{subfigure}[b]{0.3\textwidth}
        \includegraphics[width=\textwidth]{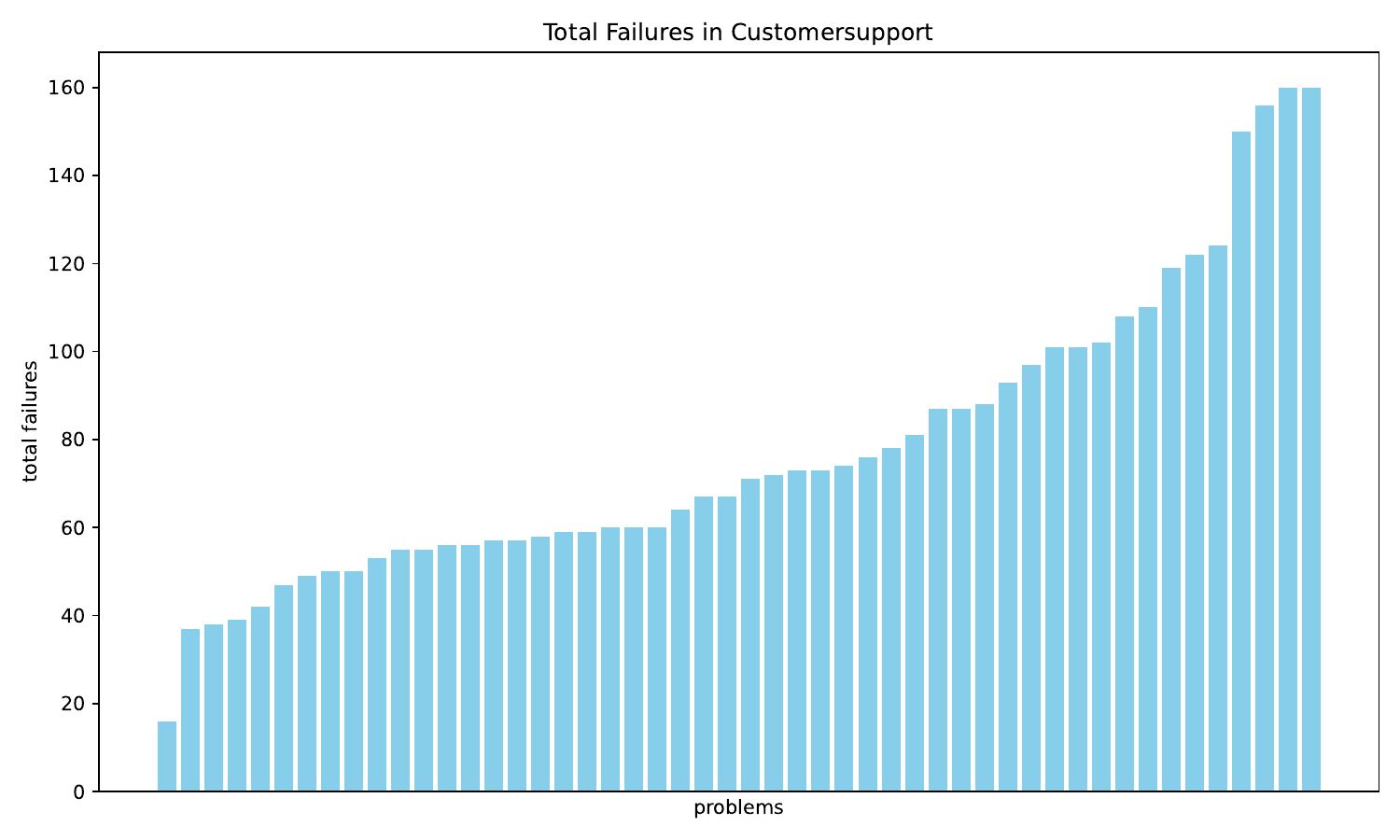}
        \caption{Customer Support}
    \end{subfigure}
    \begin{subfigure}[b]{0.3\textwidth}
        \includegraphics[width=\textwidth]{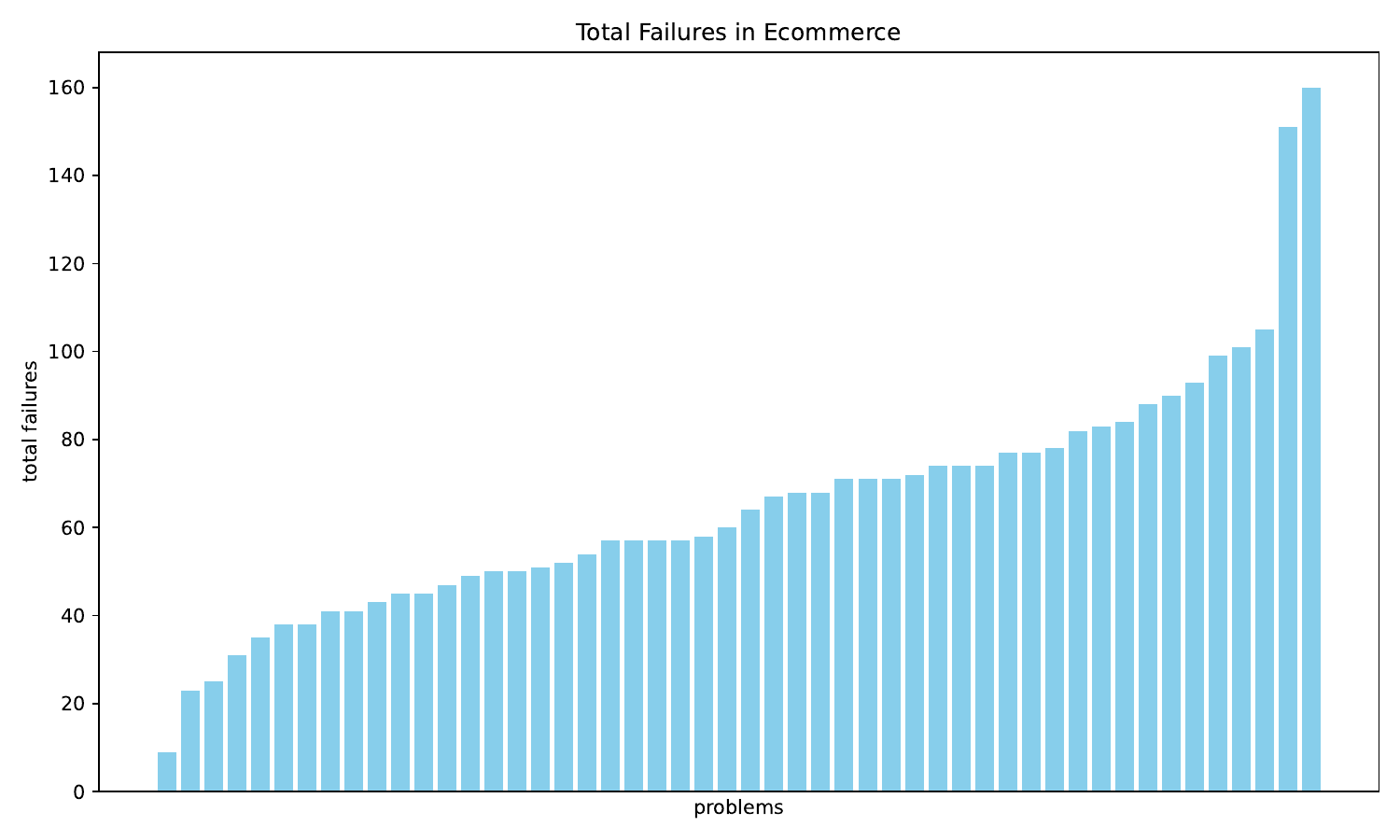}
        \caption{E-commerce}
    \end{subfigure}

    \begin{subfigure}[b]{0.3\textwidth}
        \includegraphics[width=\textwidth]{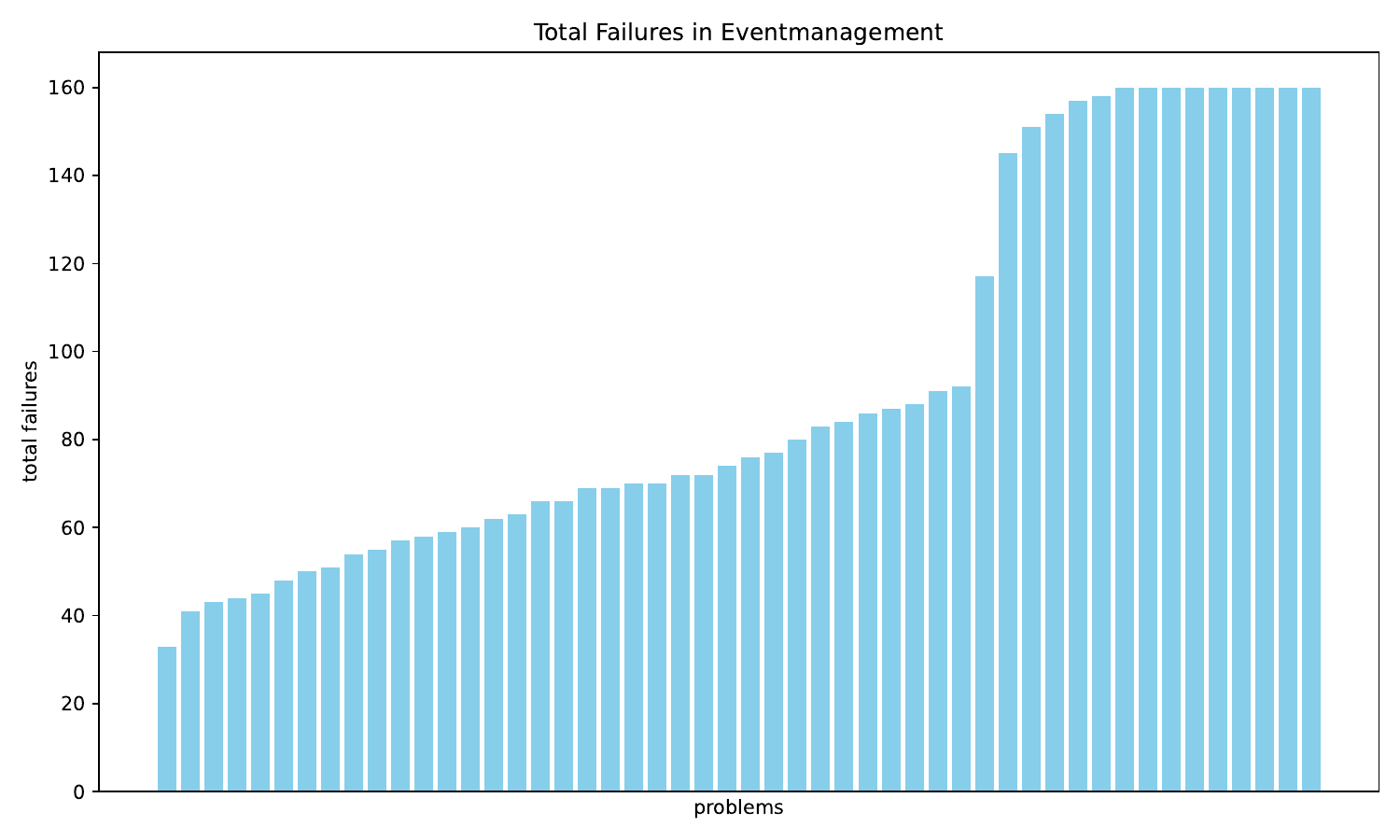}
        \caption{Event Management}
    \end{subfigure}
    \begin{subfigure}[b]{0.3\textwidth}
        \includegraphics[width=\textwidth]{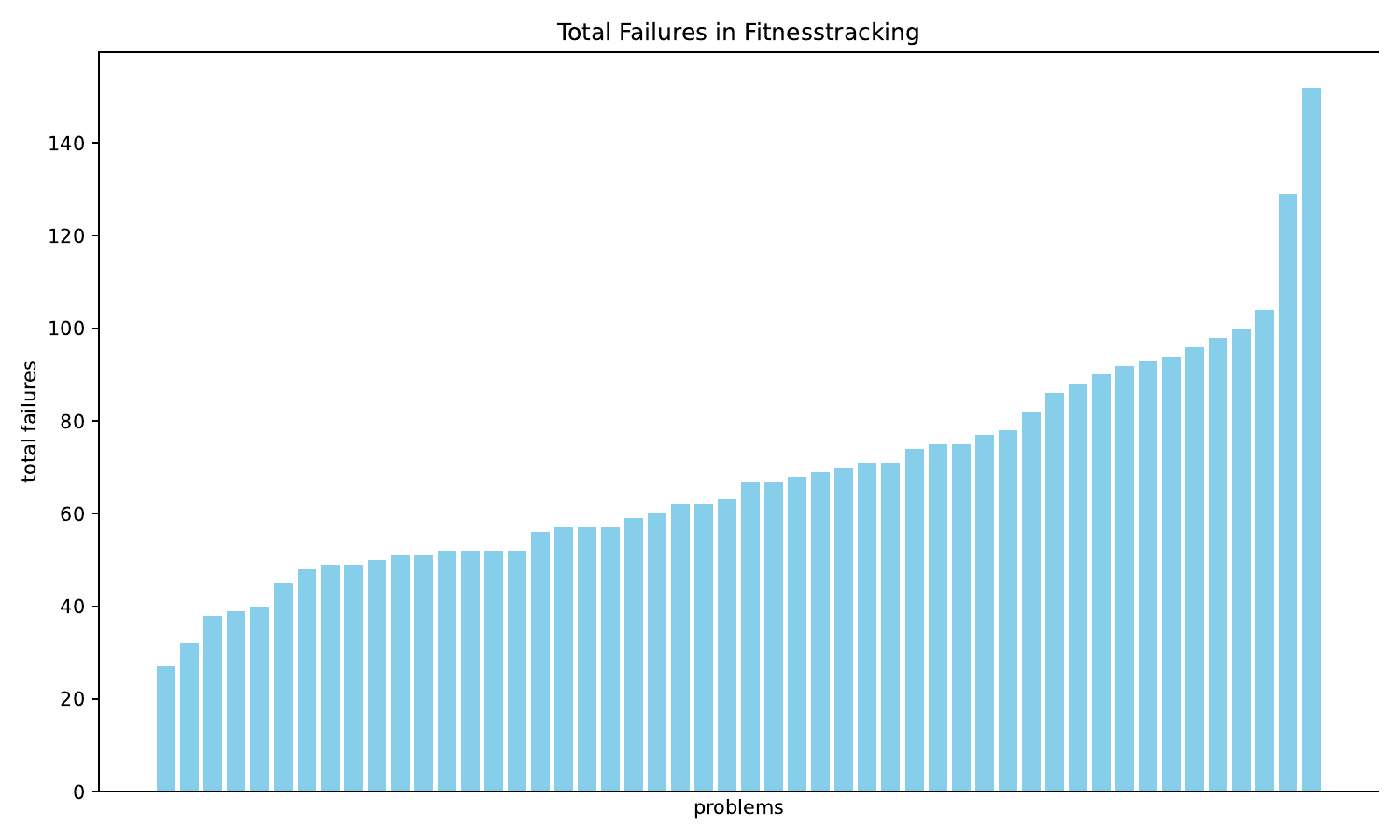}
        \caption{Fitness Tracking}
    \end{subfigure}
    \begin{subfigure}[b]{0.3\textwidth}
        \includegraphics[width=\textwidth]{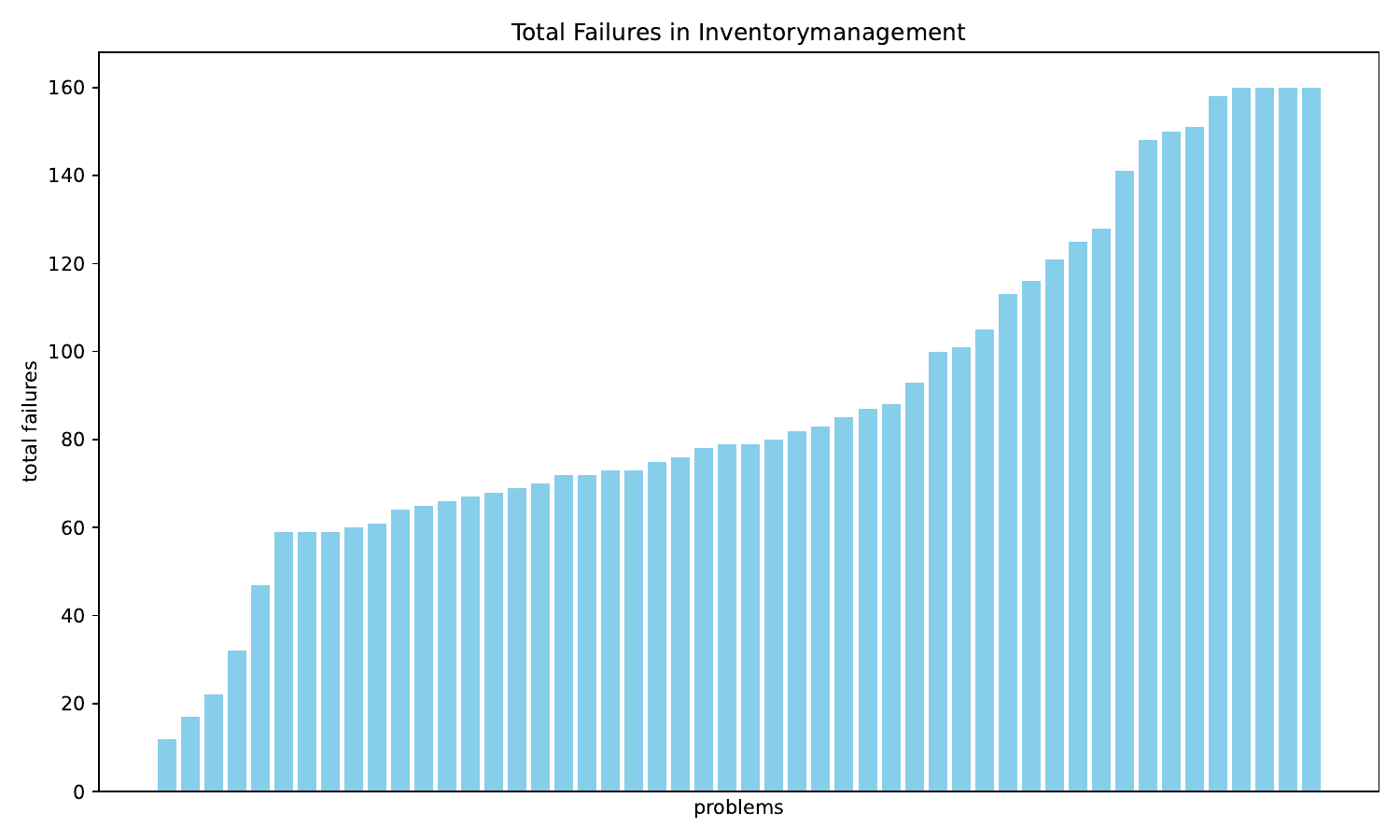}
        \caption{Inventory Management}
    \end{subfigure}

    \begin{subfigure}[b]{0.3\textwidth}
        \includegraphics[width=\textwidth]{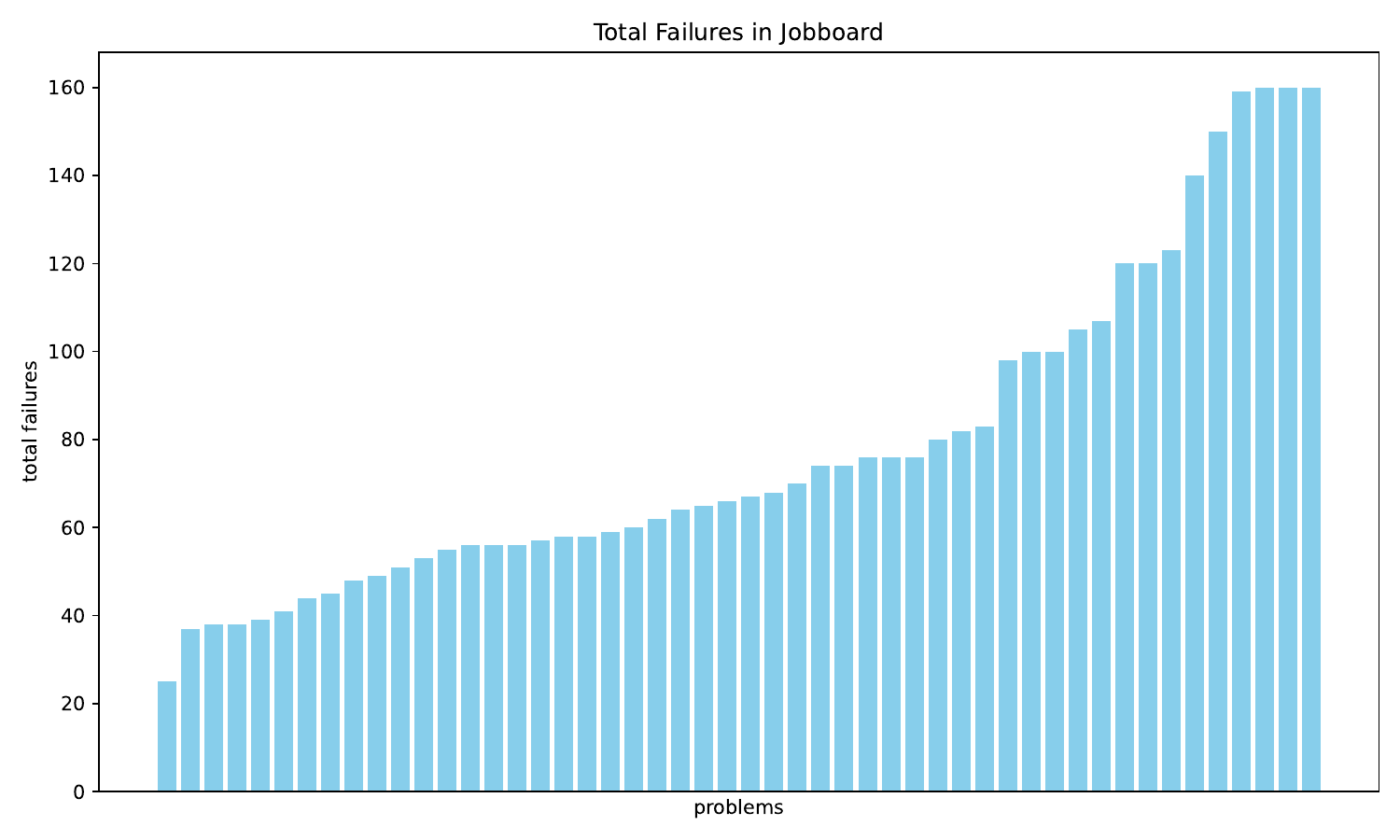}
        \caption{Job Board}
    \end{subfigure}
    \begin{subfigure}[b]{0.3\textwidth}
        \includegraphics[width=\textwidth]{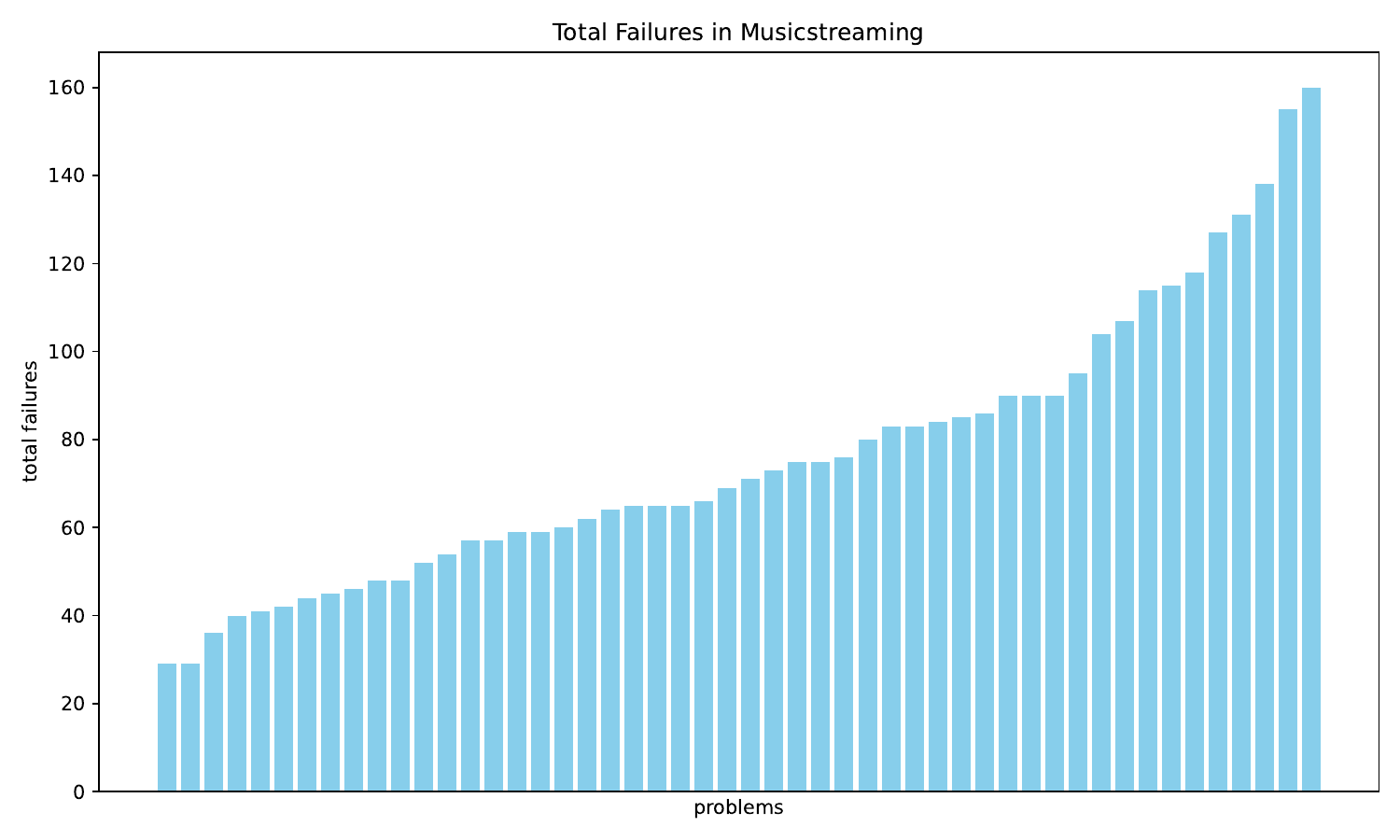}
        \caption{Music Streaming}
    \end{subfigure}
    \begin{subfigure}[b]{0.3\textwidth}
        \includegraphics[width=\textwidth]{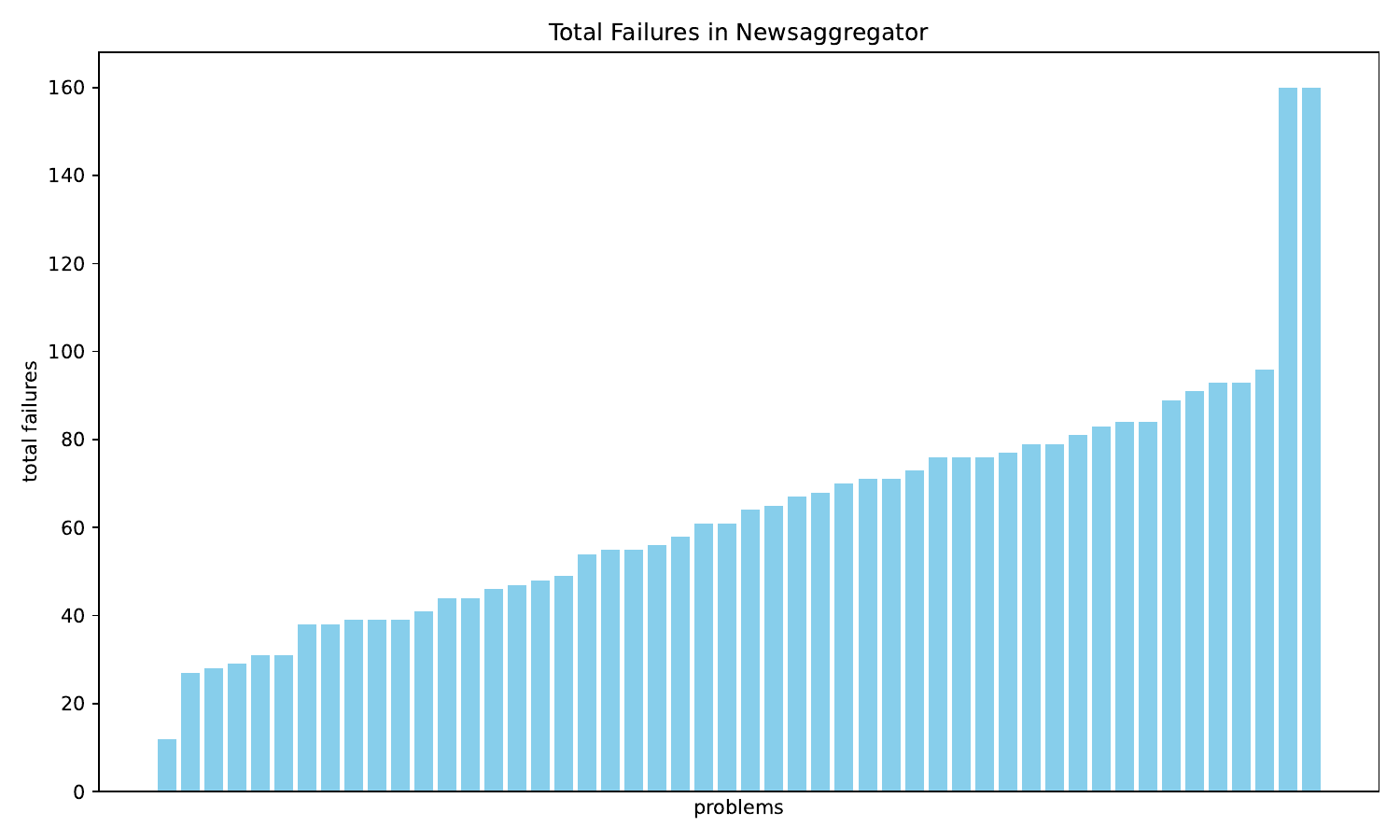}
        \caption{News Aggregator}
    \end{subfigure}

    \begin{subfigure}[b]{0.3\textwidth}
        \includegraphics[width=\textwidth]{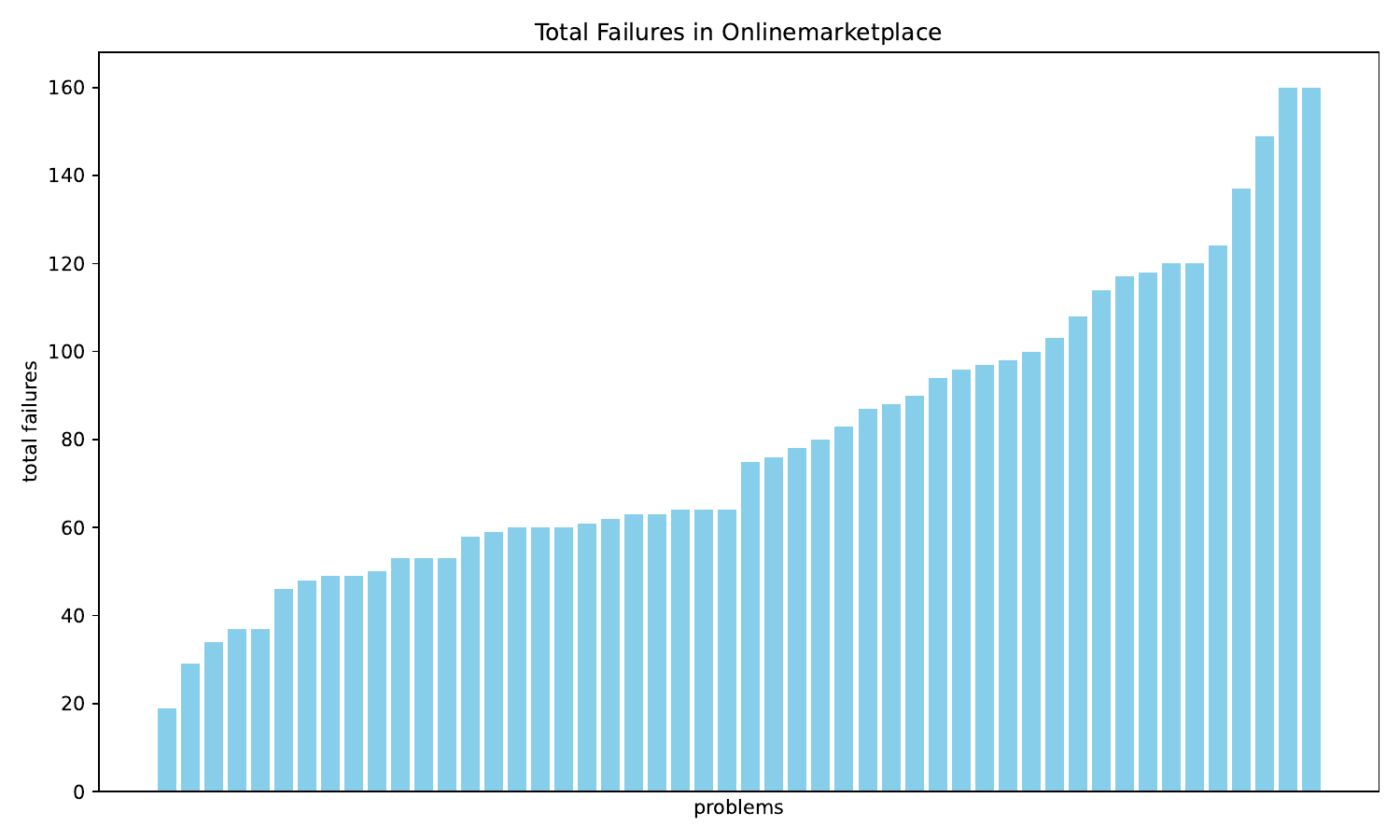}
        \caption{Online Marketplace}
    \end{subfigure}
    \begin{subfigure}[b]{0.3\textwidth}
        \includegraphics[width=\textwidth]{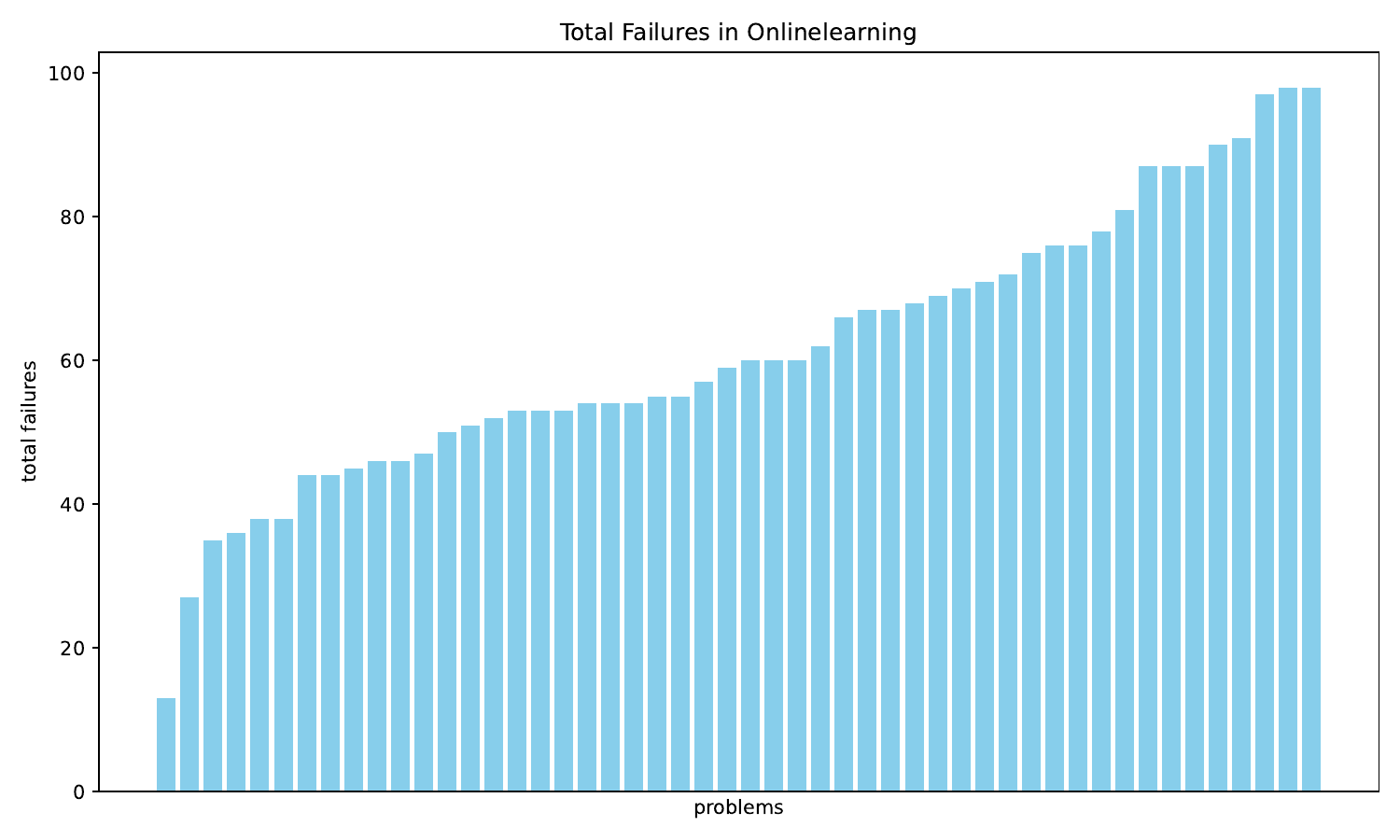}
        \caption{Online Learning}
    \end{subfigure}
    \begin{subfigure}[b]{0.3\textwidth}
        \includegraphics[width=\textwidth]{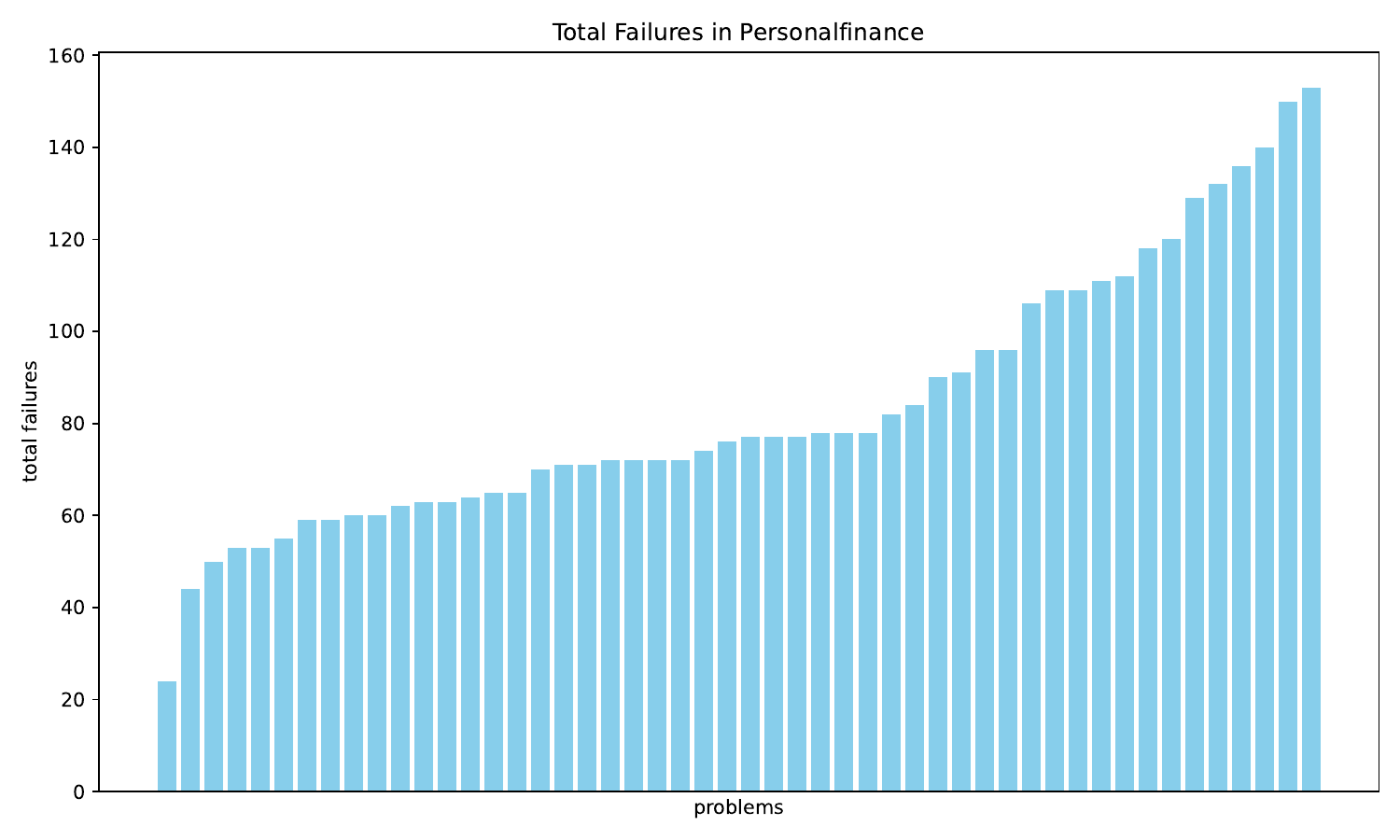}
        \caption{Personal Finance}
    \end{subfigure}

    \begin{subfigure}[b]{0.3\textwidth}
        \includegraphics[width=\textwidth]{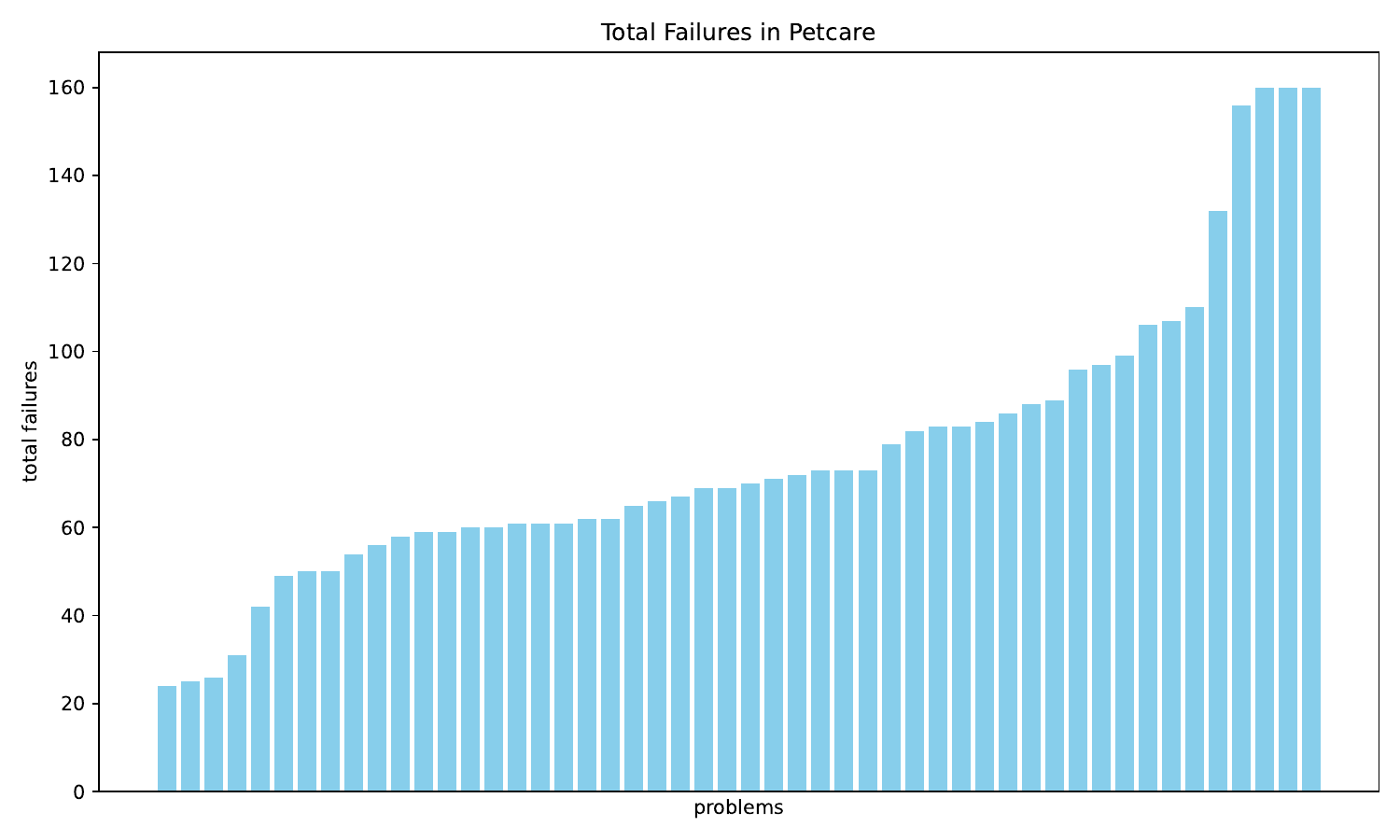}
        \caption{Pet Care}
    \end{subfigure}
    \begin{subfigure}[b]{0.3\textwidth}
        \includegraphics[width=\textwidth]{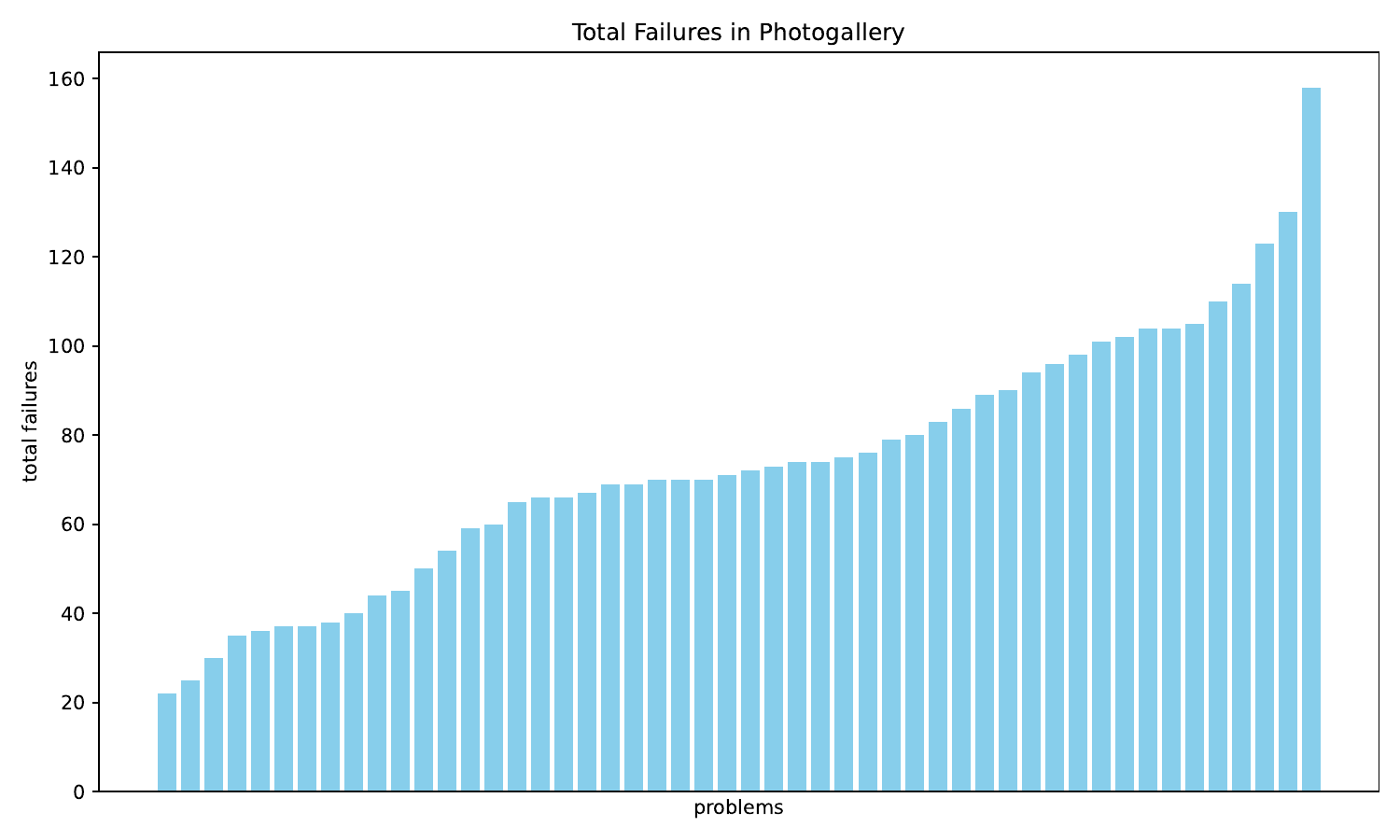}
        \caption{Photo Gallery}
    \end{subfigure}
    \begin{subfigure}[b]{0.3\textwidth}
        \includegraphics[width=\textwidth]{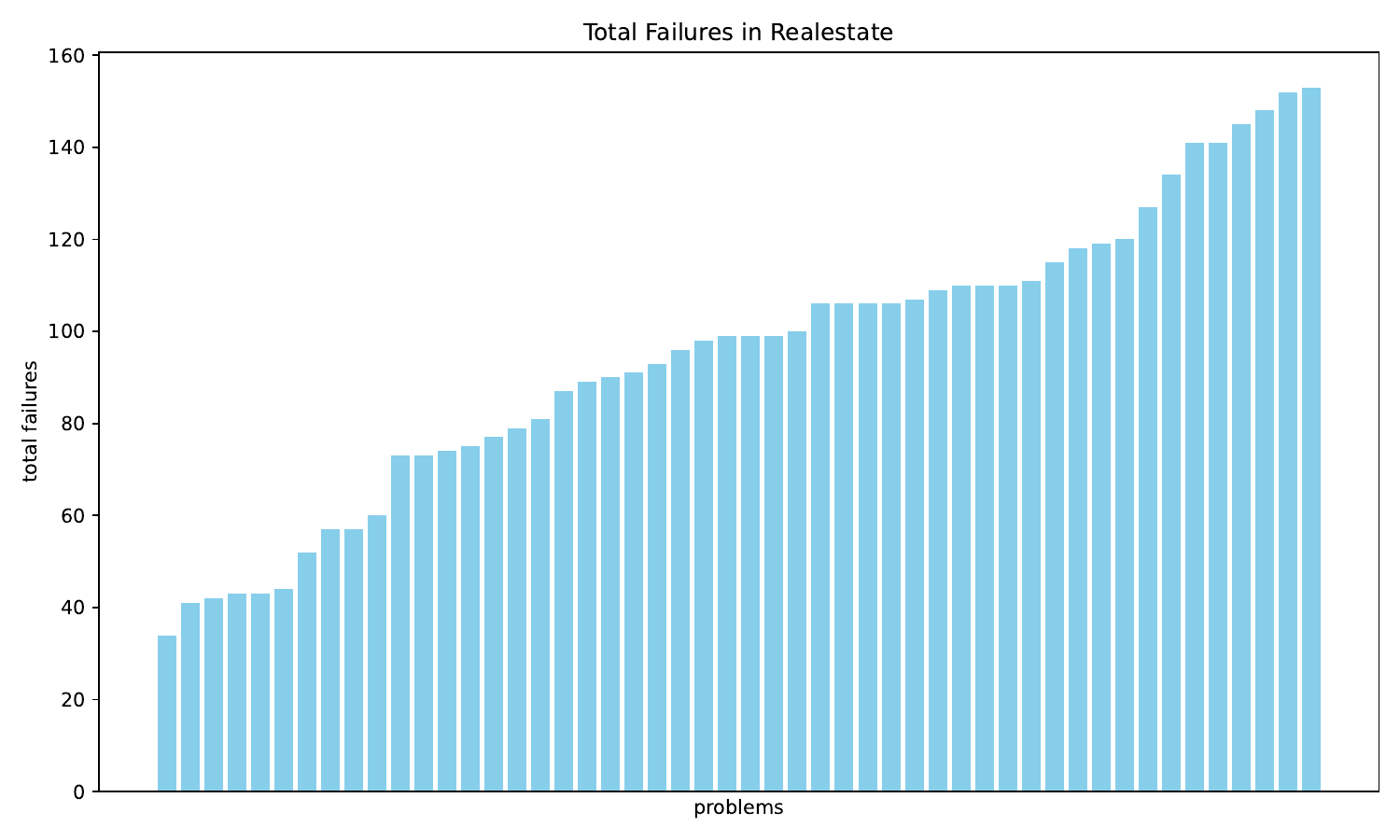}
        \caption{Real Estate}
    \end{subfigure}

    \begin{subfigure}[b]{0.3\textwidth}
        \includegraphics[width=\textwidth]{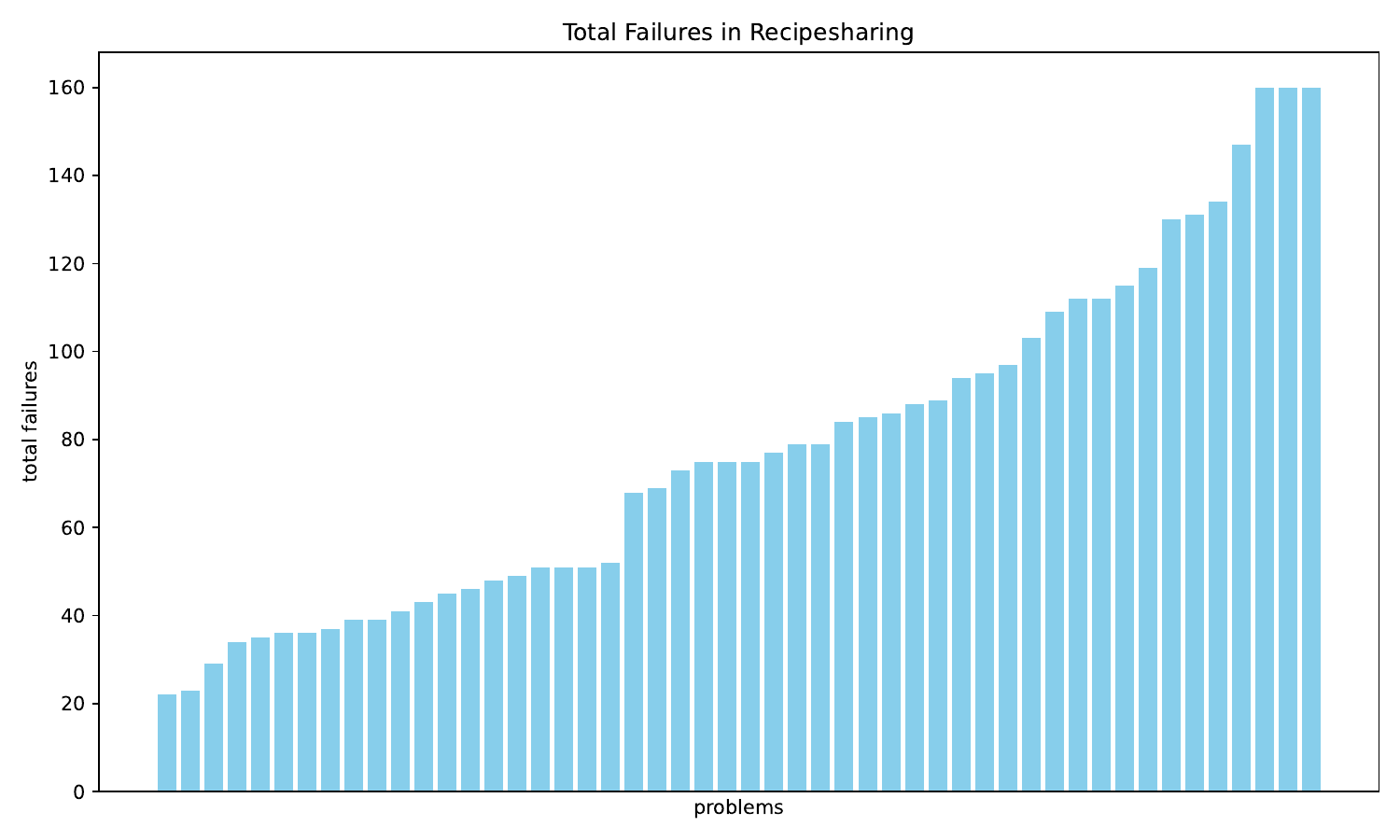}
        \caption{Recipe Sharing}
    \end{subfigure}
    \begin{subfigure}[b]{0.3\textwidth}
        \includegraphics[width=\textwidth]{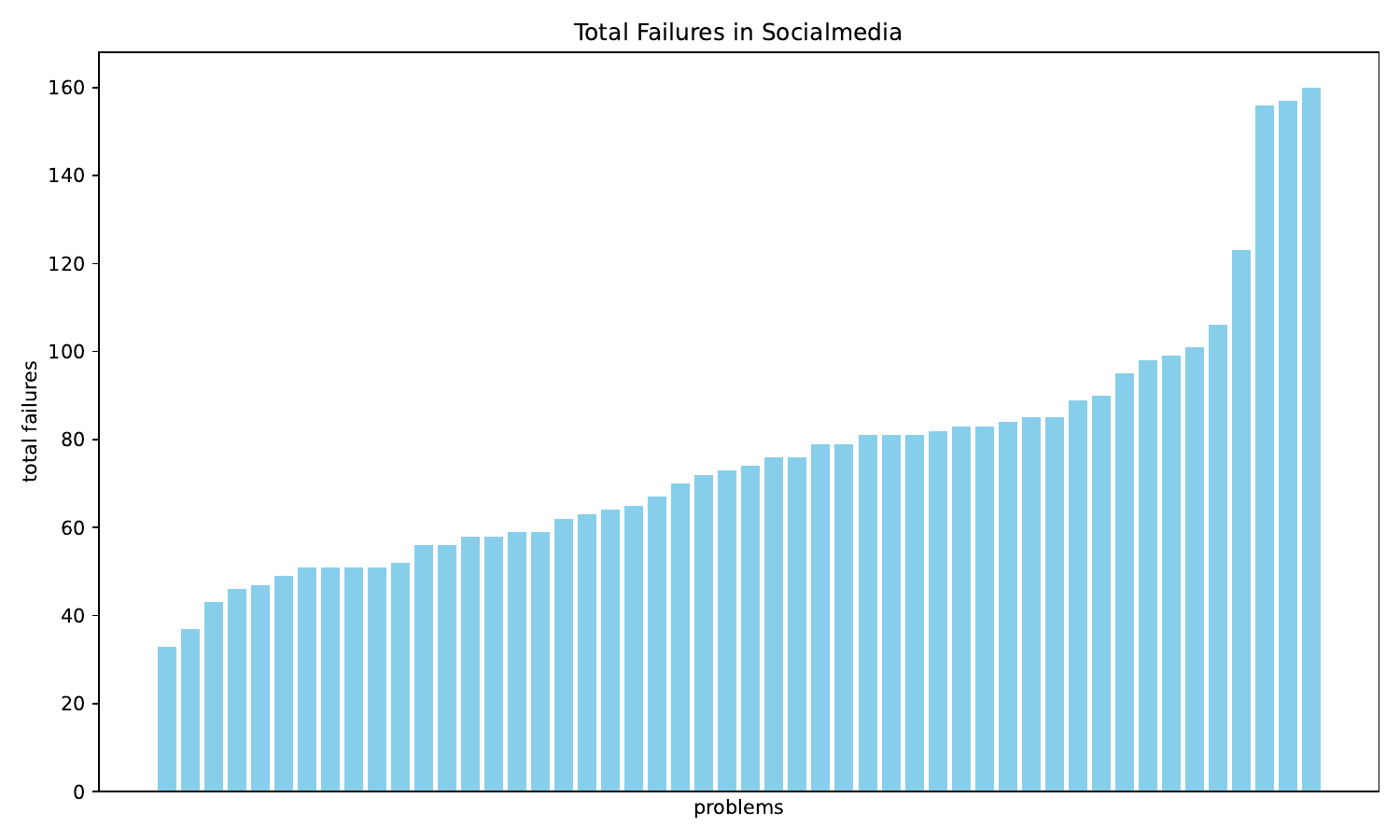}
        \caption{Social Media}
    \end{subfigure}
    \begin{subfigure}[b]{0.3\textwidth}
        \includegraphics[width=\textwidth]{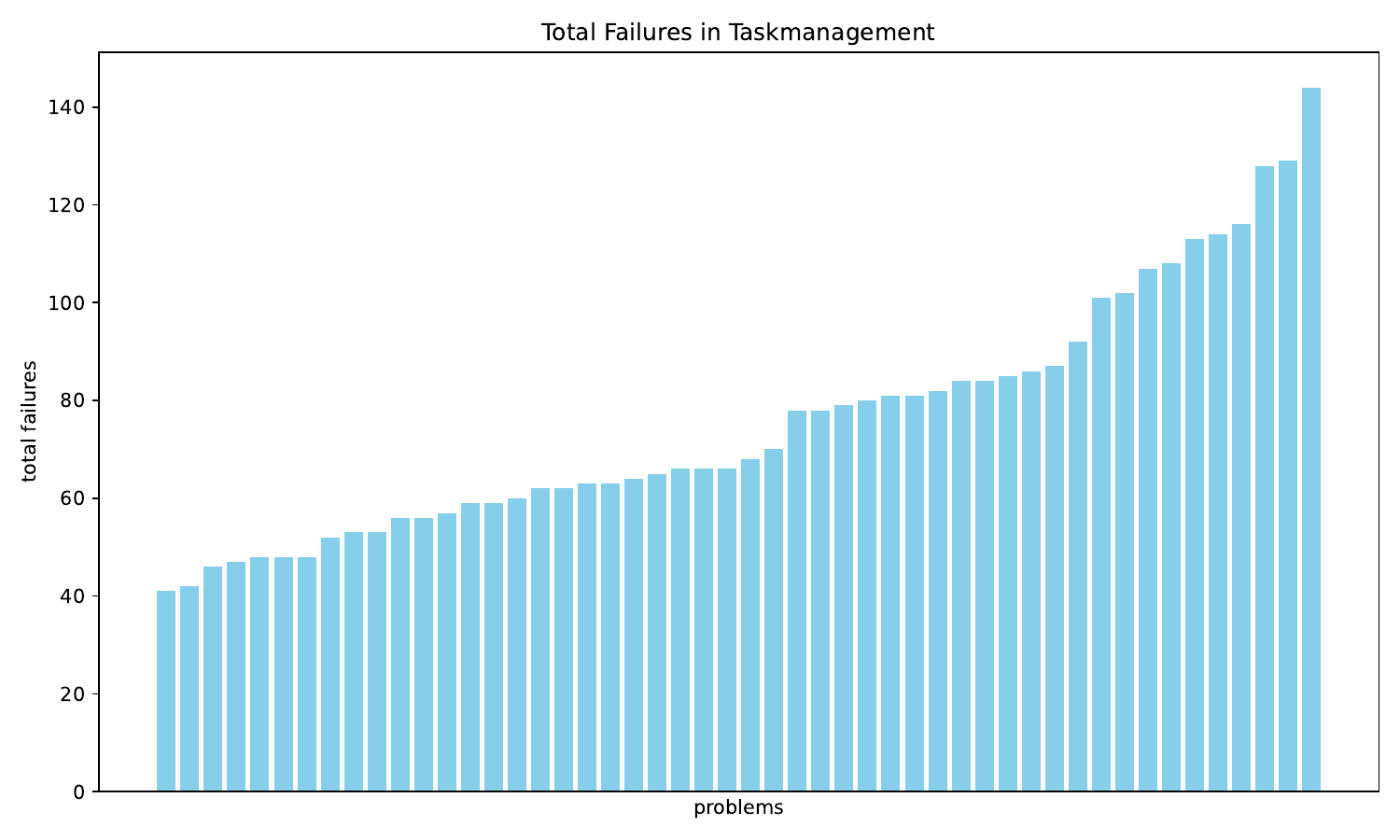}
        \caption{Task Management}
    \end{subfigure}

    \begin{subfigure}[b]{0.3\textwidth}
        \includegraphics[width=\textwidth]{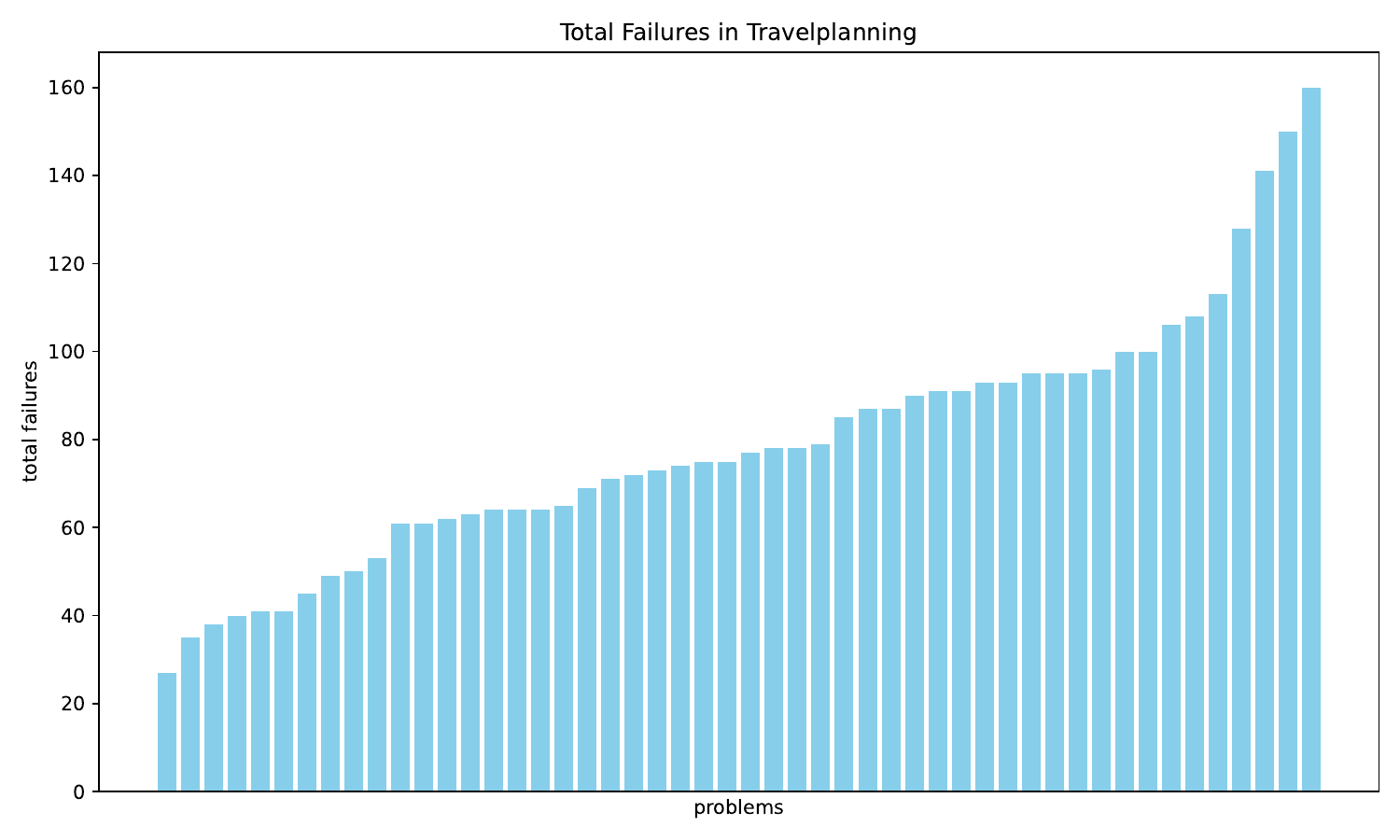}
        \caption{Travel Planning}
    \end{subfigure}
    \begin{subfigure}[b]{0.3\textwidth}
        \includegraphics[width=\textwidth]{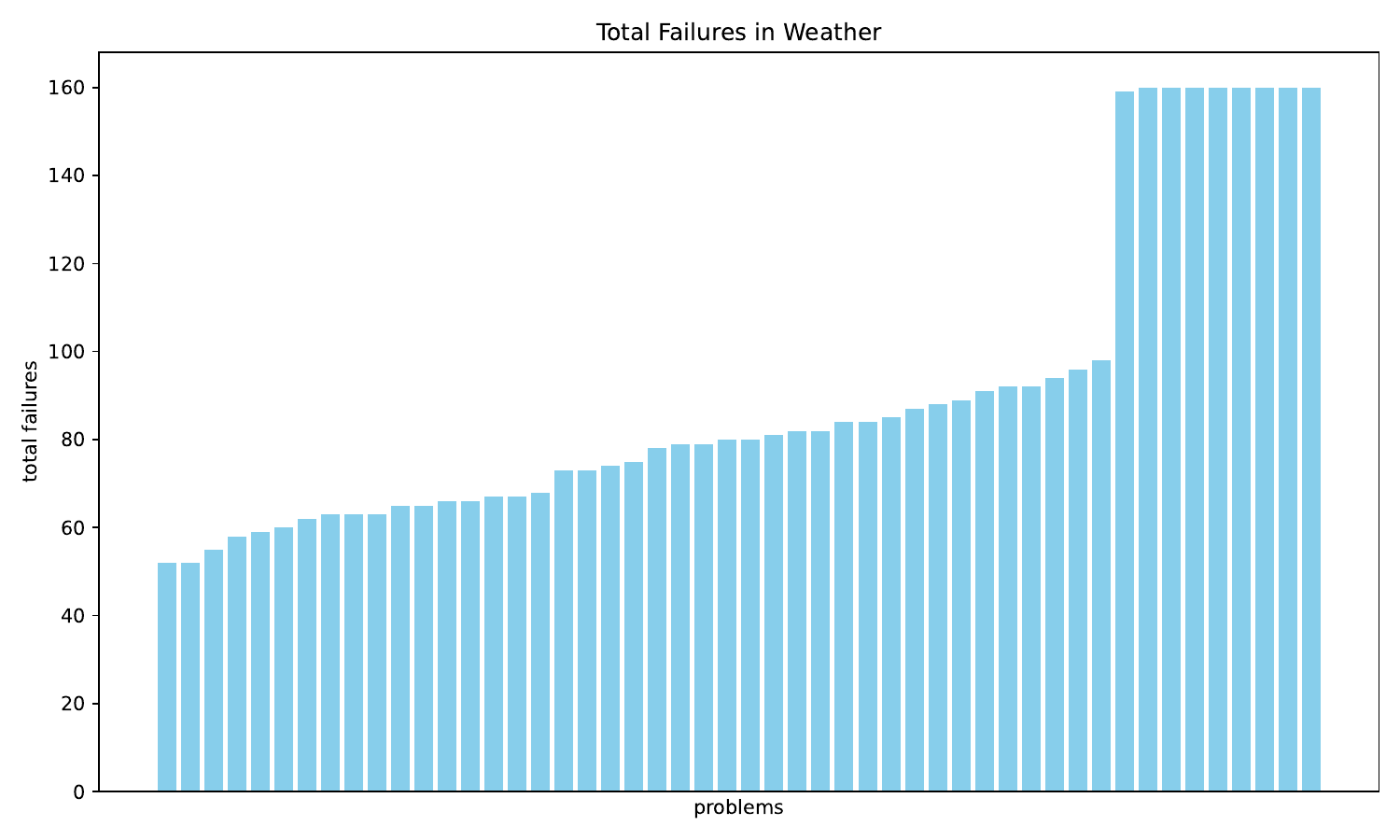}
        \caption{Weather}
    \end{subfigure}
 
    \caption{Failures per Problem by Application}
    \label{fig:failures_apps}
\end{figure}

Fig.~\ref{fig:failures_apps} shows the failure pattern broken down by applications. 
\begin{enumerate}
\item\textit{Consistency Across Applications}: All applications exhibit the same general shape—a large concentration of easier problems on the left side and a few harder problems on the right side. This consistency suggests that across different domains, there are always a few particularly challenging problems that models struggle with.

\item\textit{Variations in Skewness}: Some applications, such as Fitness Tracking and Music Streaming, show a more pronounced skew with a sharp rise in failure rates for a few problems, indicating a steeper difficulty curve. Others have a more gradual increase, indicating a more even distribution of problem difficulty.

\item\textit{Extreme Difficulty in Certain Applications}: Applications like Customer Support and Pet Care have a sharper increase towards the right, implying that these domains have a subset of problems that are especially challenging.

\item\textit{Easier Applications}: In applications like Weather and Photo Gallery, the overall number of failures seems lower compared to other appli cations, suggesting that the problems in these areas were generally easier.
\end{enumerate}
\section{Appendix: LOC Distribution by Applications}\label{sec:loc_distribution_apps}
In Tab.~\ref{tab:loc_by_apps}, we rank median LOC for each application. Consistent with the case for model ranking (Tab.~\ref{tab:loc_by_models}), the median values stay within a narrow range (37 to 46). This suggests that all models consistently produce solutions of similar length, irrespective of the task complexity or domain.

\begin{table}[h!]
\centering
\begin{tabular}{rlr}
\toprule
Application          & Mean LOC \\
\midrule
News Aggregator       & 37    \\
Music Streaming       & 37    \\
Online Marketplace    & 37    \\
E-commerce            & 37    \\
Recipe Sharing        & 38    \\
Fitness Tracking      & 38    \\
Online Learning       & 38    \\
Blogging              & 39    \\
Weather               & 40    \\
Real Estate           & 42    \\
Social Media          & 42    \\
Job Board             & 42    \\
Inventory Management  & 42    \\
Pet Care              & 42    \\
Travel Planning       & 42    \\
Personal Finance      & 43    \\
Customer Support      & 44    \\
Photo Gallery         & 44    \\
Event Management      & 45    \\
Task Management       & 46    \\
\bottomrule
\end{tabular}
\caption{Applications Ranked by Mean LOC}
\label{tab:loc_by_apps}
\end{table}

\begin{figure}[h]
    \centering
    \begin{subfigure}{0.3\textwidth}
        \centering
        \includegraphics[width=\textwidth]{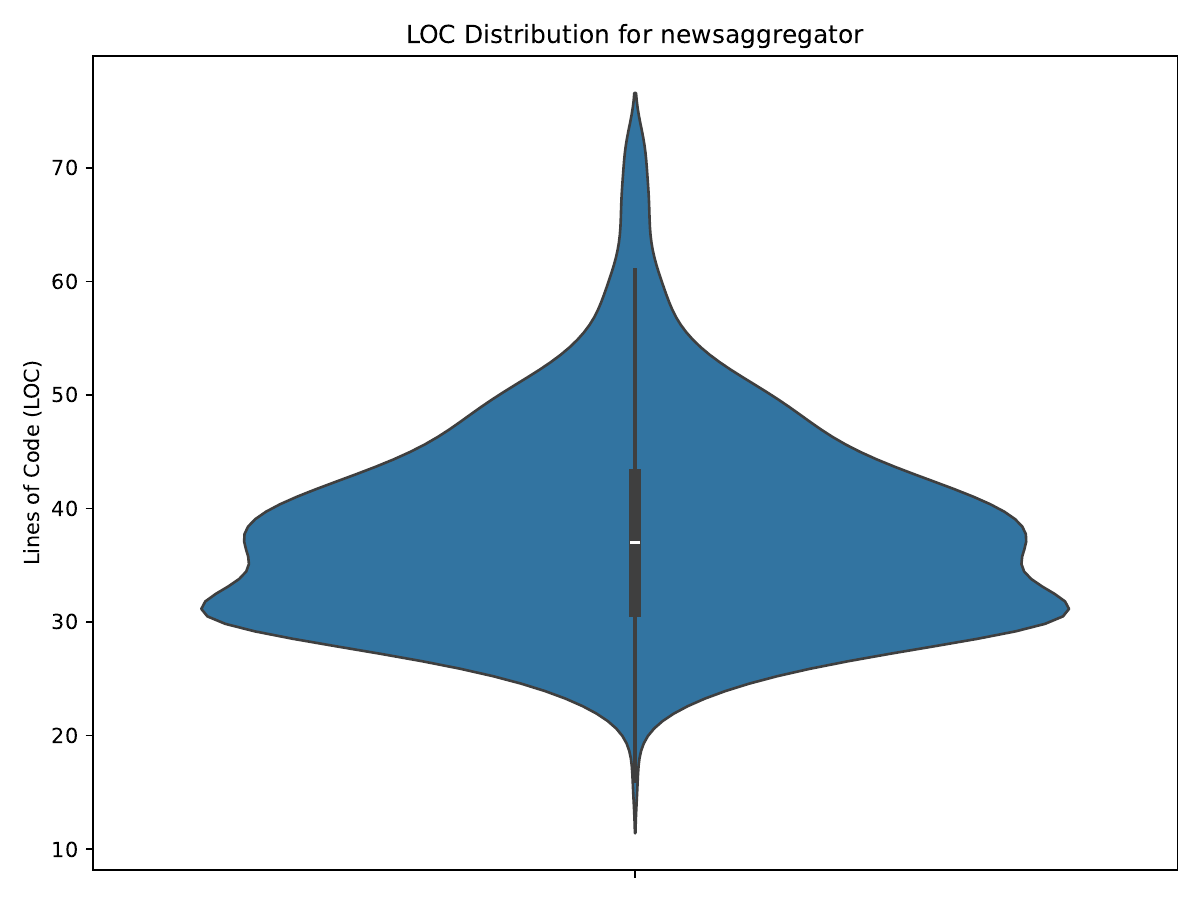}
        \caption{News Aggregator}
    \end{subfigure}
    \begin{subfigure}{0.3\textwidth}
        \centering
        \includegraphics[width=\textwidth]{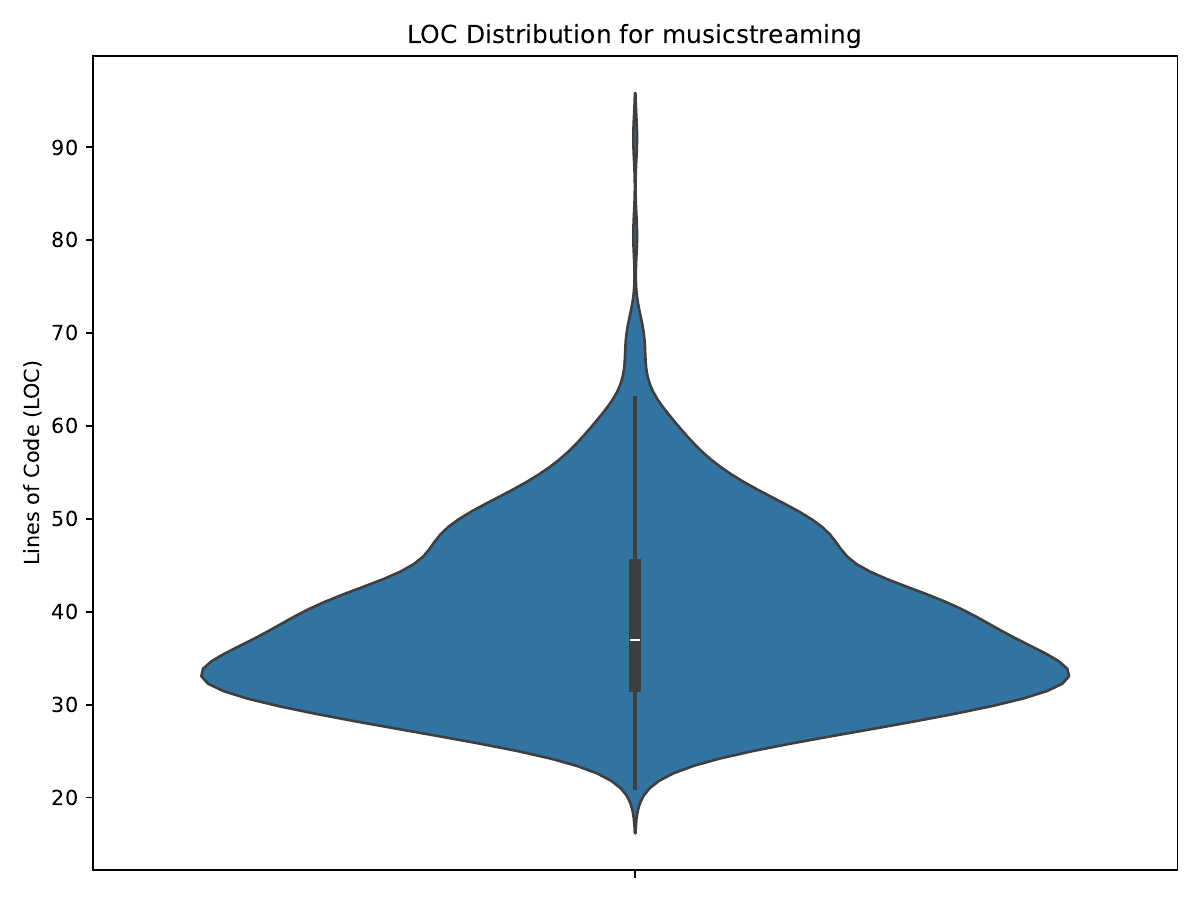}
        \caption{Music Streaming}
    \end{subfigure}
    \begin{subfigure}{0.3\textwidth}
        \centering
        \includegraphics[width=\textwidth]{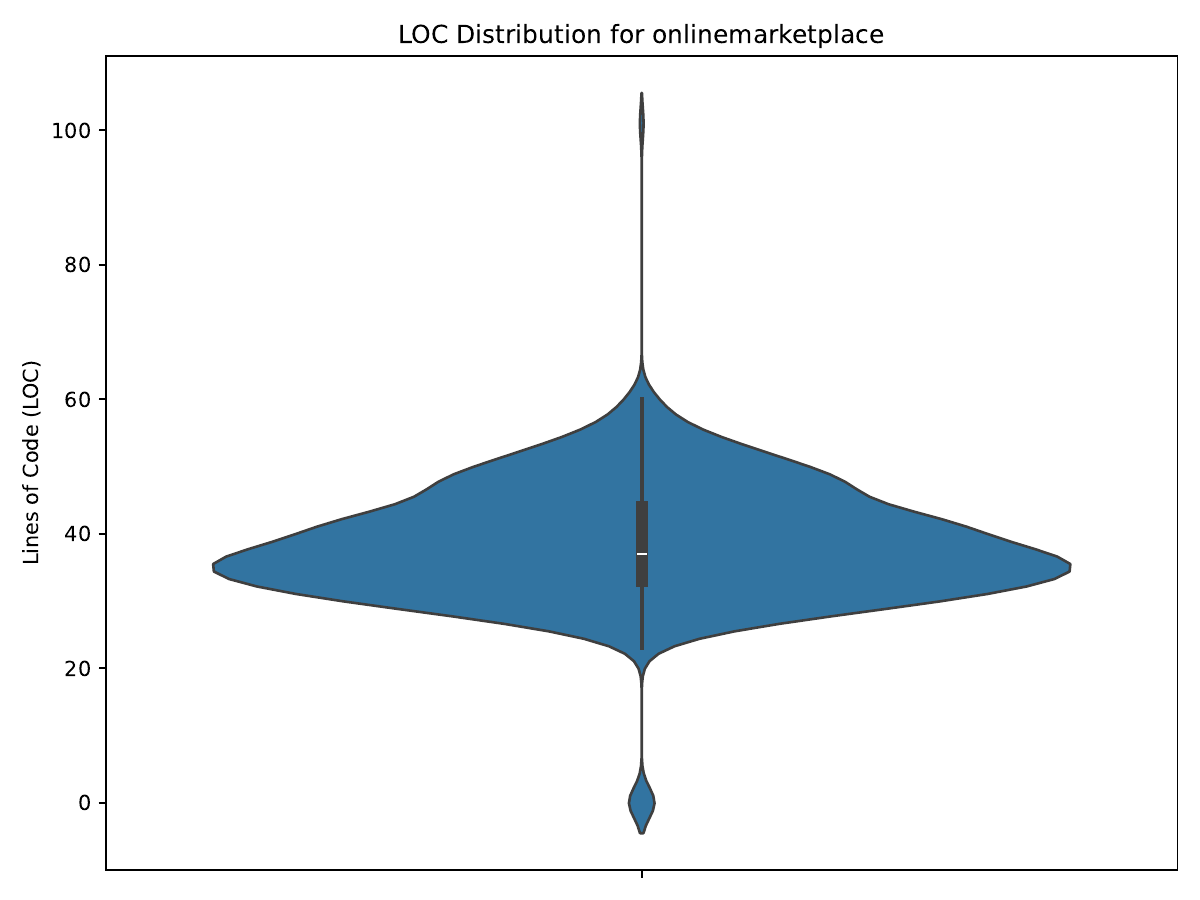}
        \caption{Online Marketplace}
    \end{subfigure}
    
    \begin{subfigure}{0.3\textwidth}
        \centering
        \includegraphics[width=\textwidth]{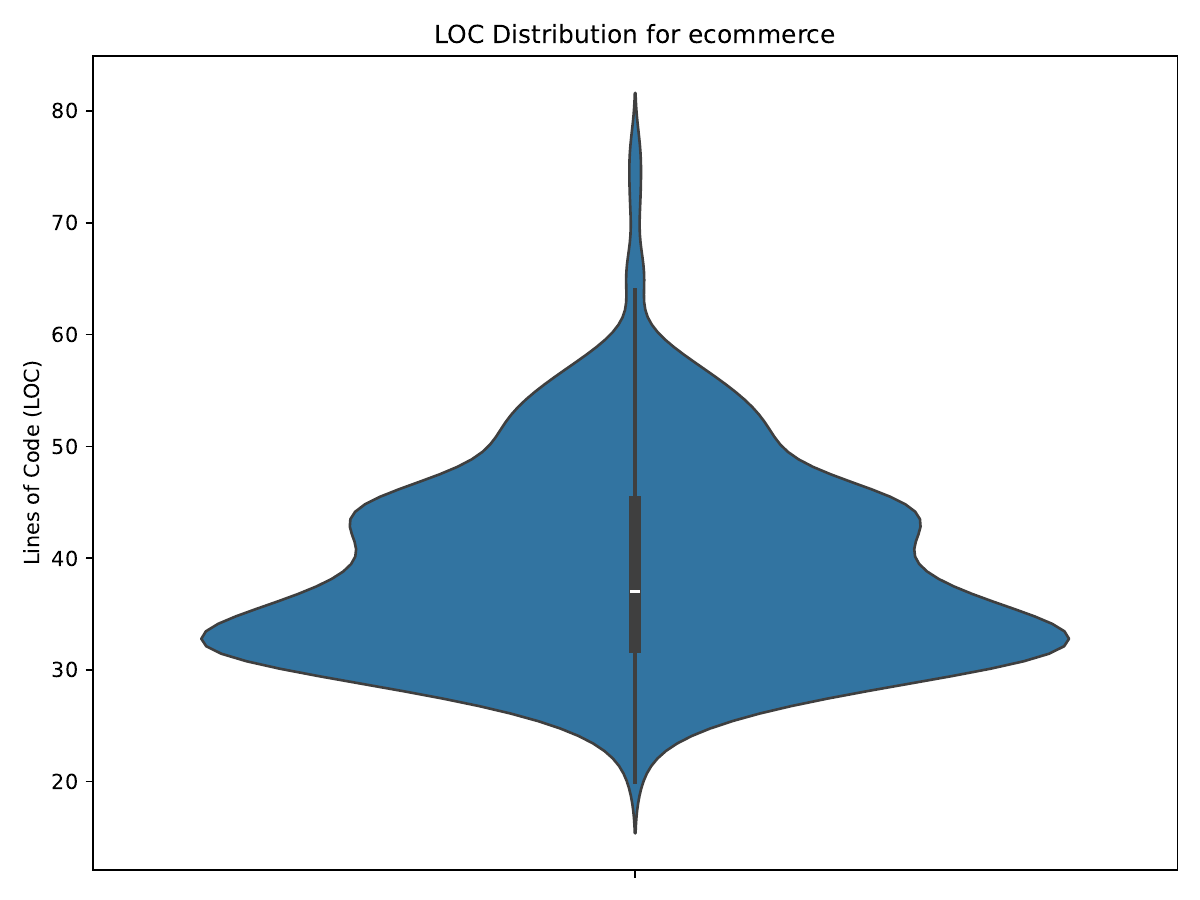}
        \caption{E-commerce}
    \end{subfigure}
    \begin{subfigure}{0.3\textwidth}
        \centering
        \includegraphics[width=\textwidth]{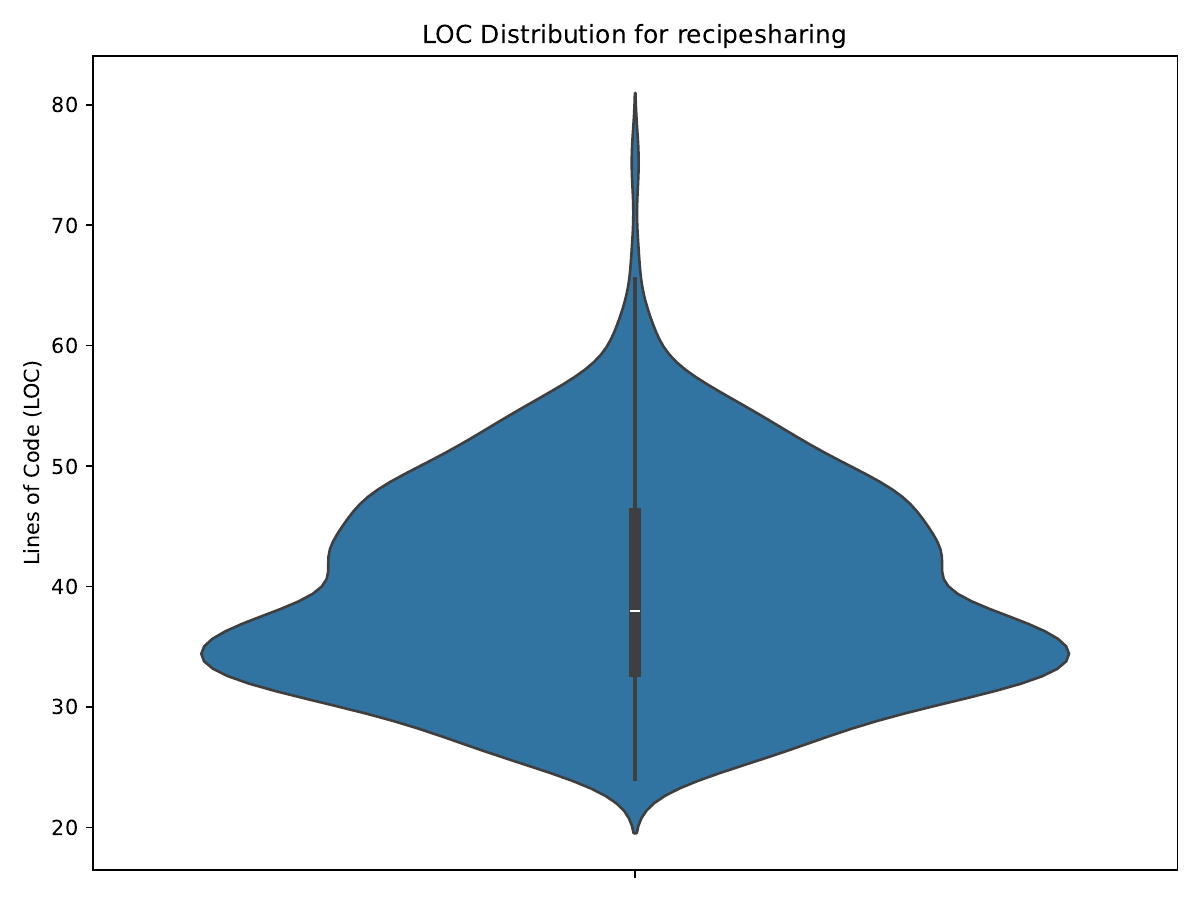}
        \caption{Recipe Sharing}
    \end{subfigure}
    \begin{subfigure}{0.3\textwidth}
        \centering
        \includegraphics[width=\textwidth]{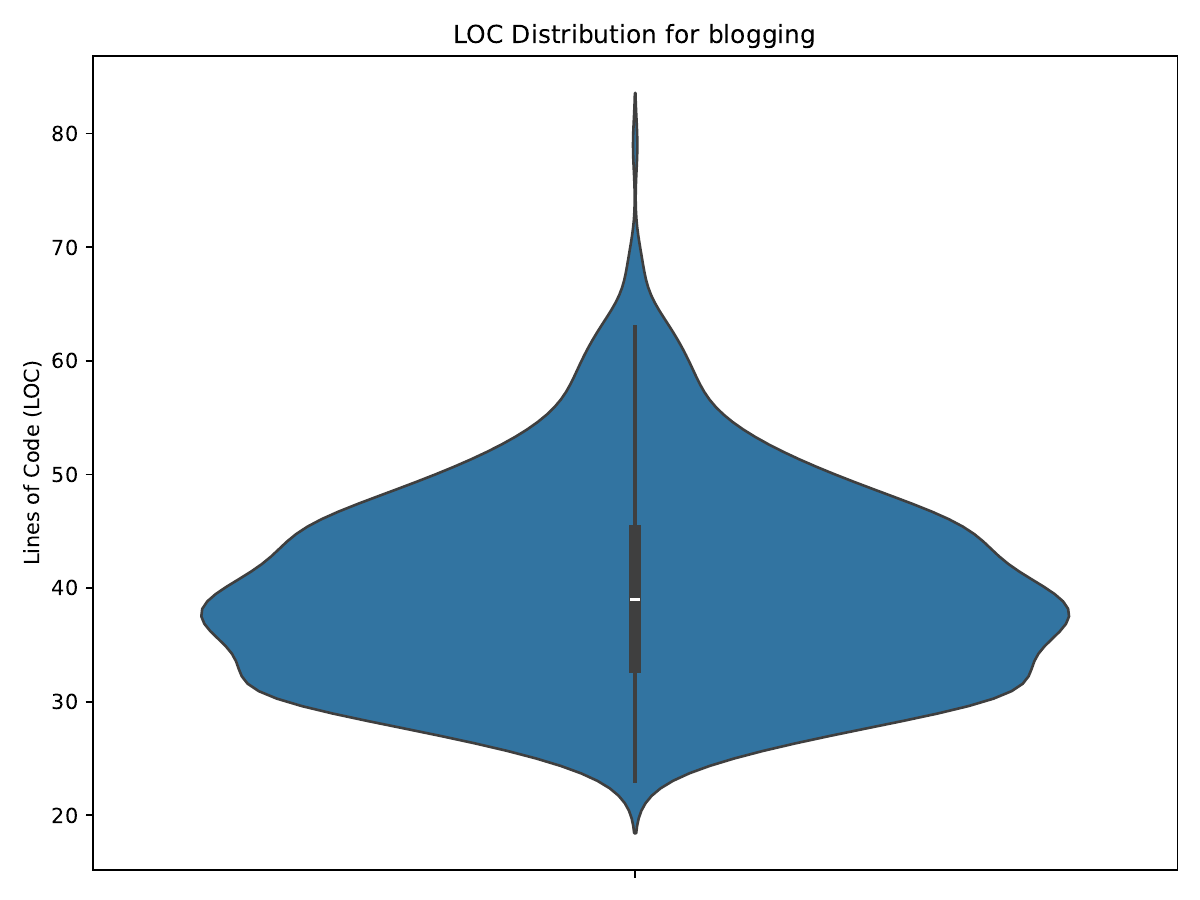}
        \caption{Blogging}
    \end{subfigure}
    
    \begin{subfigure}{0.3\textwidth}
        \centering
        \includegraphics[width=\textwidth]{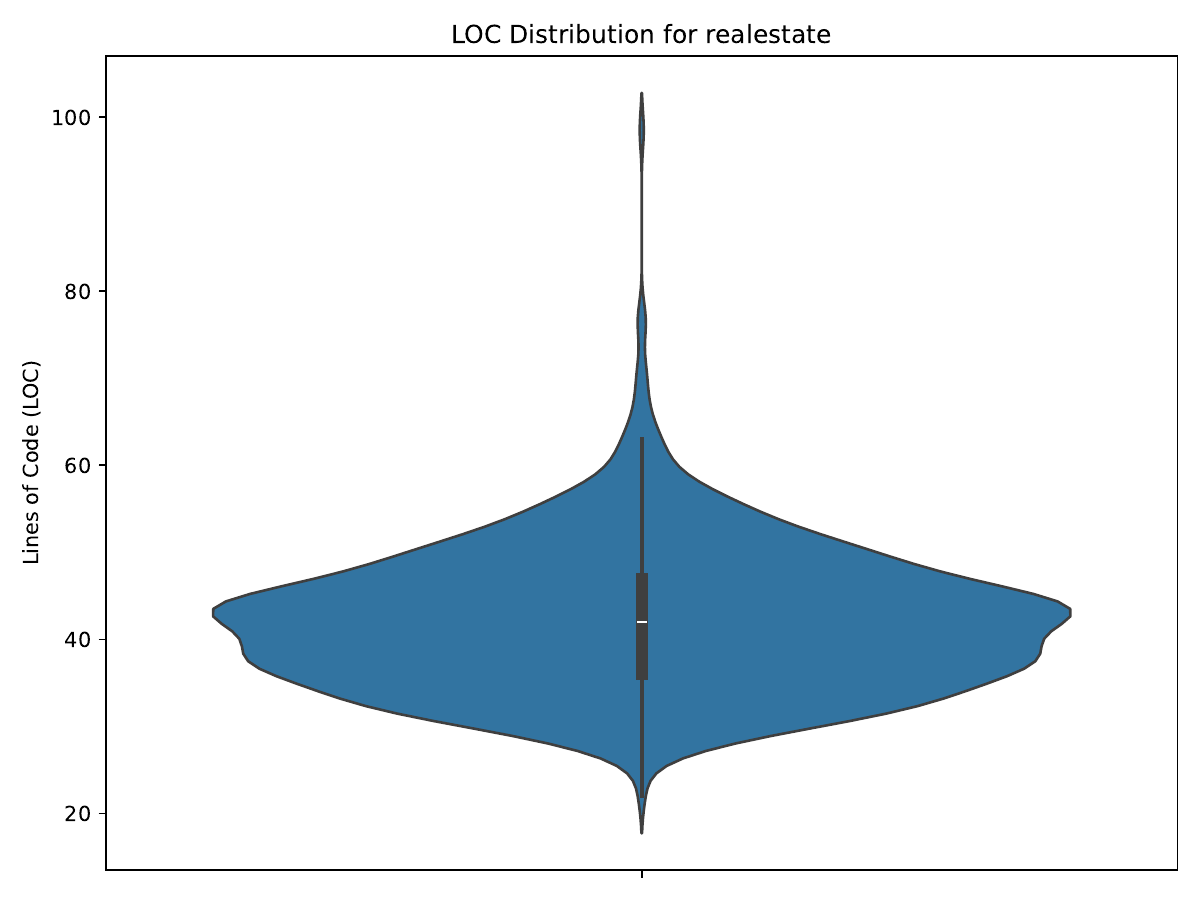}
        \caption{Real Estate}
    \end{subfigure}
    \begin{subfigure}{0.3\textwidth}
        \centering
        \includegraphics[width=\textwidth]{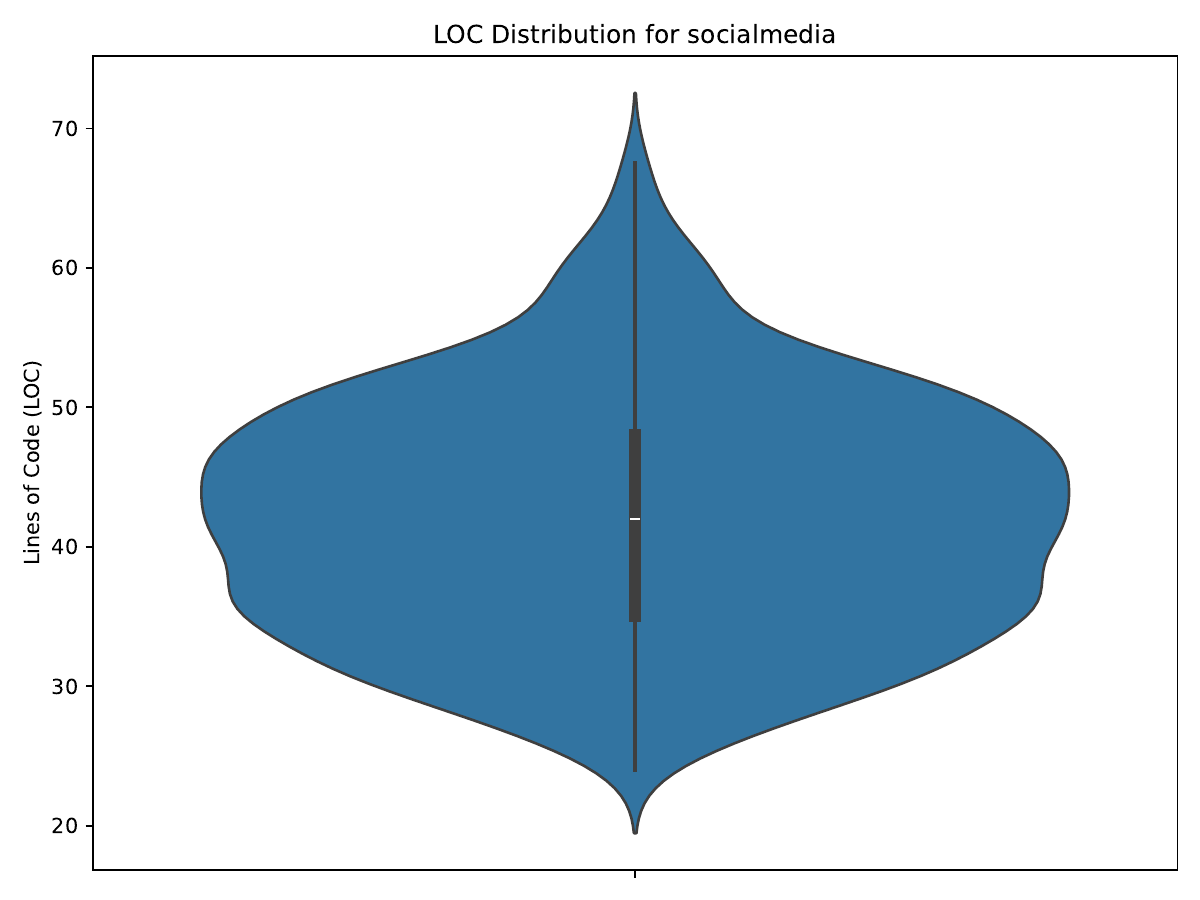}
        \caption{Social Media}
    \end{subfigure}
    \begin{subfigure}{0.3\textwidth}
        \centering
        \includegraphics[width=\textwidth]{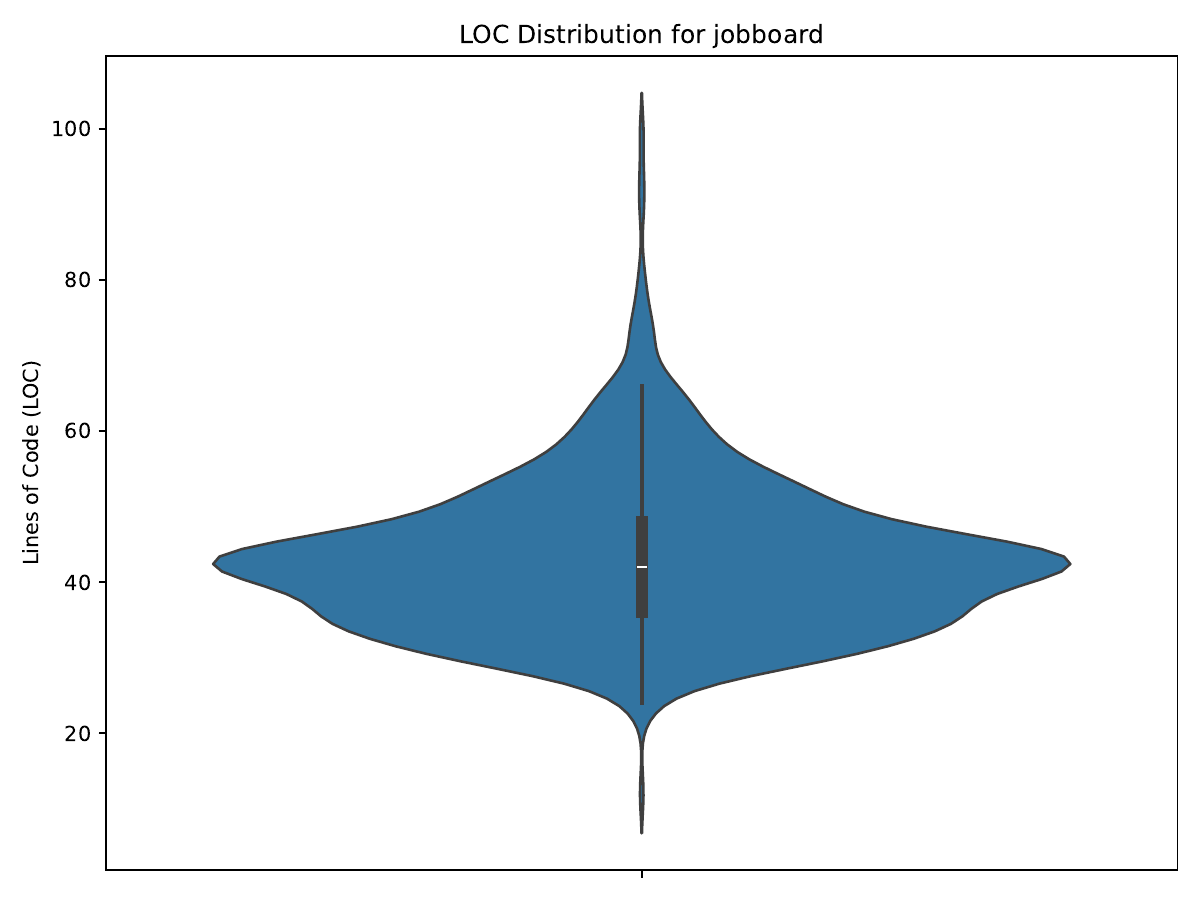}
        \caption{Job Board}
    \end{subfigure}
 
    \begin{subfigure}{0.3\textwidth}
        \centering
        \includegraphics[width=\textwidth]{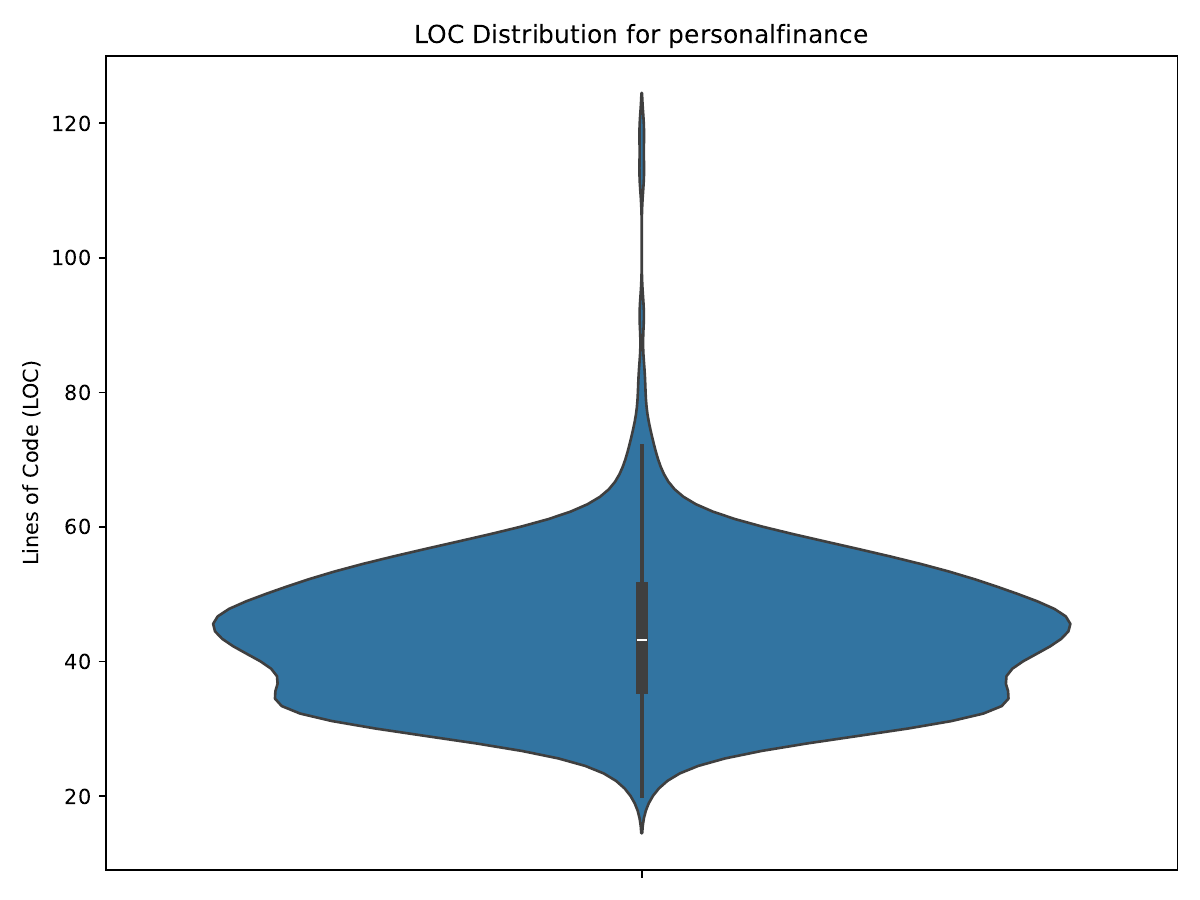}
        \caption{Personal Finance}
    \end{subfigure}
    \begin{subfigure}{0.3\textwidth}
        \centering
        \includegraphics[width=\textwidth]{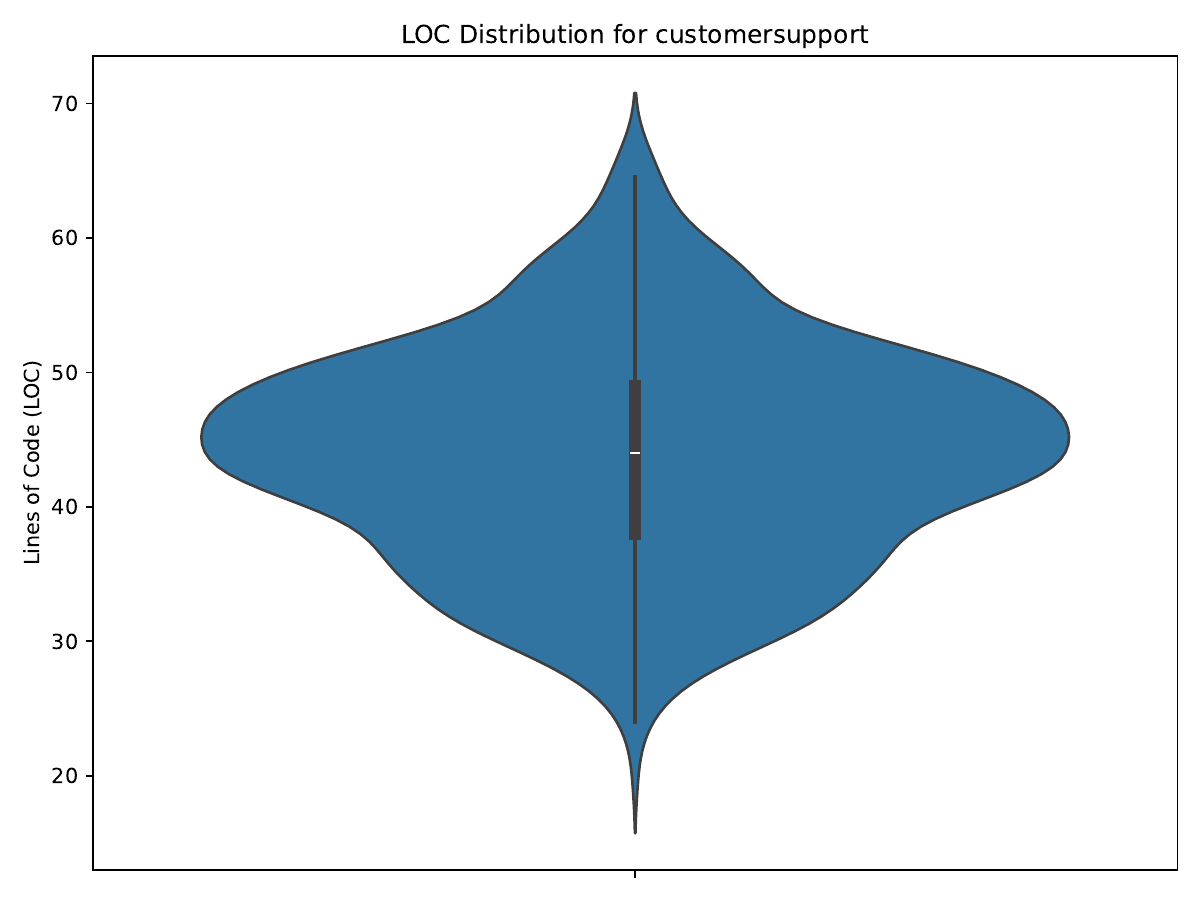}
        \caption{Customer Support}
    \end{subfigure}
        \begin{subfigure}{0.3\textwidth}
        \centering
        \includegraphics[width=\textwidth]{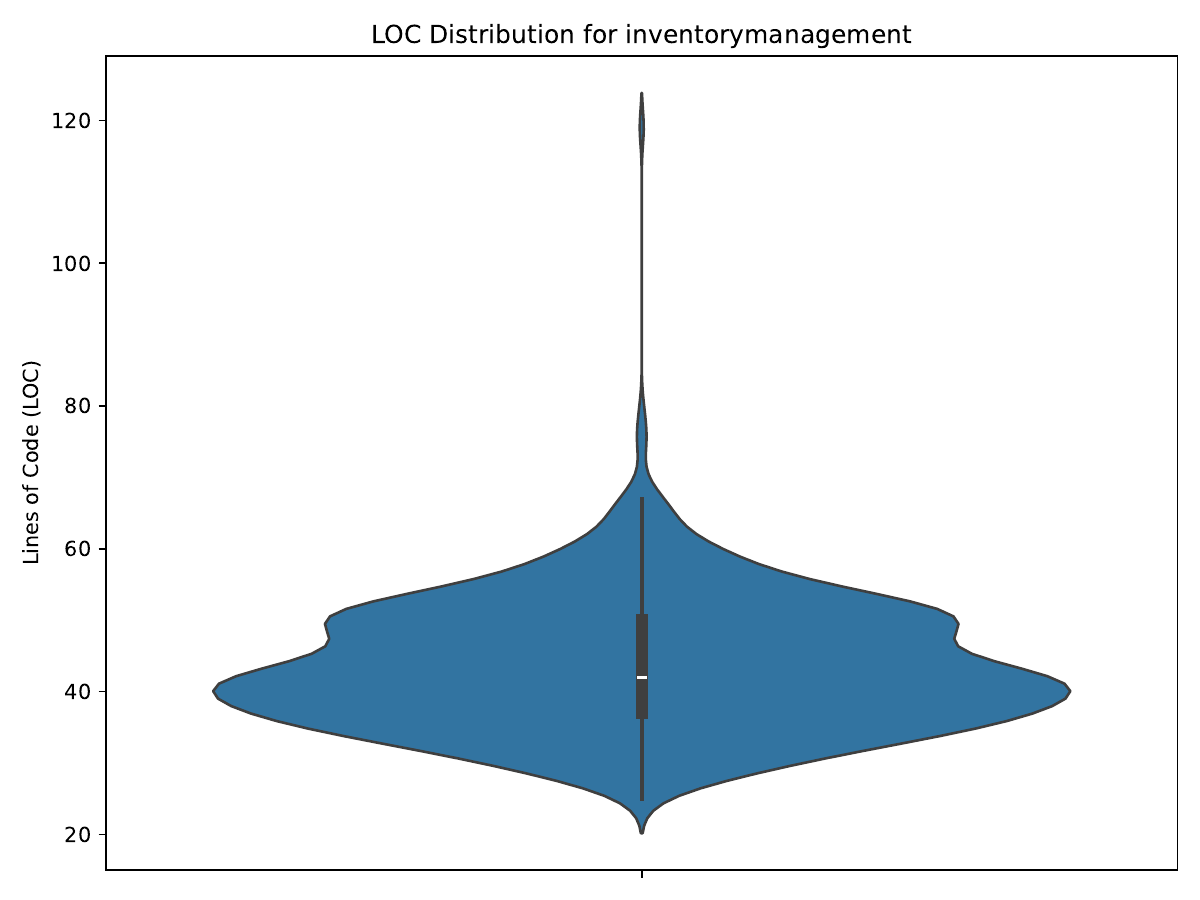}
        \caption{Inventory Management}
    \end{subfigure}

    \begin{subfigure}{0.3\textwidth}
        \centering
        \includegraphics[width=\textwidth]{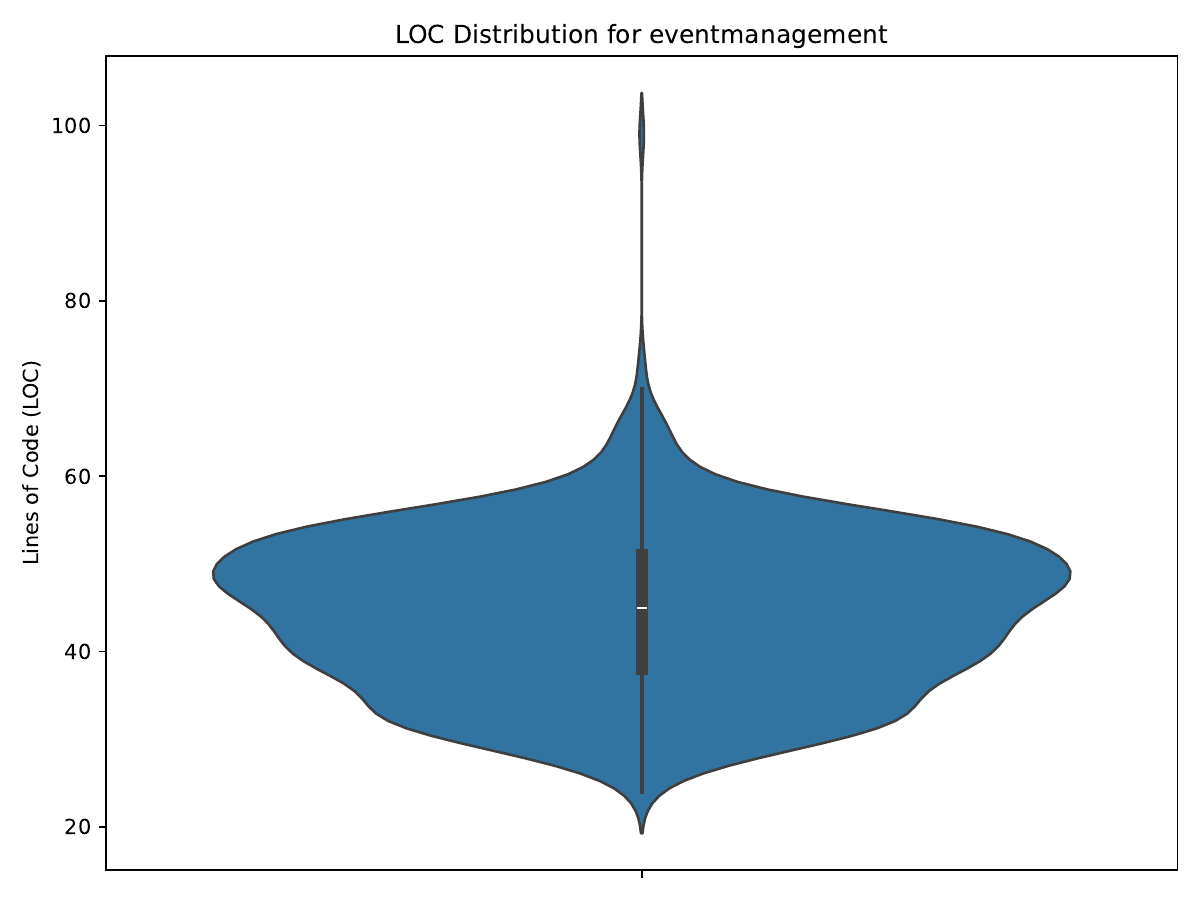}
        \caption{Event Management}
    \end{subfigure}
    \begin{subfigure}{0.3\textwidth}
        \centering
        \includegraphics[width=\textwidth]{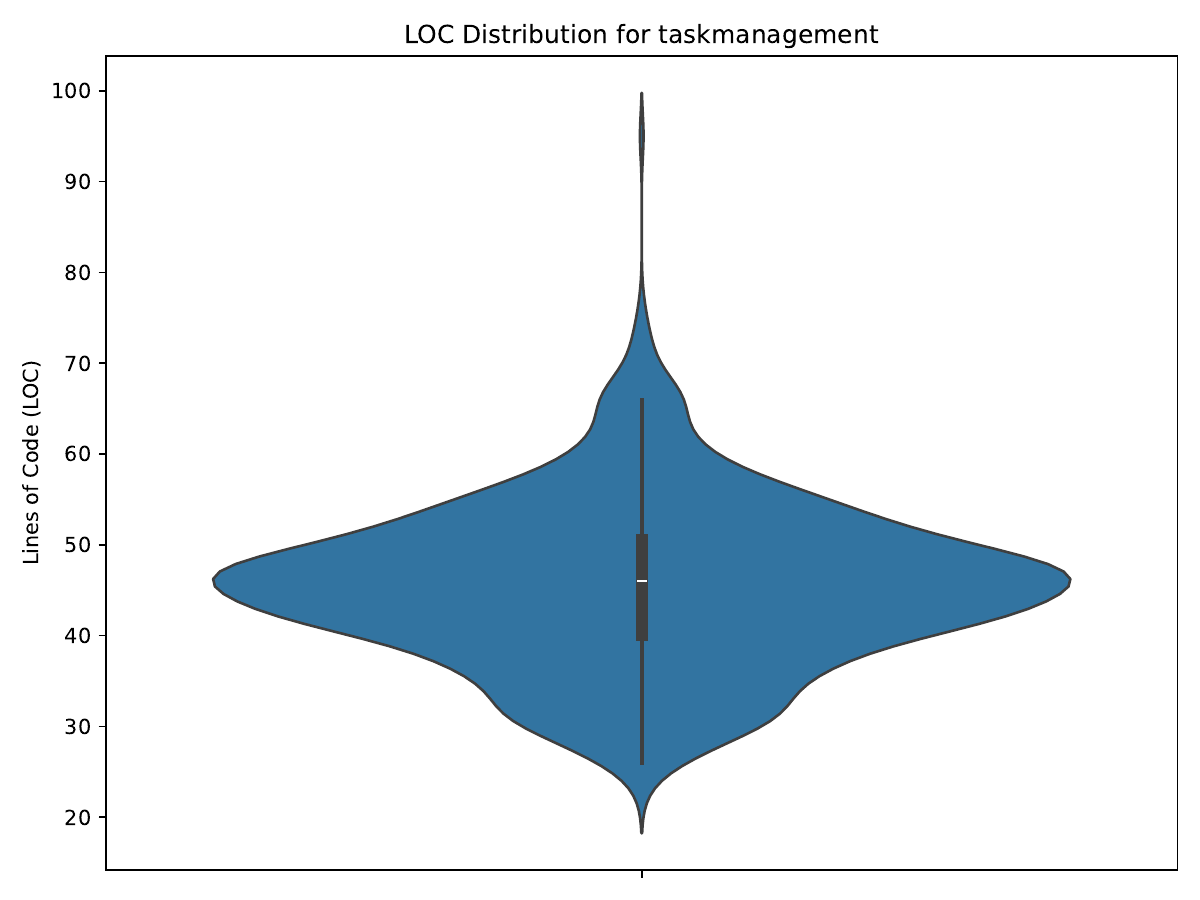}
        \caption{Task Management}
    \end{subfigure}
    \caption{LOC Distribution by Applications: UniModal}
    \label{fig:loc_distribution_apps_unimodal}
\end{figure}

\begin{figure}[h]
    \centering
    \begin{subfigure}{0.3\textwidth}
        \centering
        \includegraphics[width=\textwidth]{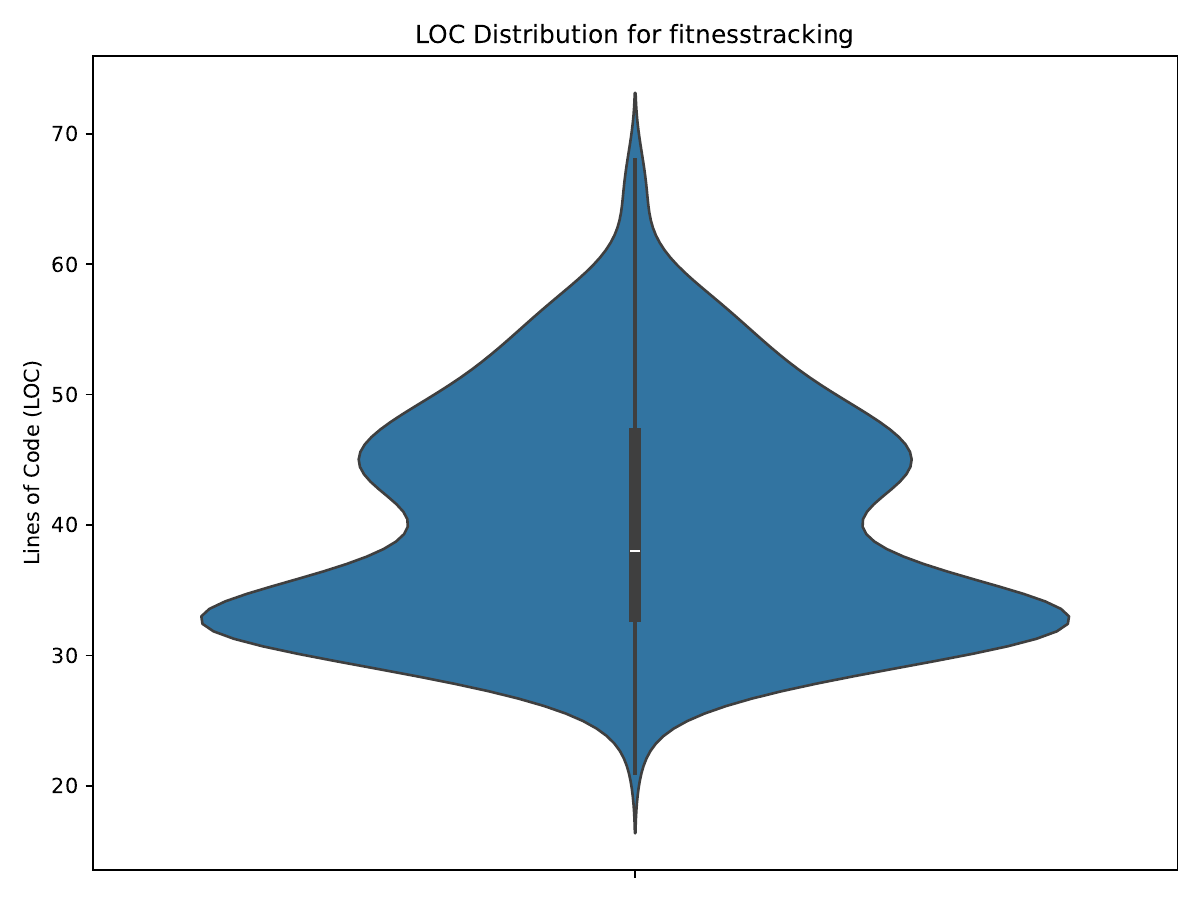}
        \caption{Fitness Tracking}
    \end{subfigure}
    \begin{subfigure}{0.3\textwidth}
        \centering
        \includegraphics[width=\textwidth]{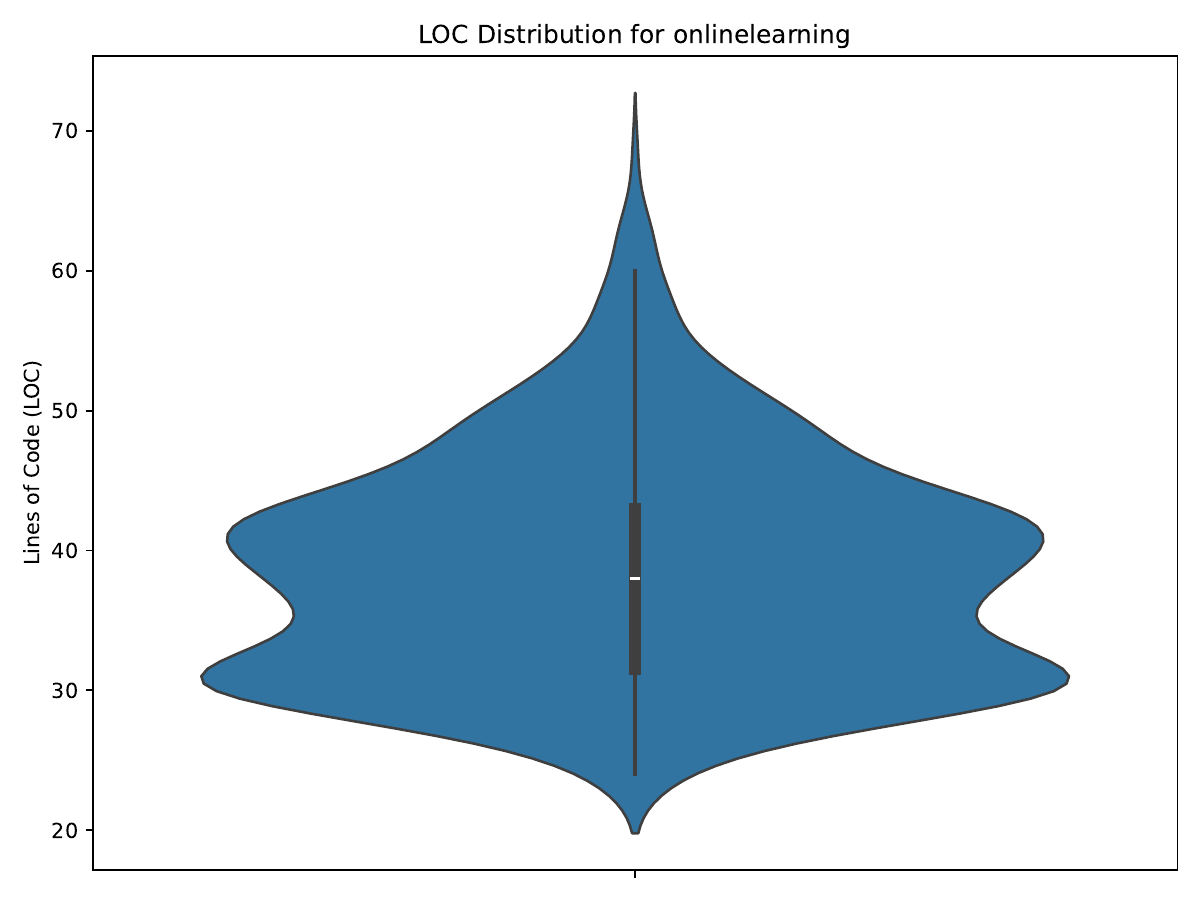}
        \caption{Online Learning}
    \end{subfigure}
    \begin{subfigure}{0.3\textwidth}
        \centering
        \includegraphics[width=\textwidth]{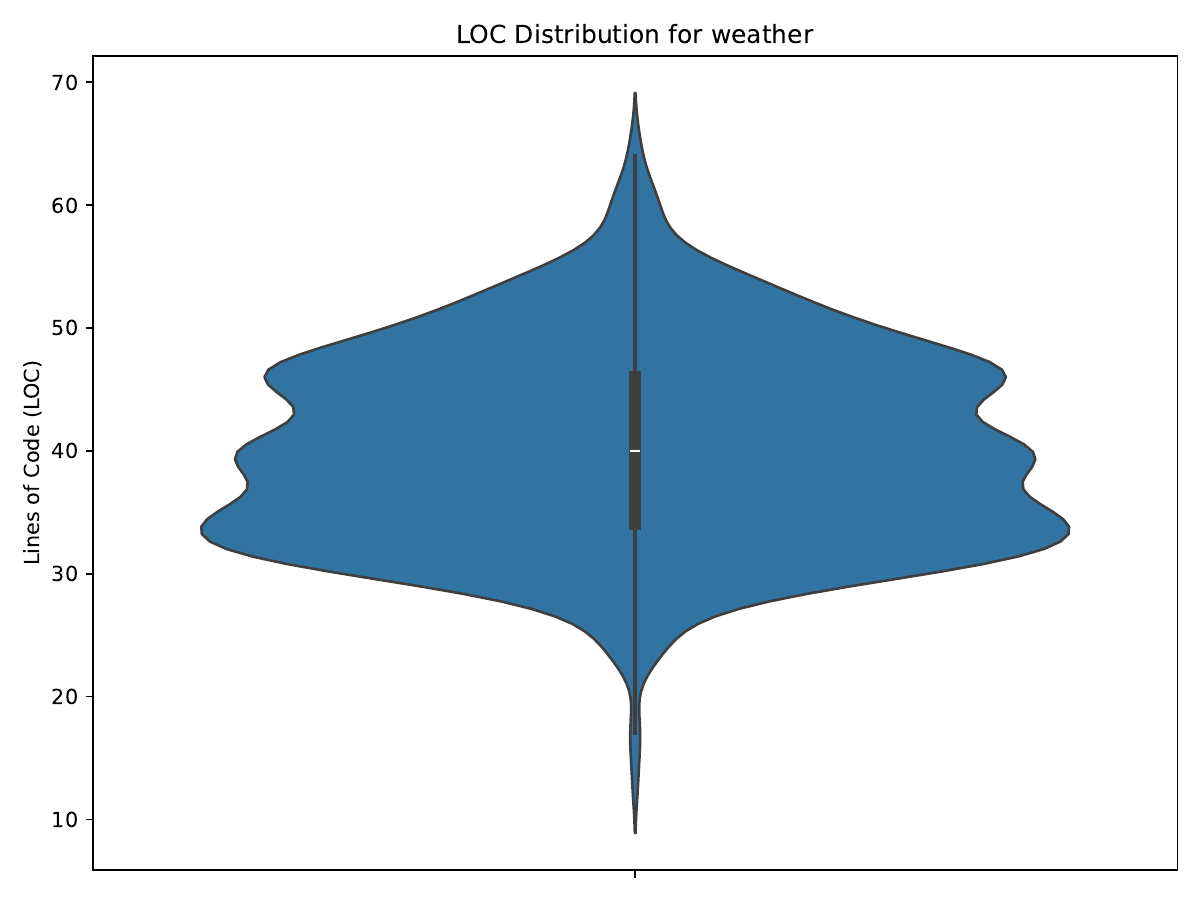}
        \caption{Weather}
    \end{subfigure}
    
    \begin{subfigure}{0.3\textwidth}
        \centering
        \includegraphics[width=\textwidth]{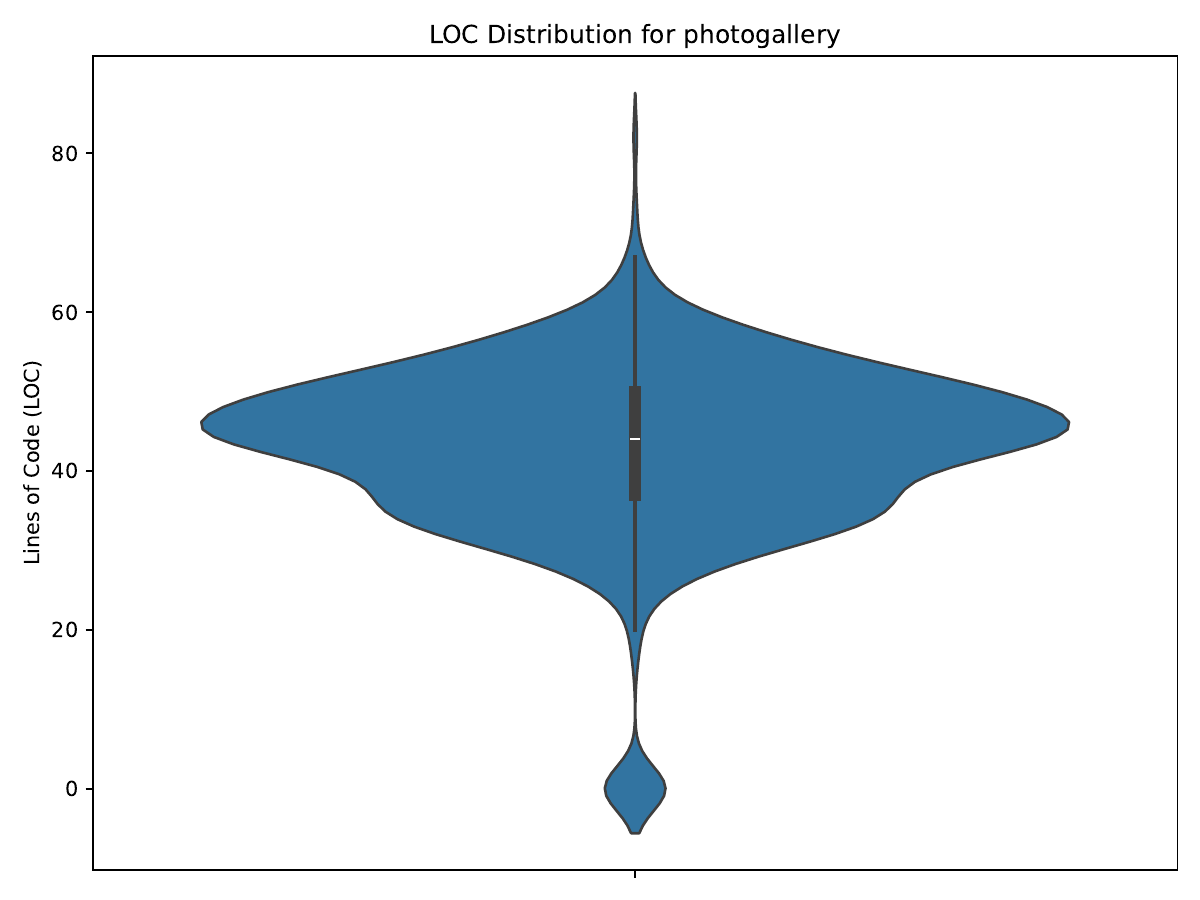}
        \caption{Photo Gallery}
    \end{subfigure}
    \begin{subfigure}{0.3\textwidth}
        \centering
        \includegraphics[width=\textwidth]{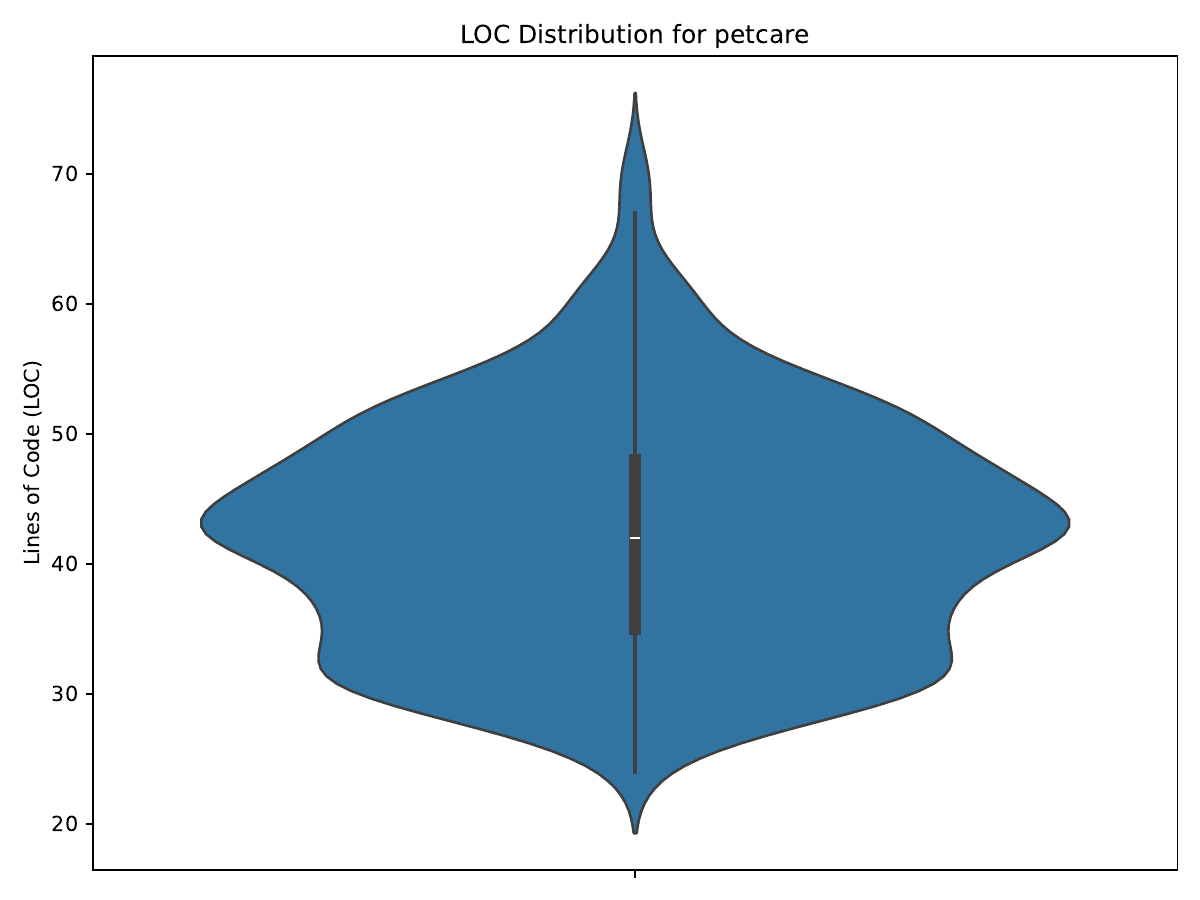}
        \caption{Pet Care}
    \end{subfigure}
    \begin{subfigure}{0.3\textwidth}
        \centering
        \includegraphics[width=\textwidth]{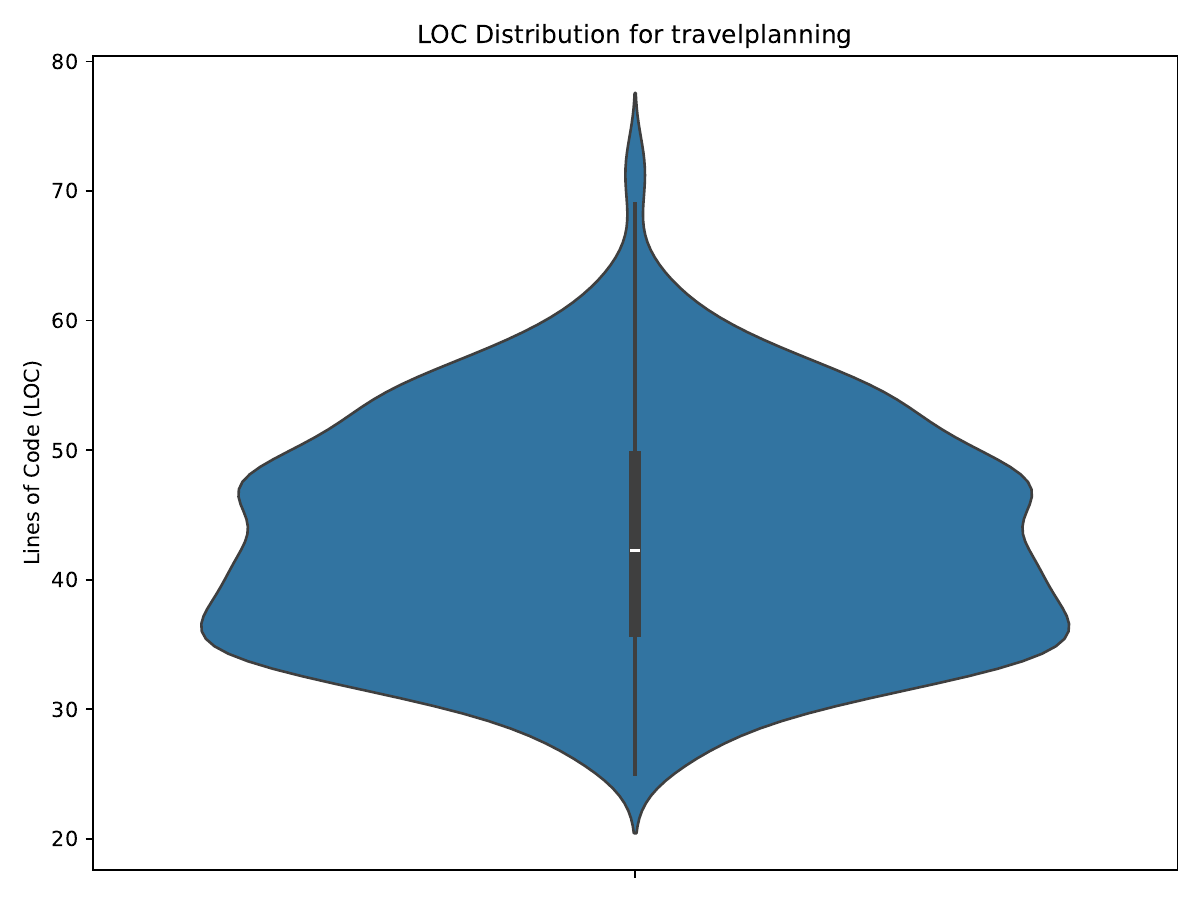}
        \caption{Travel Planning}
    \end{subfigure}
    \caption{LOC Distribution by Applications: MultiModal}
    \label{fig:loc_distribution_apps_multimodal}
\end{figure}

Fig.~\ref{fig:loc_distribution_apps_unimodal} collects violin charts of 14 applications following unimodal distribution, where the model outputs are centered around a common length, with less variation between extremes. The remaining 6 applications are in Fig.~\ref{fig:loc_distribution_apps_multimodal}, following multimodal distribution. In both cases, the median LOC is always positioned centrally in each distribution, which suggests that the code generation is stable across applications. Applications in Fig.~\ref{fig:loc_distribution_apps_multimodal} exhibit more complex patterns, but the distributions remain balanced with the median value positioned at the center of the distribution.

\section{Appendix: LOC Distribution by Models: Success vs Failure}\label{sec:loc_successfail_distribution_models}
\begin{figure}[h!]
    \centering
    \begin{subfigure}{0.49\textwidth}
        \centering
        \includegraphics[width=\linewidth]{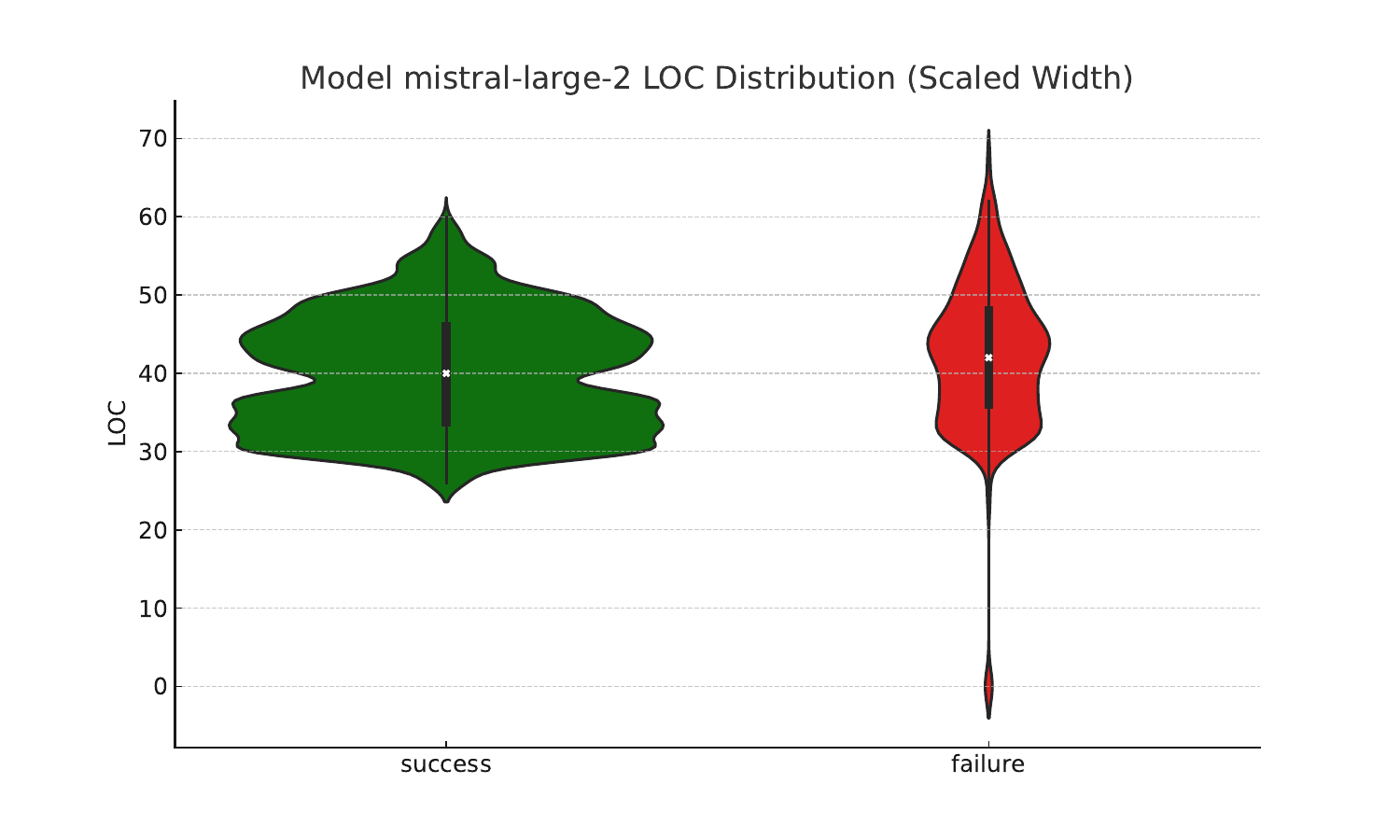}
        \caption{mistral-large-2 ($pass@1$ = 0.7804)}
    \end{subfigure}
    \begin{subfigure}{0.49\textwidth}
        \centering
        \includegraphics[width=\linewidth]{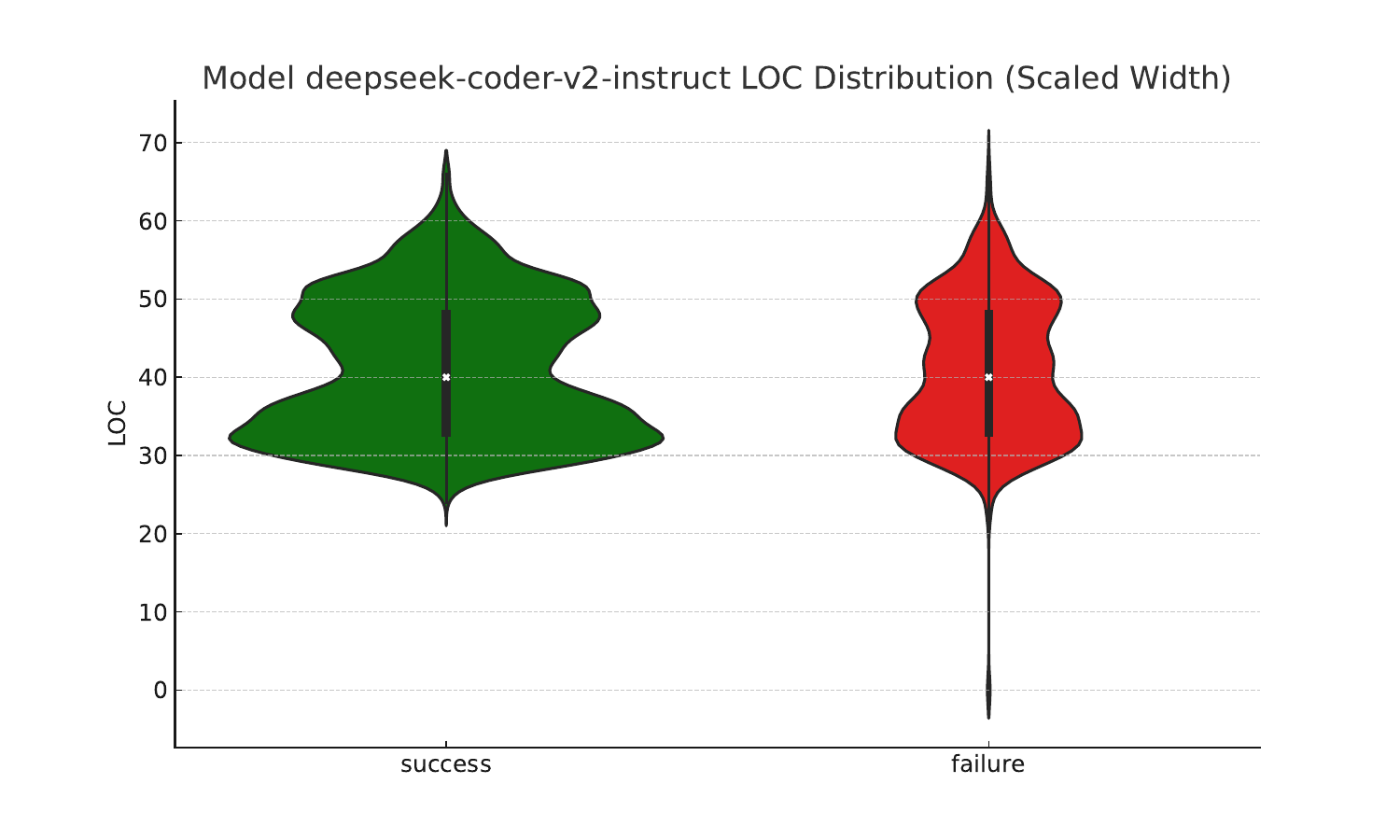}
        \caption{deepseek-coder-v2-instruct ($pass@1$ = 0.7002)}
    \end{subfigure}

    \begin{subfigure}{0.49\textwidth}
        \centering
        \includegraphics[width=\linewidth]{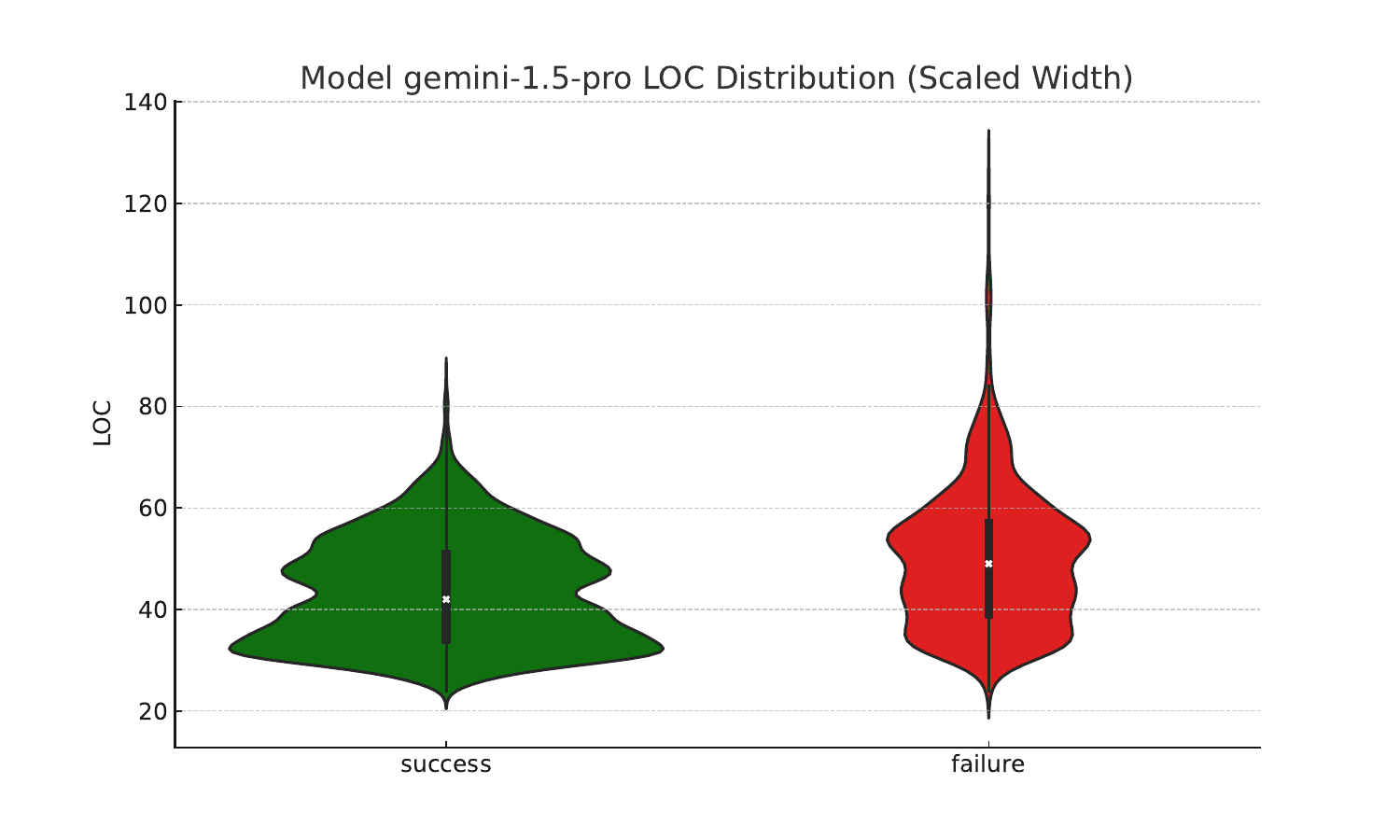}
        \caption{gemini-1.5-pro ($pass@1$ = 0.6813)}
    \end{subfigure}
    \begin{subfigure}{0.49\textwidth}
        \centering
        \includegraphics[width=\linewidth]{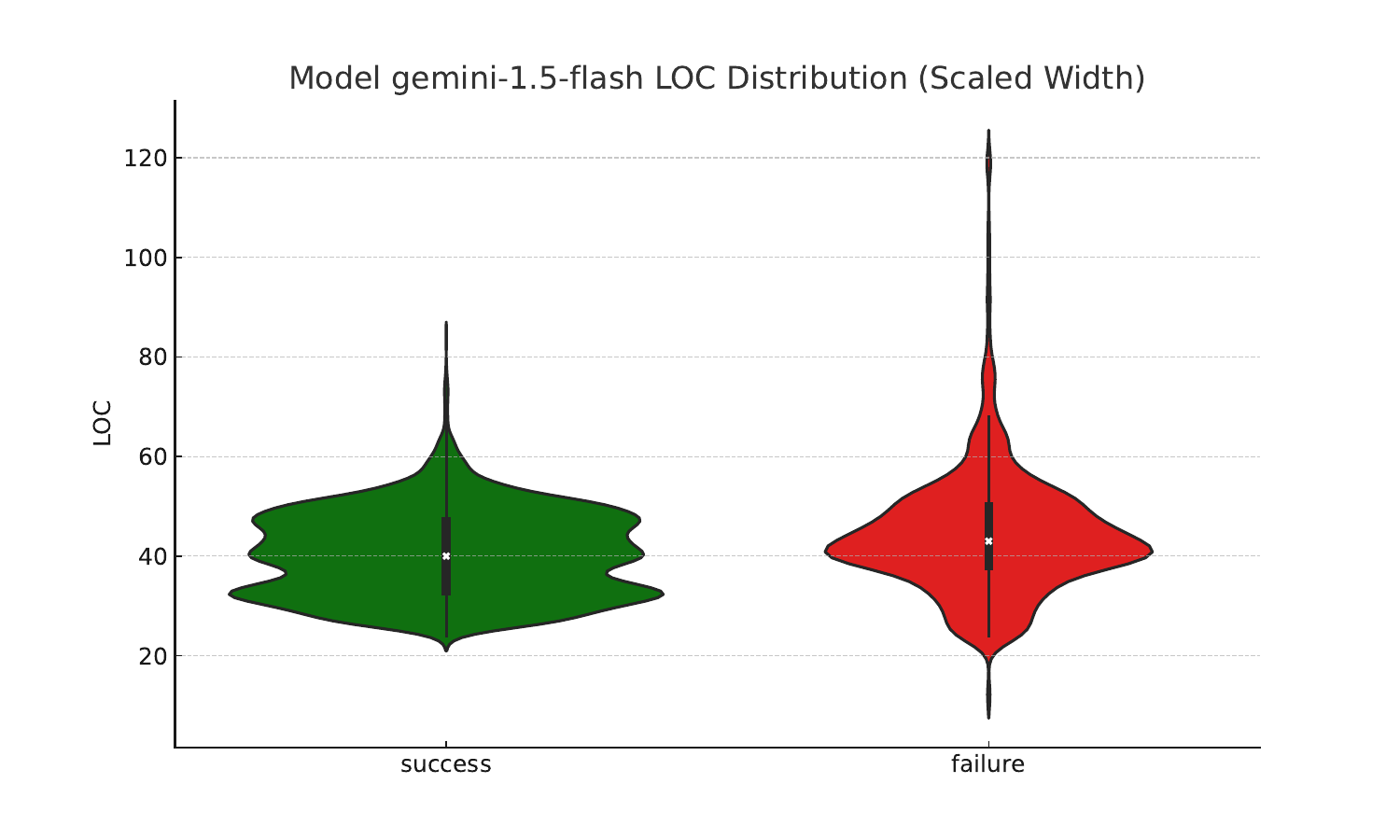}
        \caption{gemini-1.5-flash ($pass@1$ = 0.57)}
    \end{subfigure}

    \begin{subfigure}{0.49\textwidth}
        \centering
        \includegraphics[width=\linewidth]{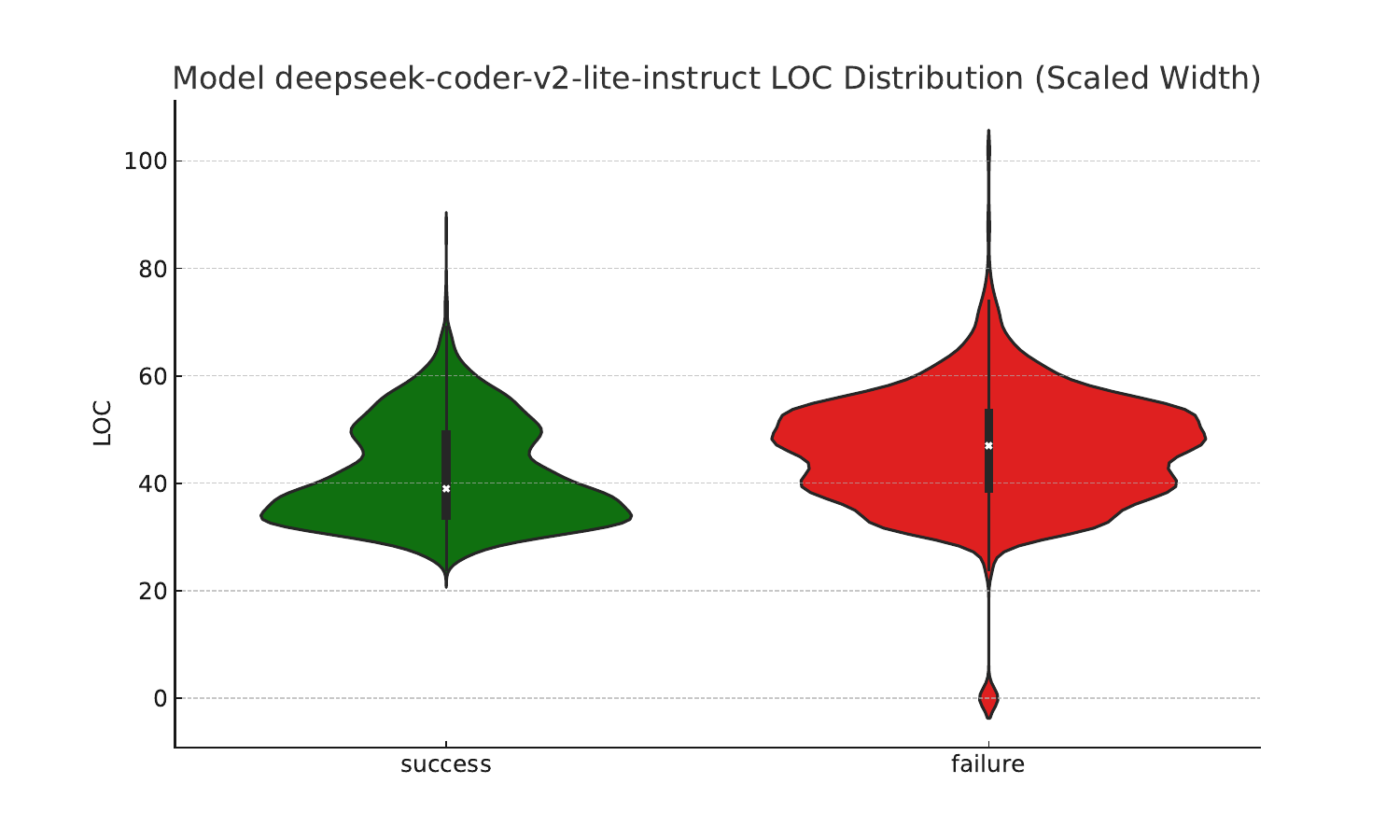}
        \caption{deepseek-coder-v2-lite-instruct ($pass@1$ = 0.4606)}
    \end{subfigure}
    \begin{subfigure}{0.49\textwidth}
        \centering
        \includegraphics[width=\linewidth]{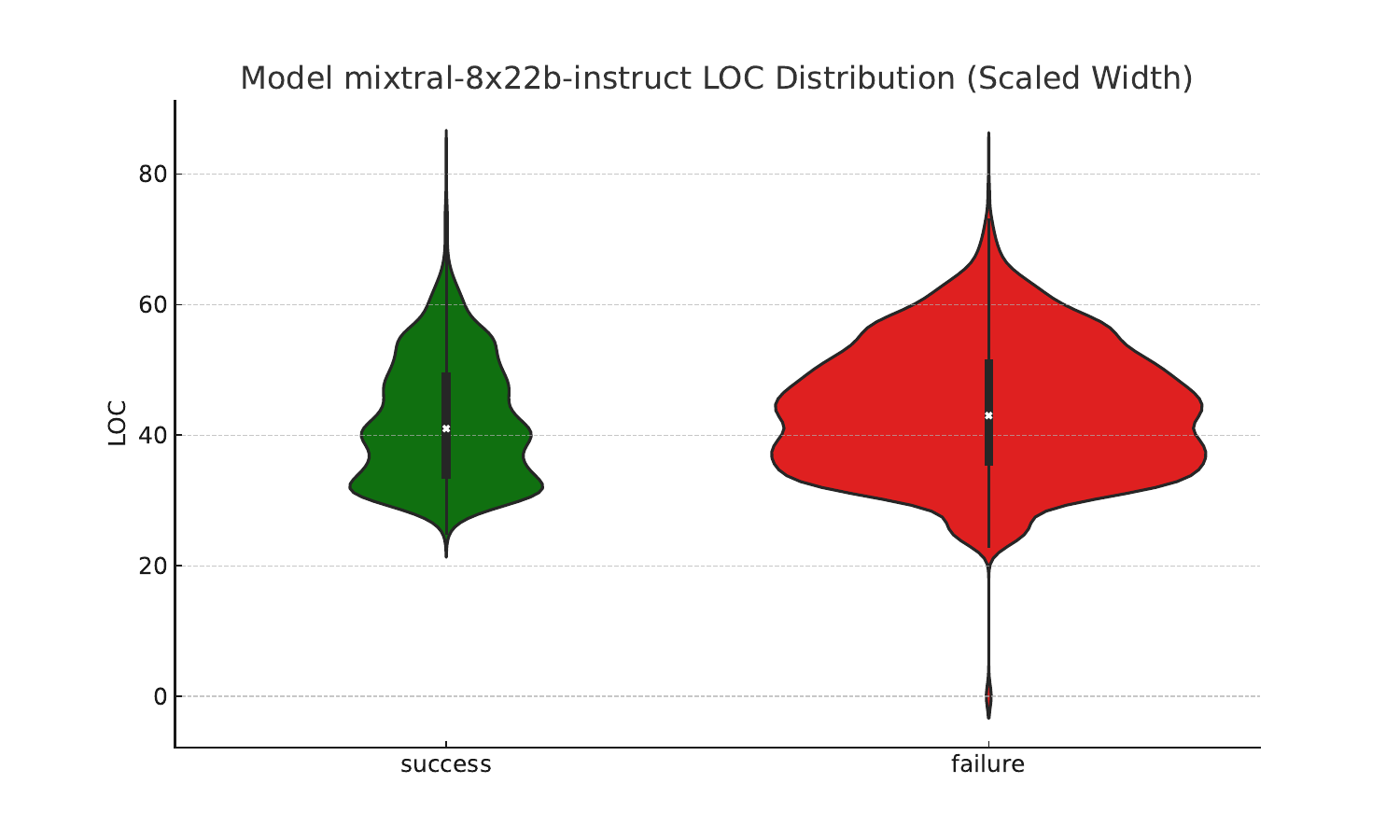}
        \caption{mixtral-8x22b-instruct ($pass@1$ = 0.3074)}
    \end{subfigure}

    \begin{subfigure}{0.49\textwidth}
        \centering
        \includegraphics[width=\linewidth]{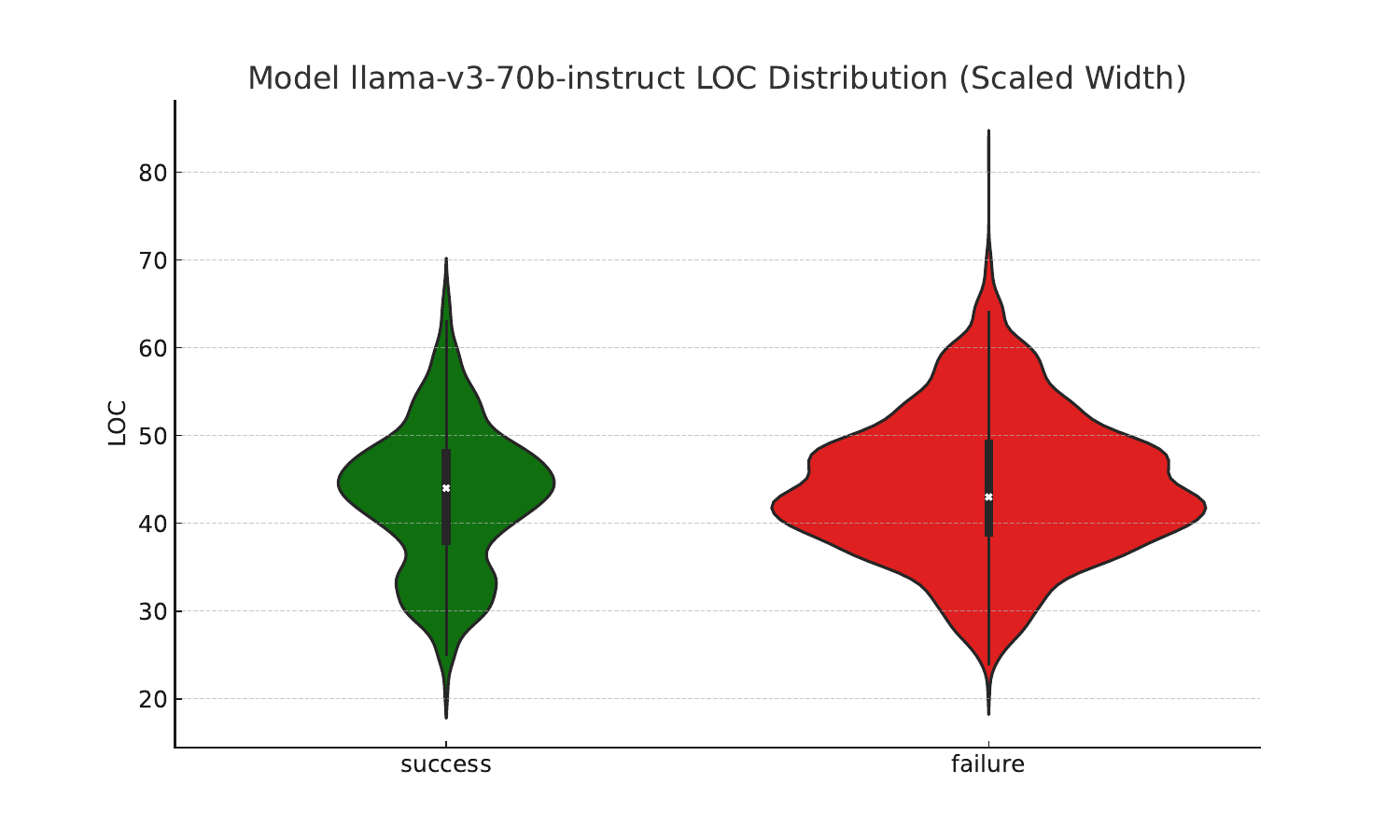}
        \caption{llama-v3-70b-instruct (pass@1 = 0.3323)}
    \end{subfigure}
    \begin{subfigure}{0.49\textwidth}
        \centering
        \includegraphics[width=\linewidth]{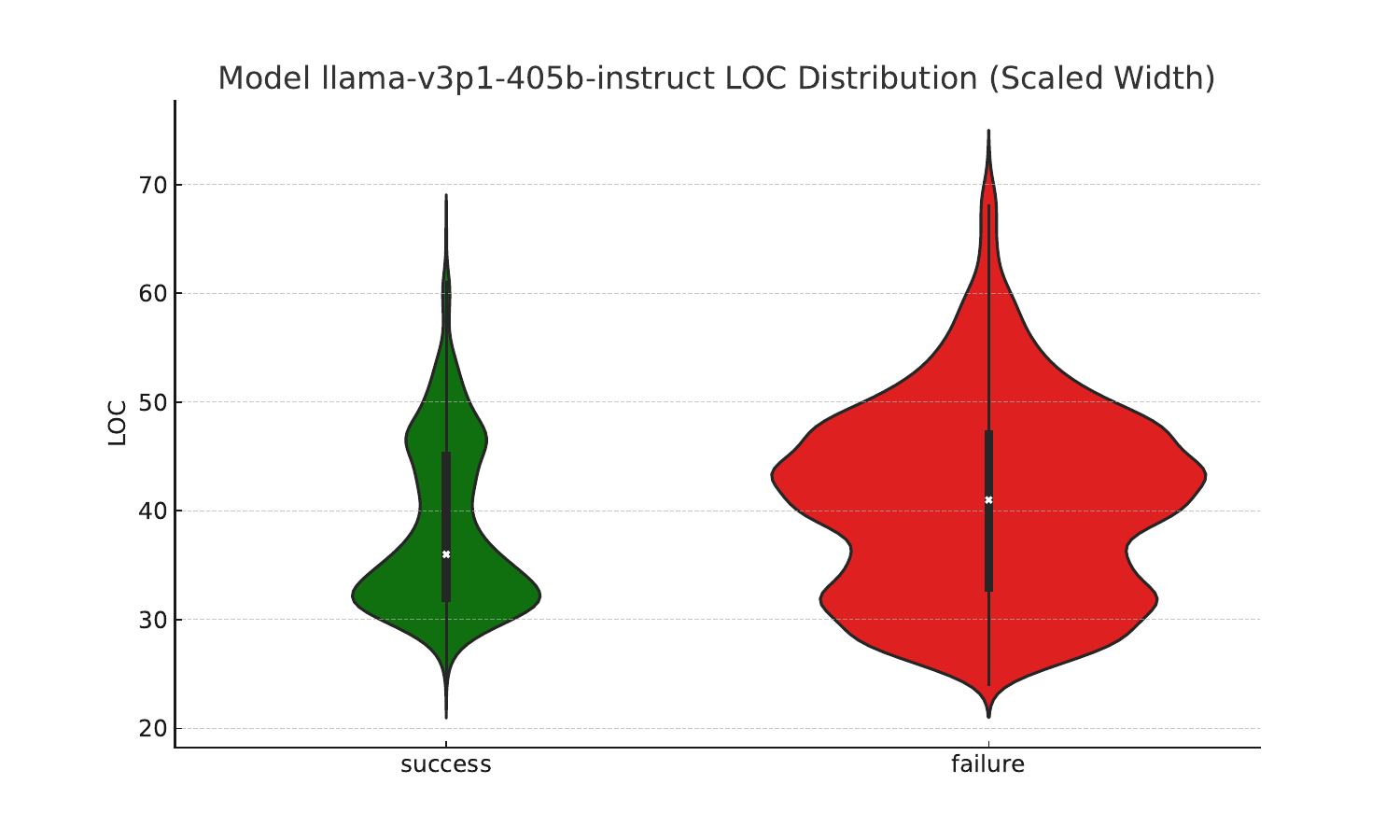}
        \caption{llama-v3p1-405b-instruct (pass@1 = 0.302)}
    \end{subfigure}
    \caption{LOC Distribution by Model: Success and Failure}
    \label{fig:loc_successfail_distribution_models}
\end{figure}

Continuing to Fig.~\ref{fig:loc_success_distribution_models} and \ref{fig:loc_fail_distribution_models}, Fig.~\ref{fig:loc_successfail_distribution_models} shows the success/fail LOC distribution of remaining 8 models.

\section{Appendix: LOC Distribution by Applications: Success vs Failure}\label{sec:loc_successfail_distribution_apps}
We conduct the same study described in Sec.~\ref{sec:loc_successfail}, except we shard the LOC distribution across applications instead of models. The results are collected in Fig.~\ref{fig:loc_successfail_distribution_apps}.
\begin{figure}[h!]
    \centering
    \begin{subfigure}{0.49\textwidth}
        \centering
        \includegraphics[width=\linewidth]{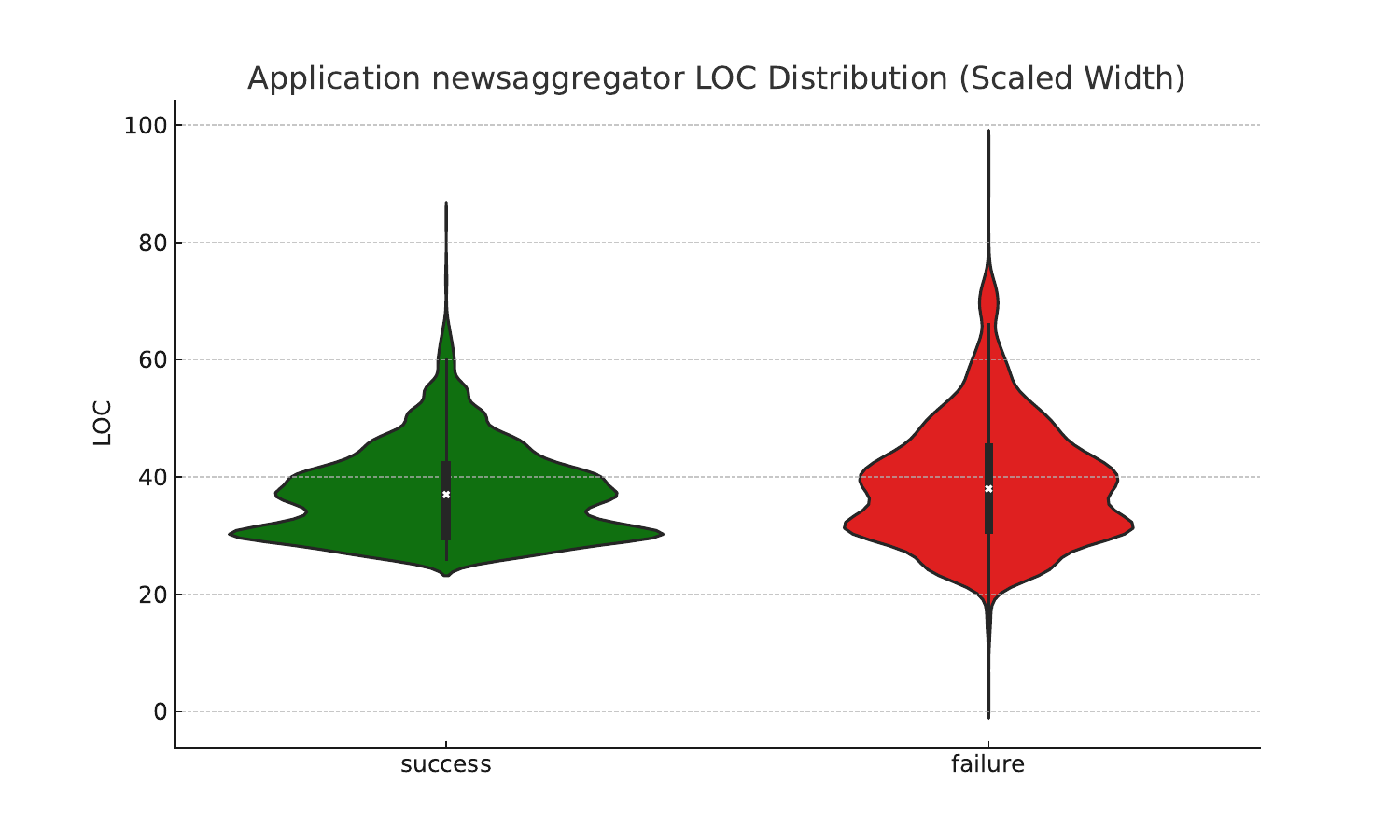}
        \caption{News Aggregator (Mean LOC = 37)}
    \end{subfigure}
    \begin{subfigure}{0.49\textwidth}
        \centering
        \includegraphics[width=\linewidth]{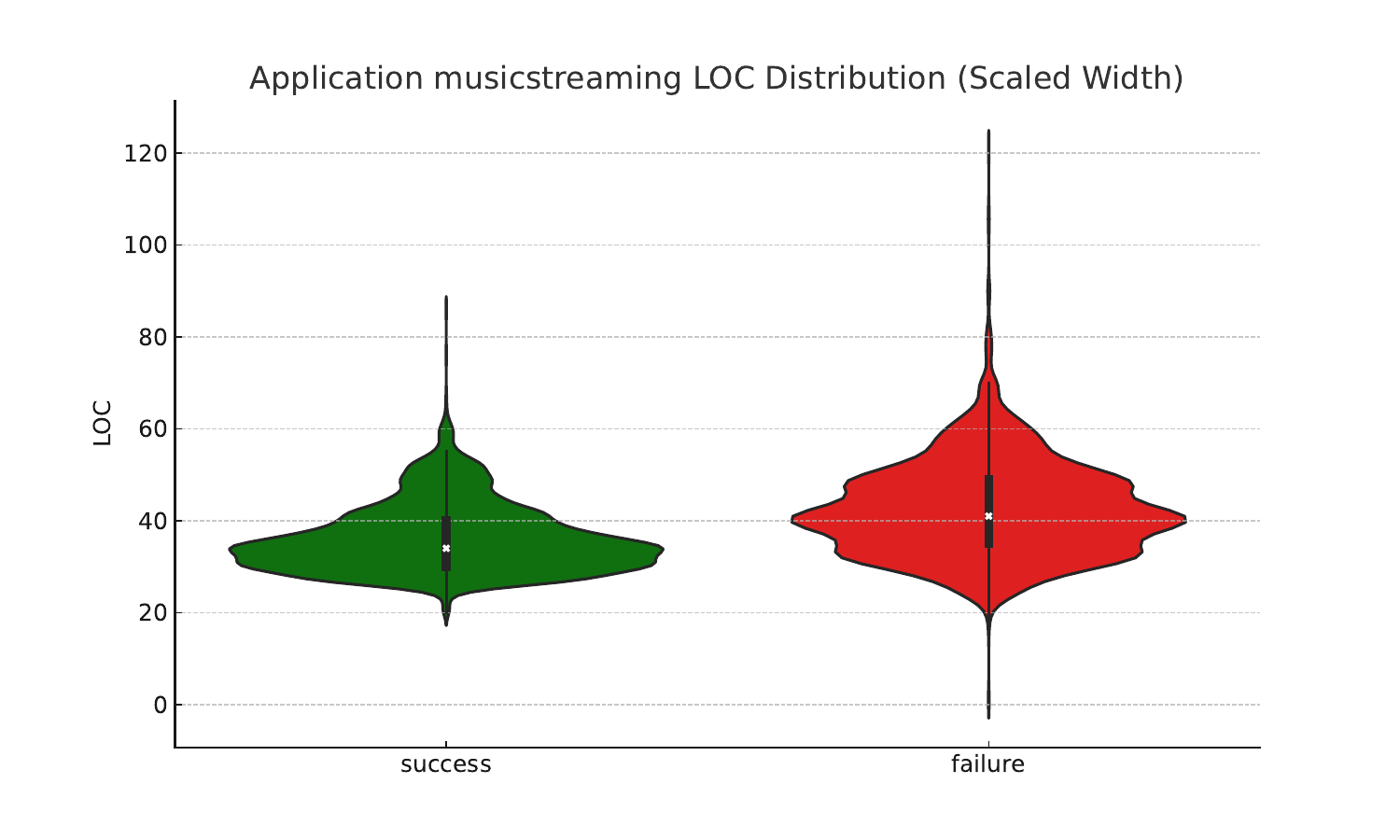}
        \caption{Music Streaming (Mean LOC = 37)}
    \end{subfigure}

    \begin{subfigure}{0.49\textwidth}
        \centering
        \includegraphics[width=\linewidth]{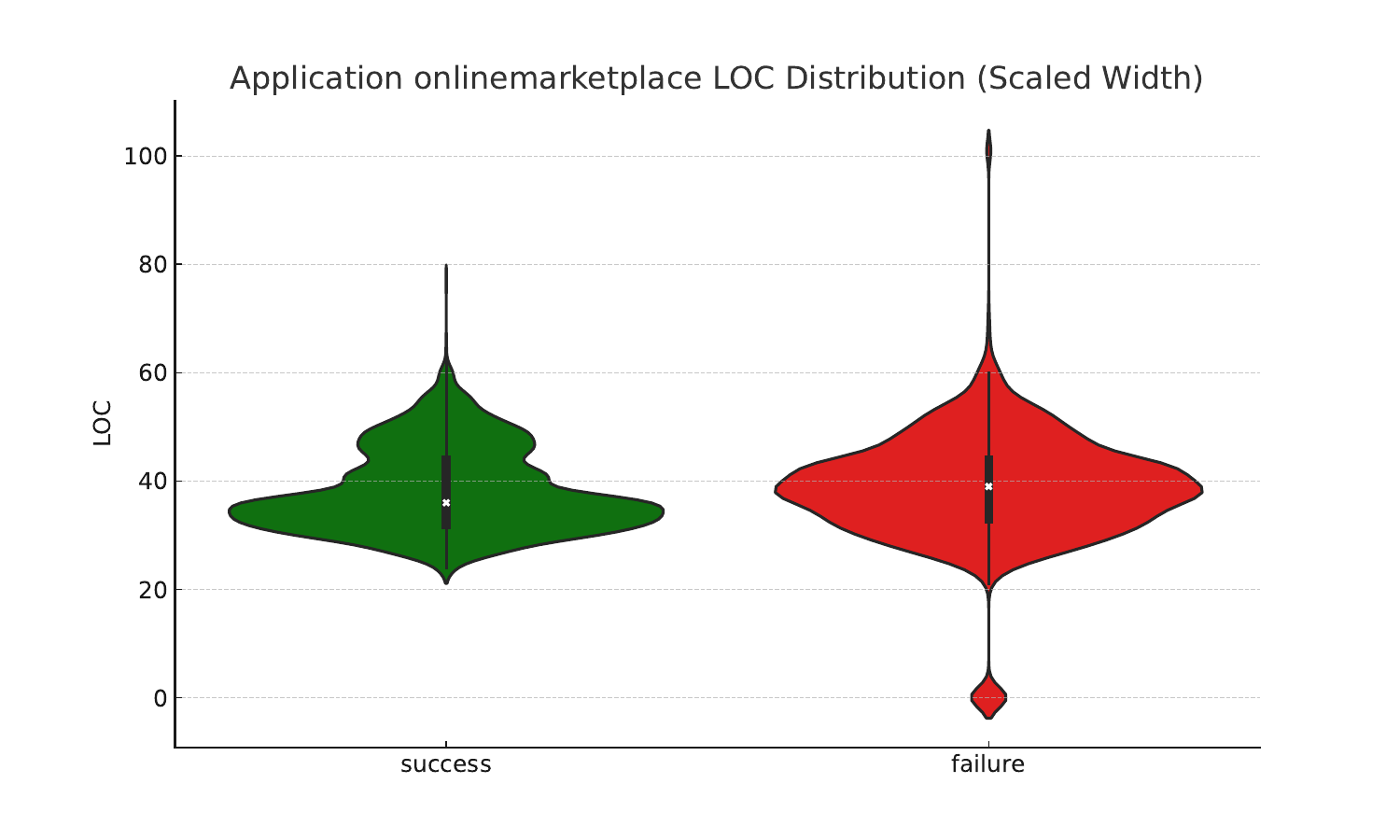}
        \caption{Online Marketplace (Mean LOC = 37)}
    \end{subfigure}
    \begin{subfigure}{0.49\textwidth}
        \centering
        \includegraphics[width=\linewidth]{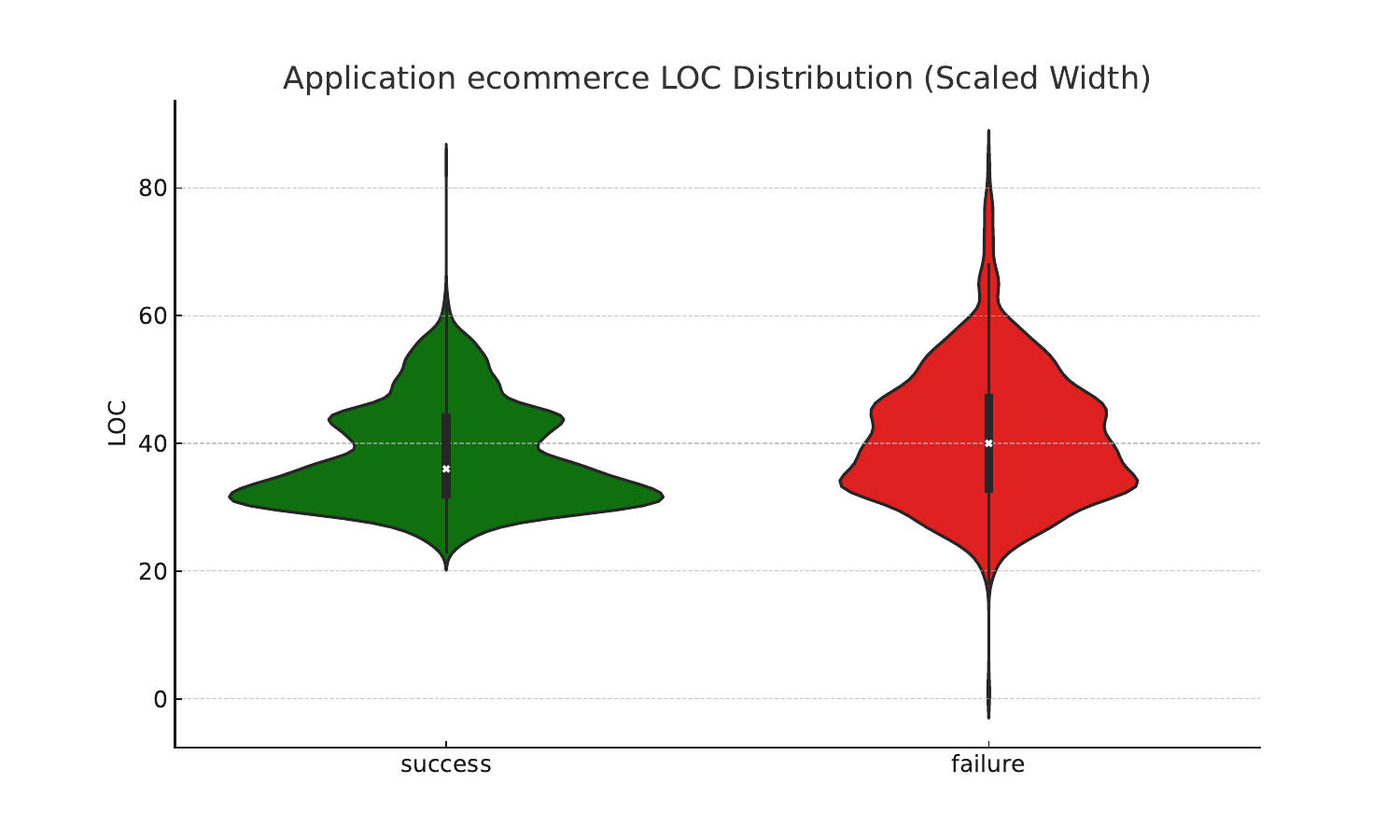}
        \caption{E-commerce (Mean LOC = 37)}
    \end{subfigure}

    \begin{subfigure}{0.49\textwidth}
        \centering
        \includegraphics[width=\linewidth]{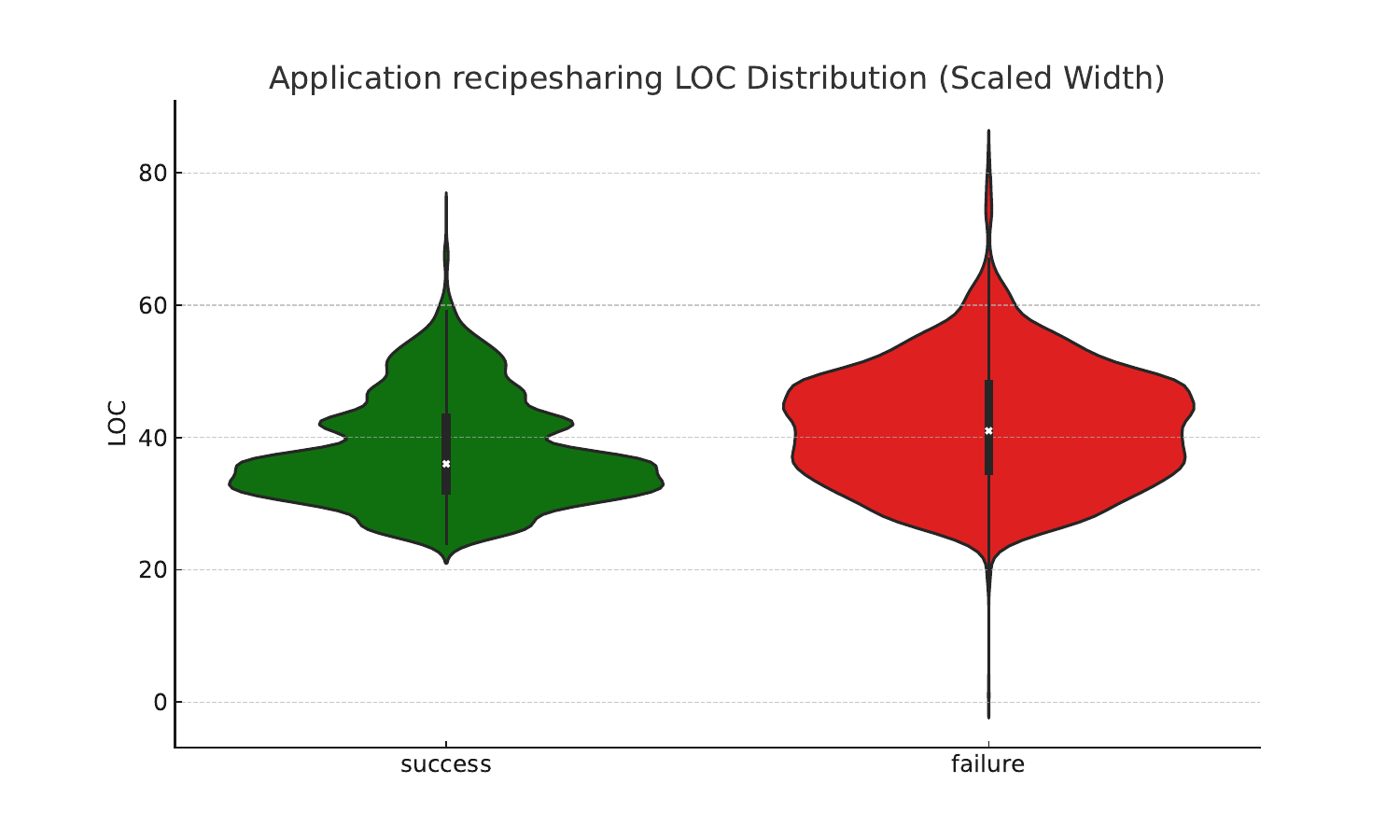}
        \caption{Recipe Sharing (Mean LOC = 38)}
    \end{subfigure}
    \begin{subfigure}{0.49\textwidth}
        \centering
        \includegraphics[width=\linewidth]{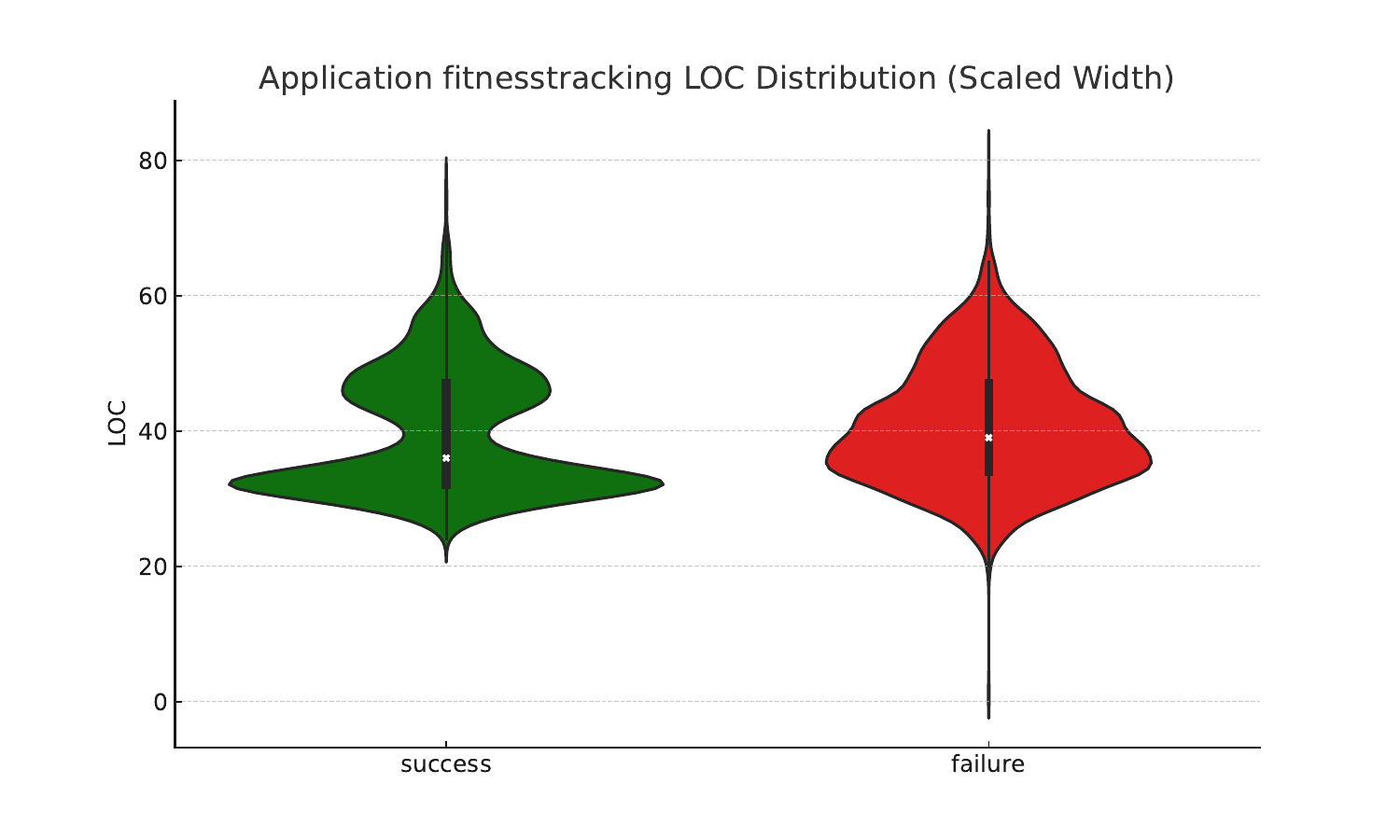}
        \caption{Fitness Tracking (Mean LOC = 38)}
    \end{subfigure}

    \begin{subfigure}{0.49\textwidth}
        \centering
        \includegraphics[width=\linewidth]{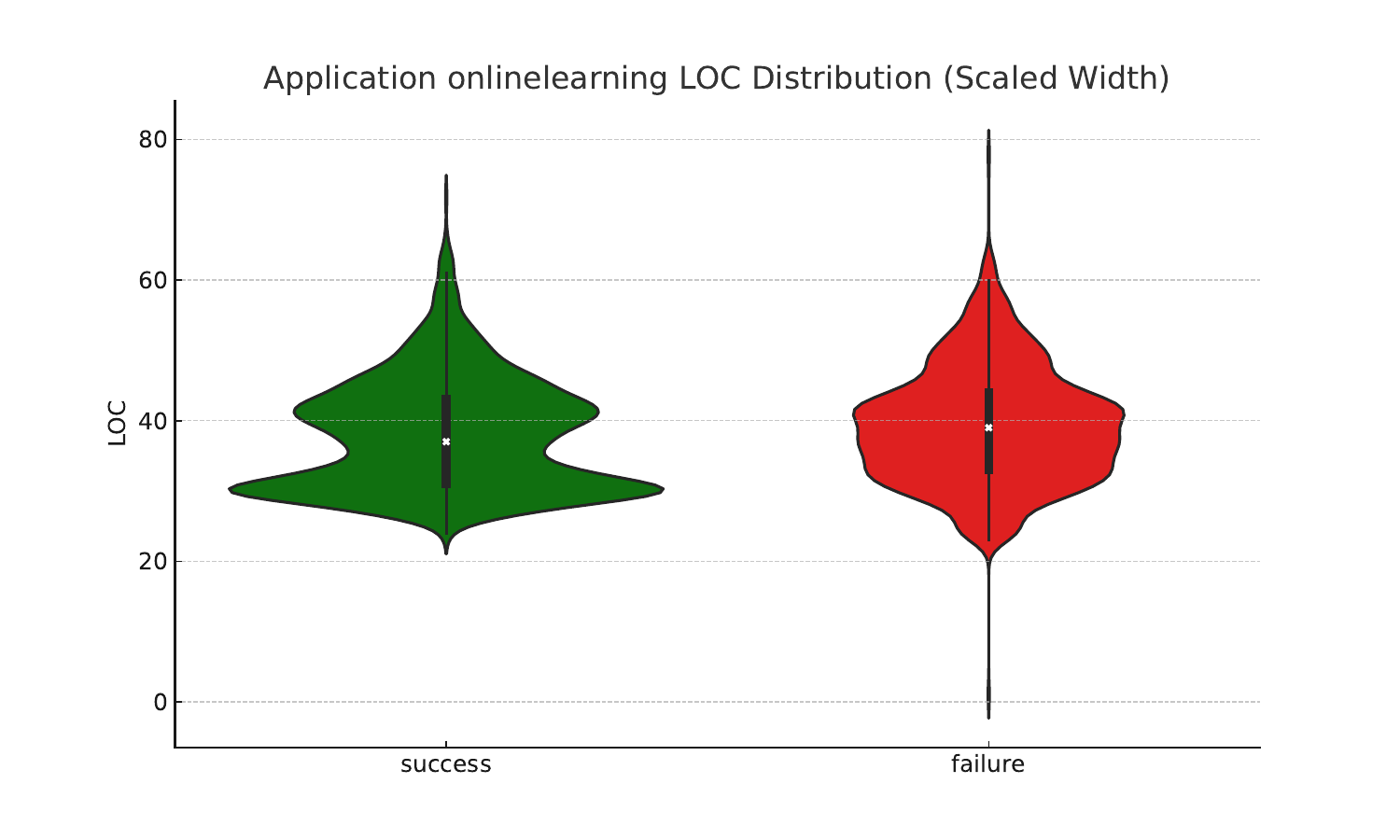}
        \caption{Online Learning (Mean LOC = 38)}
    \end{subfigure}
    \begin{subfigure}{0.49\textwidth}
        \centering
        \includegraphics[width=\linewidth]{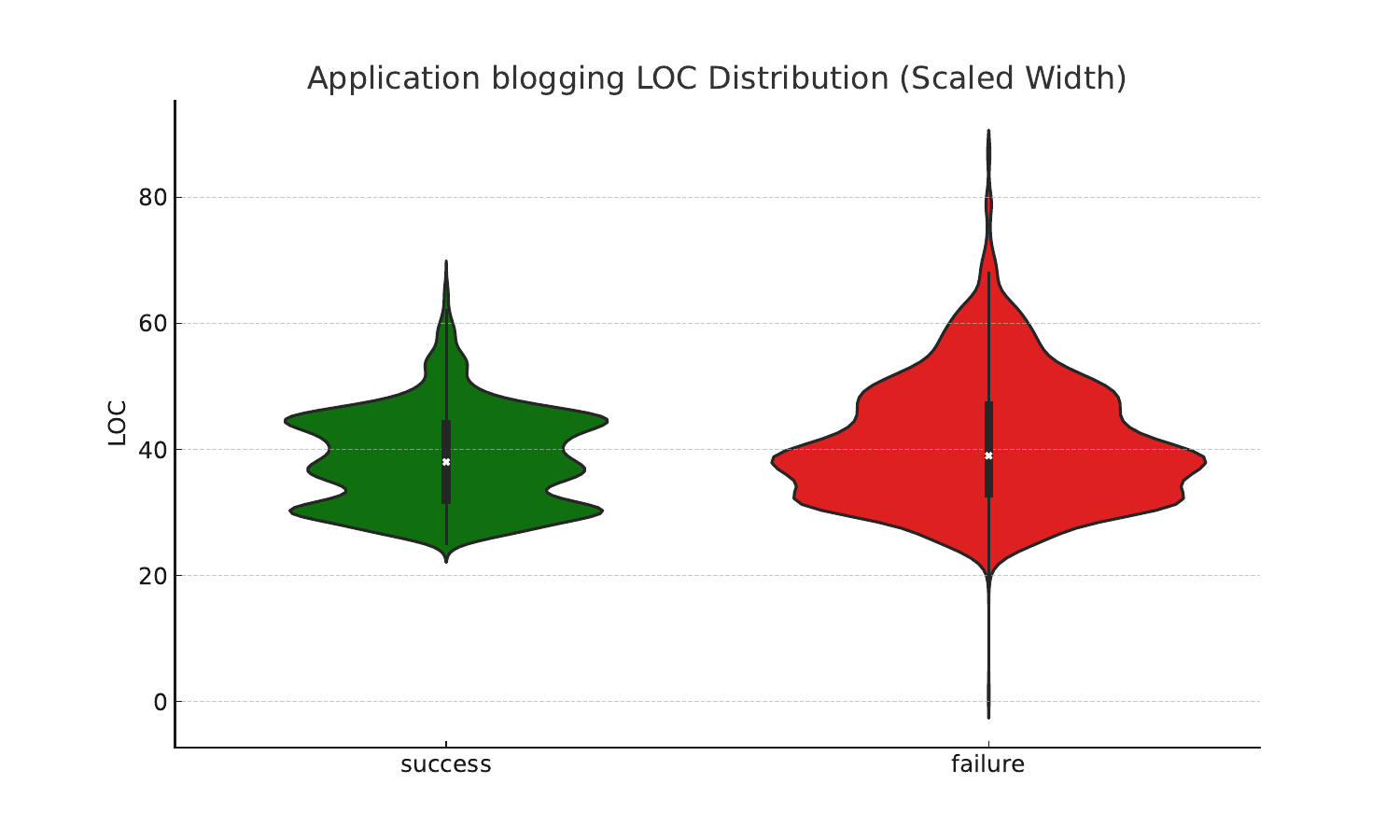}
        \caption{Blogging (Mean LOC = 39)}
    \end{subfigure}

    \begin{subfigure}{0.49\textwidth}
        \centering
        \includegraphics[width=\linewidth]{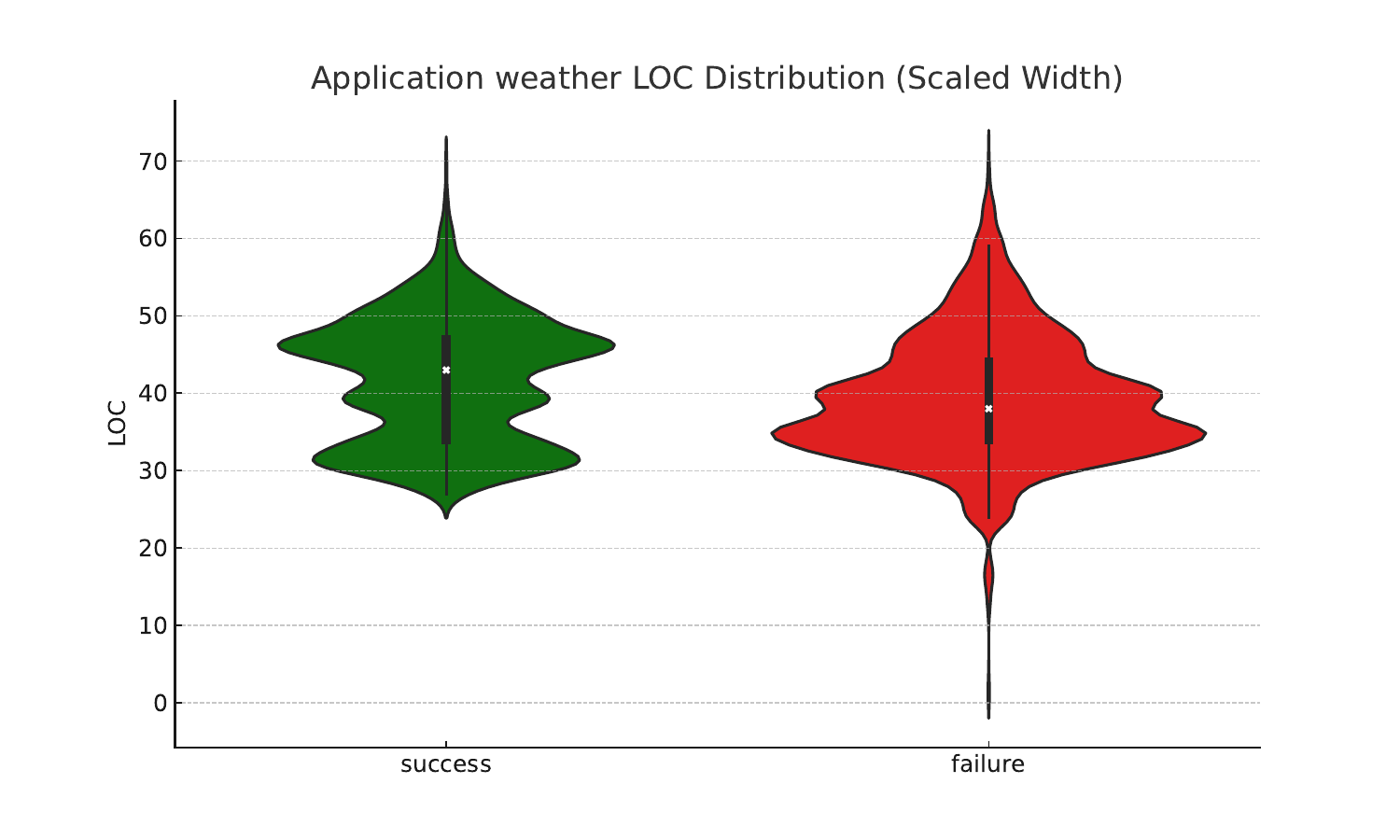}
        \caption{Weather (Mean LOC = 40)}
    \end{subfigure}
    \begin{subfigure}{0.49\textwidth}
        \centering
        \includegraphics[width=\linewidth]{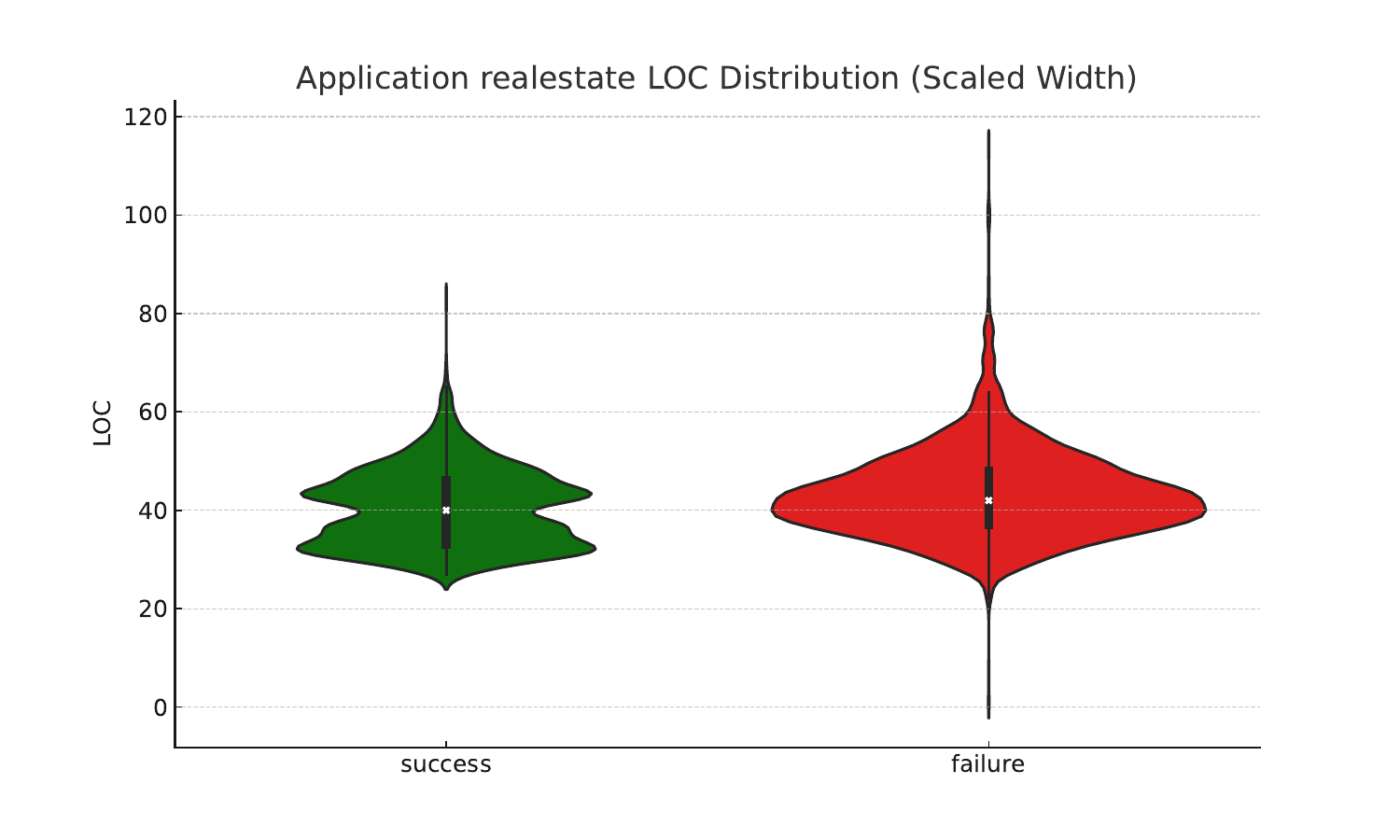}
        \caption{Real Estate (Mean LOC = 42)}
    \end{subfigure}
\end{figure}

\begin{figure}[h]
    \ContinuedFloat
    \centering
    \begin{subfigure}{0.49\textwidth}
        \centering
        \includegraphics[width=\linewidth]{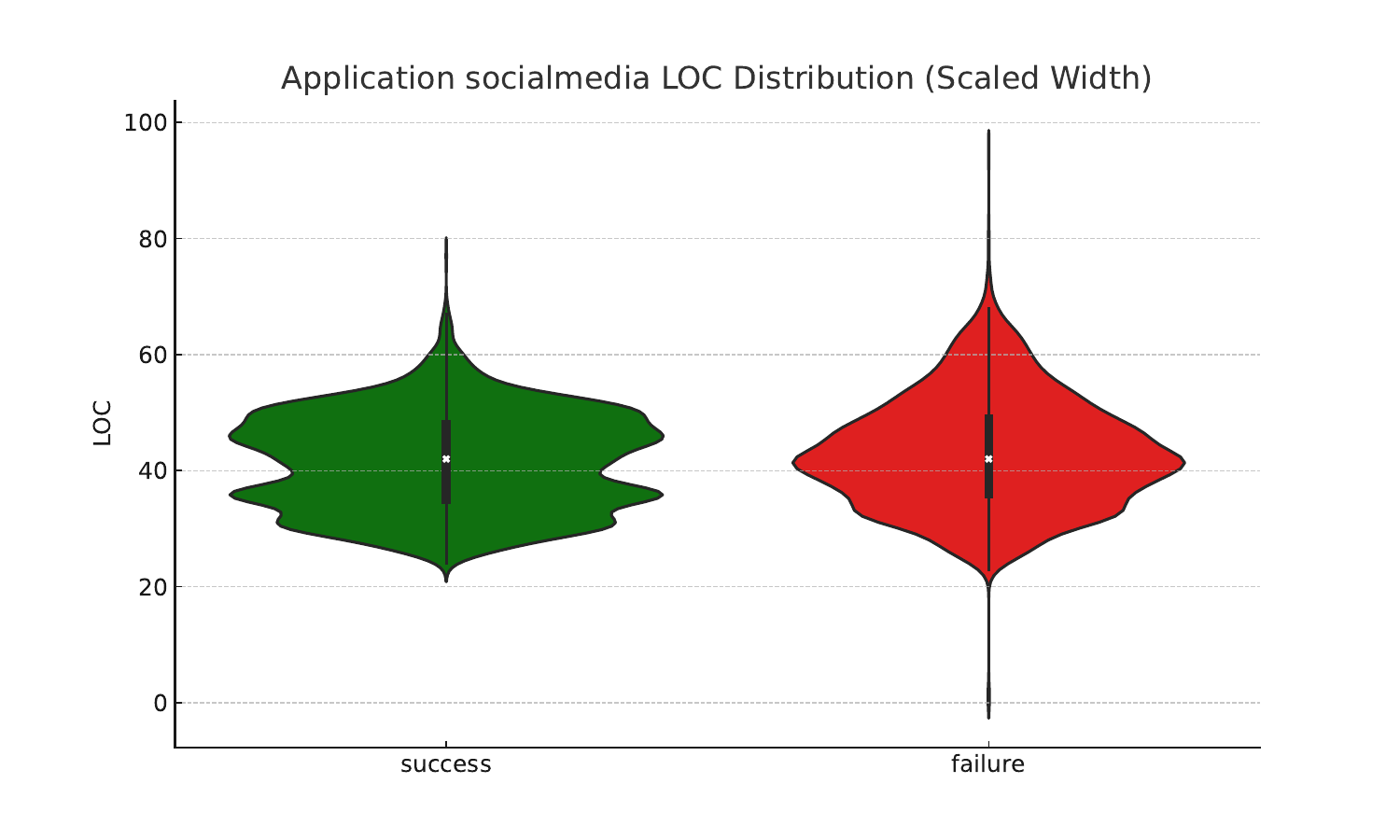}
        \caption{Social Media (Mean LOC = 42)}
    \end{subfigure}
    \begin{subfigure}{0.49\textwidth}
        \centering
        \includegraphics[width=\linewidth]{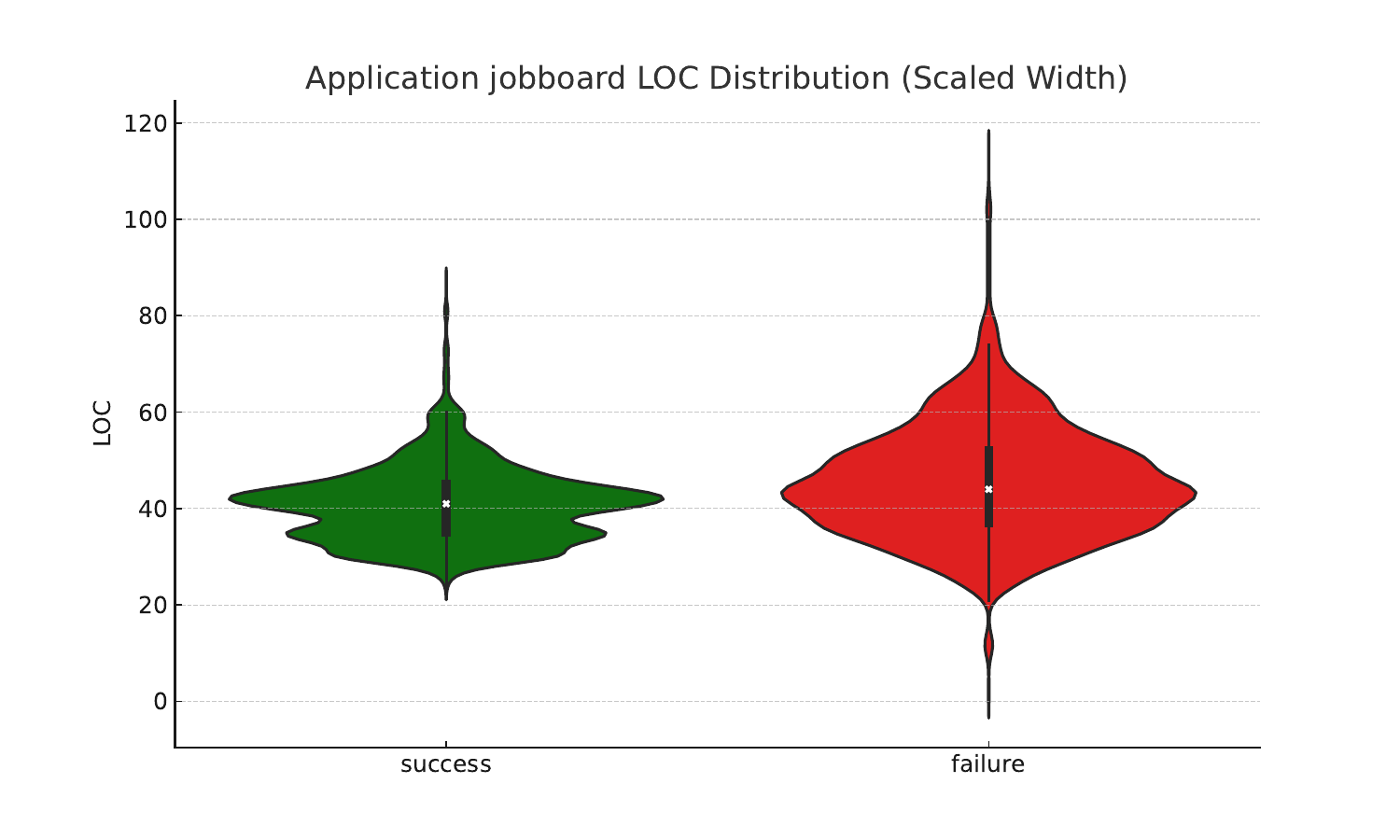}
        \caption{Job Board (Mean LOC = 42)}
    \end{subfigure}

    \begin{subfigure}{0.49\textwidth}
        \centering
        \includegraphics[width=\linewidth]{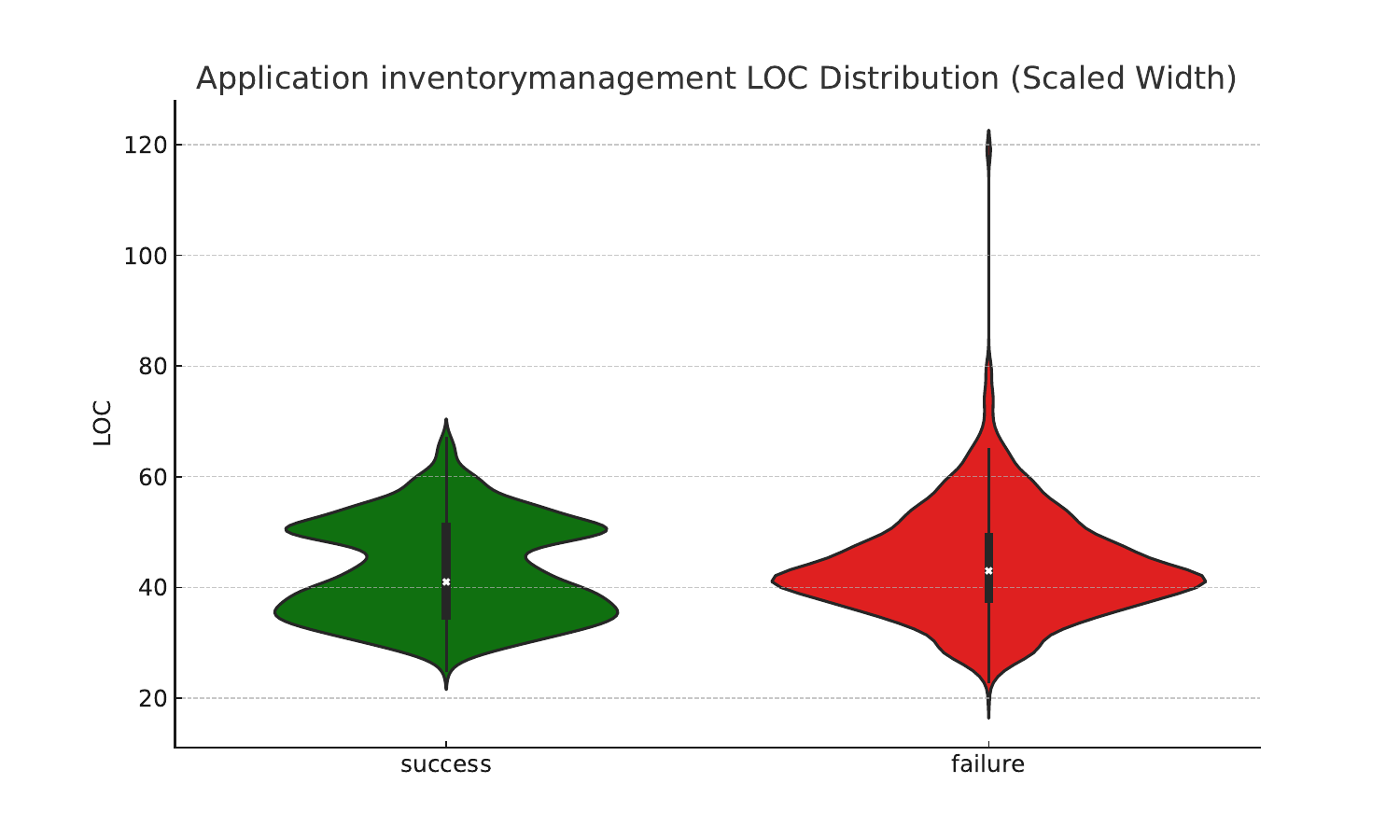}
        \caption{Inventory Management (Mean LOC = 42)}
    \end{subfigure}
    \begin{subfigure}{0.49\textwidth}
        \centering
        \includegraphics[width=\linewidth]{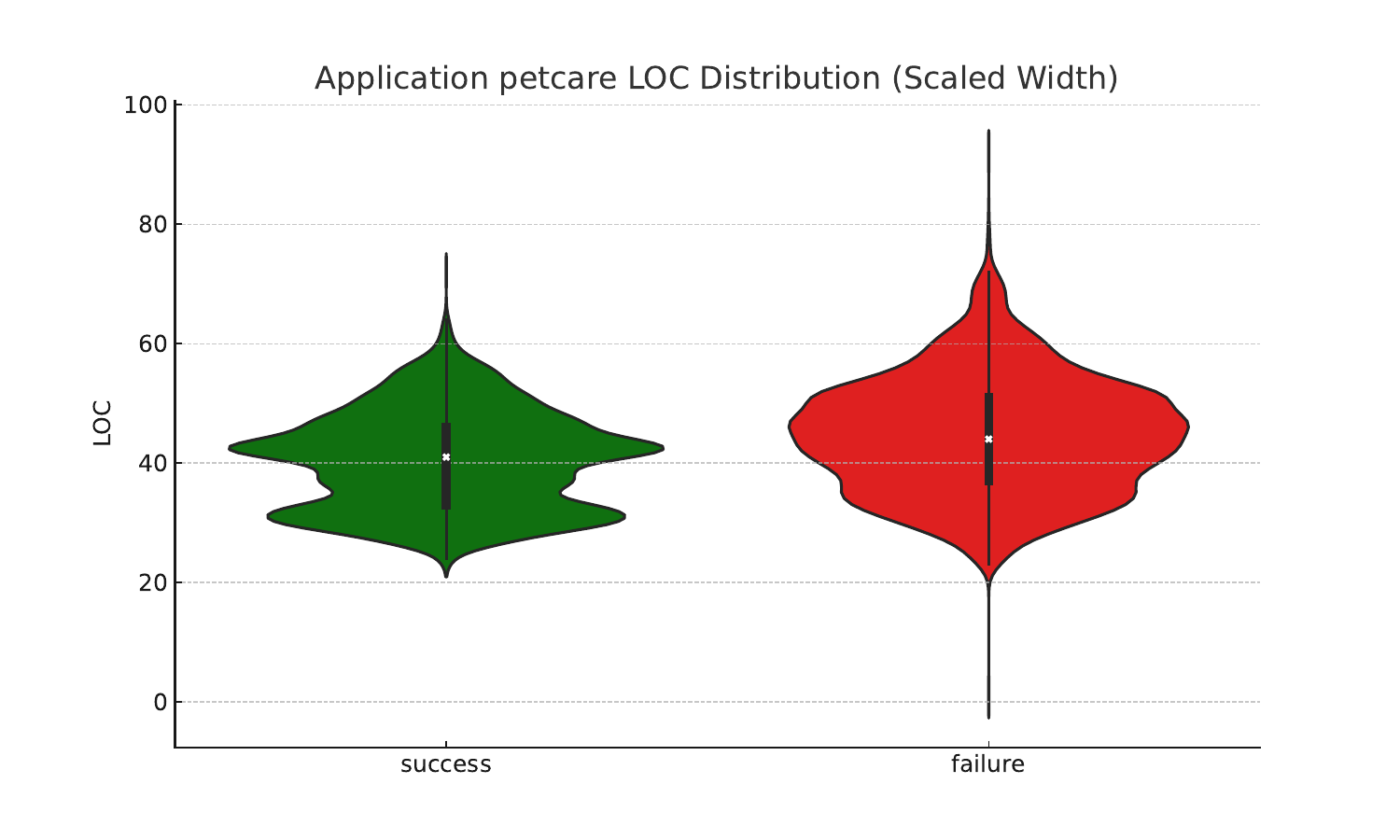}
        \caption{Pet Care (Mean LOC = 42)}
    \end{subfigure}

    \begin{subfigure}{0.49\textwidth}
        \centering
        \includegraphics[width=\linewidth]{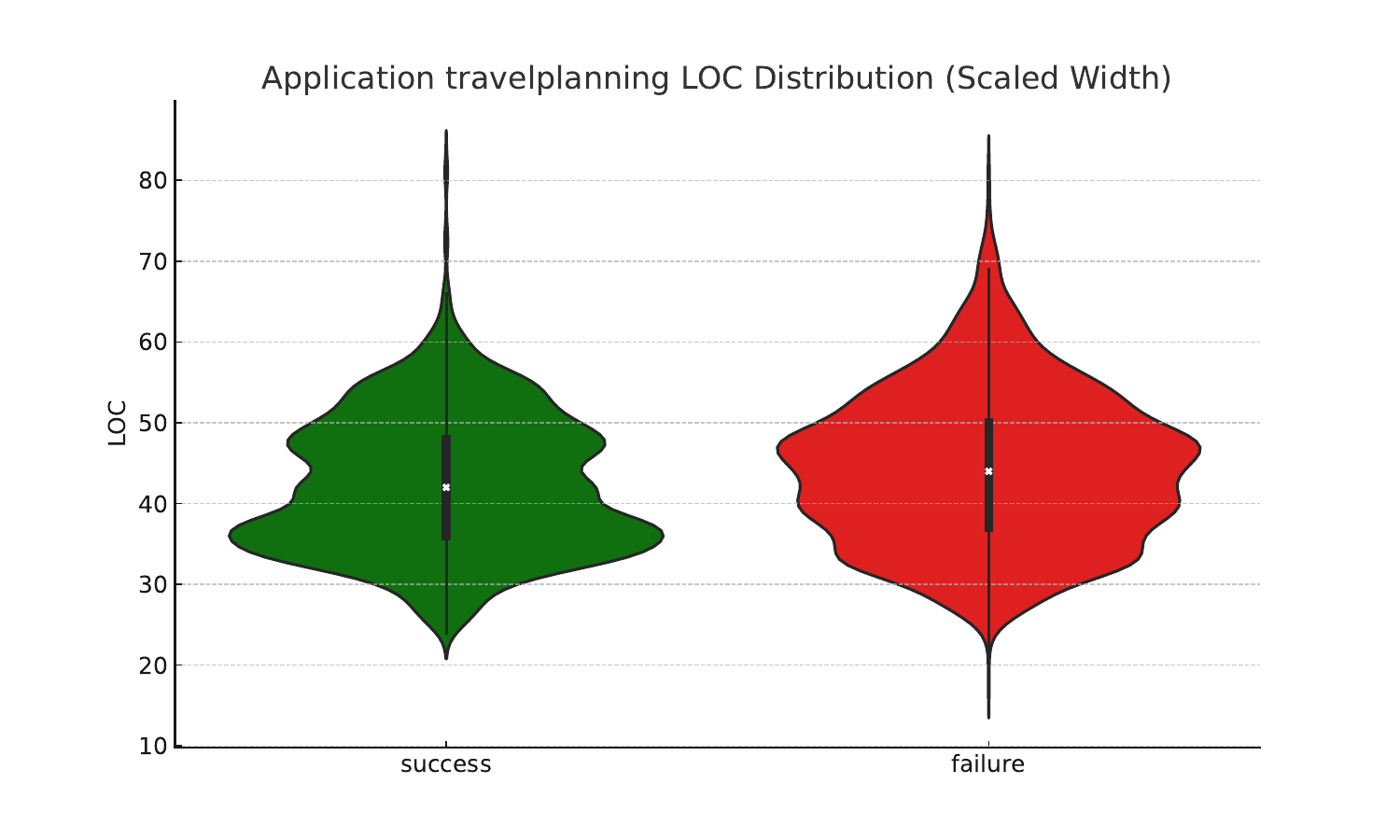}
        \caption{Travel Planning (Mean LOC = 42)}
    \end{subfigure}
    \begin{subfigure}{0.49\textwidth}
        \centering
        \includegraphics[width=\linewidth]{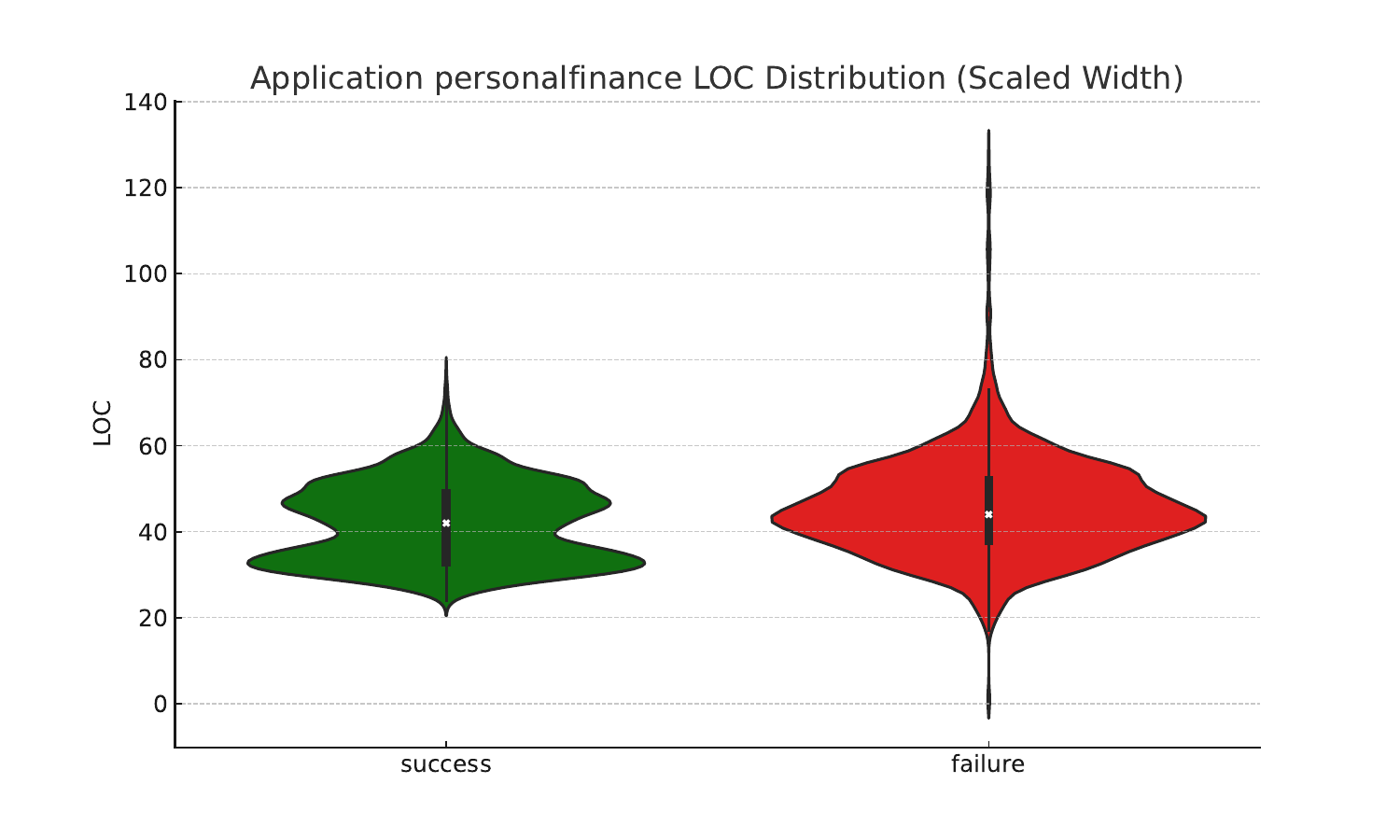}
        \caption{Personal Finance (Mean LOC = 43)}
    \end{subfigure}

    \begin{subfigure}{0.49\textwidth}
        \centering
        \includegraphics[width=\linewidth]{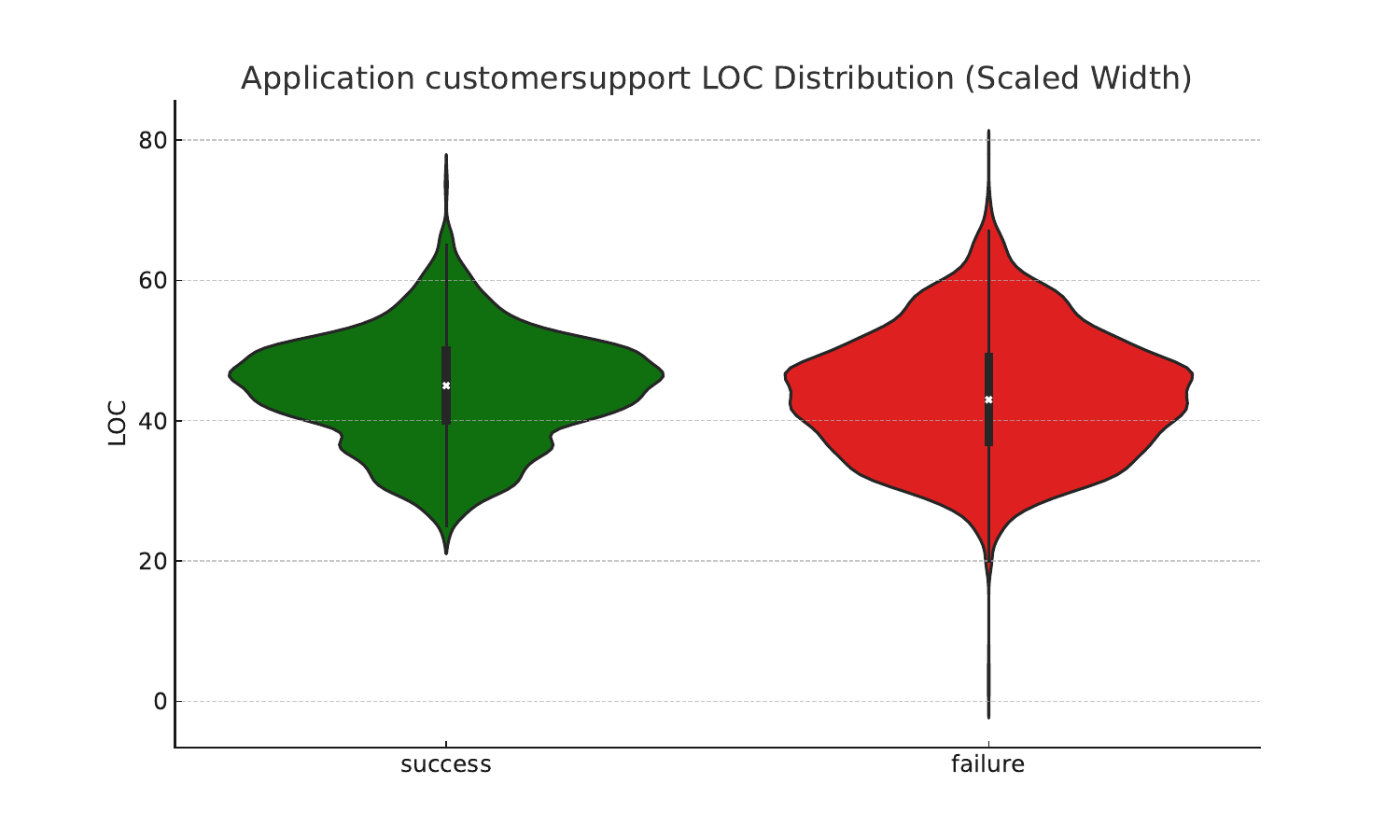}
        \caption{Customer Support (Mean LOC = 44)}
    \end{subfigure}
    \begin{subfigure}{0.49\textwidth}
        \centering
        \includegraphics[width=\linewidth]{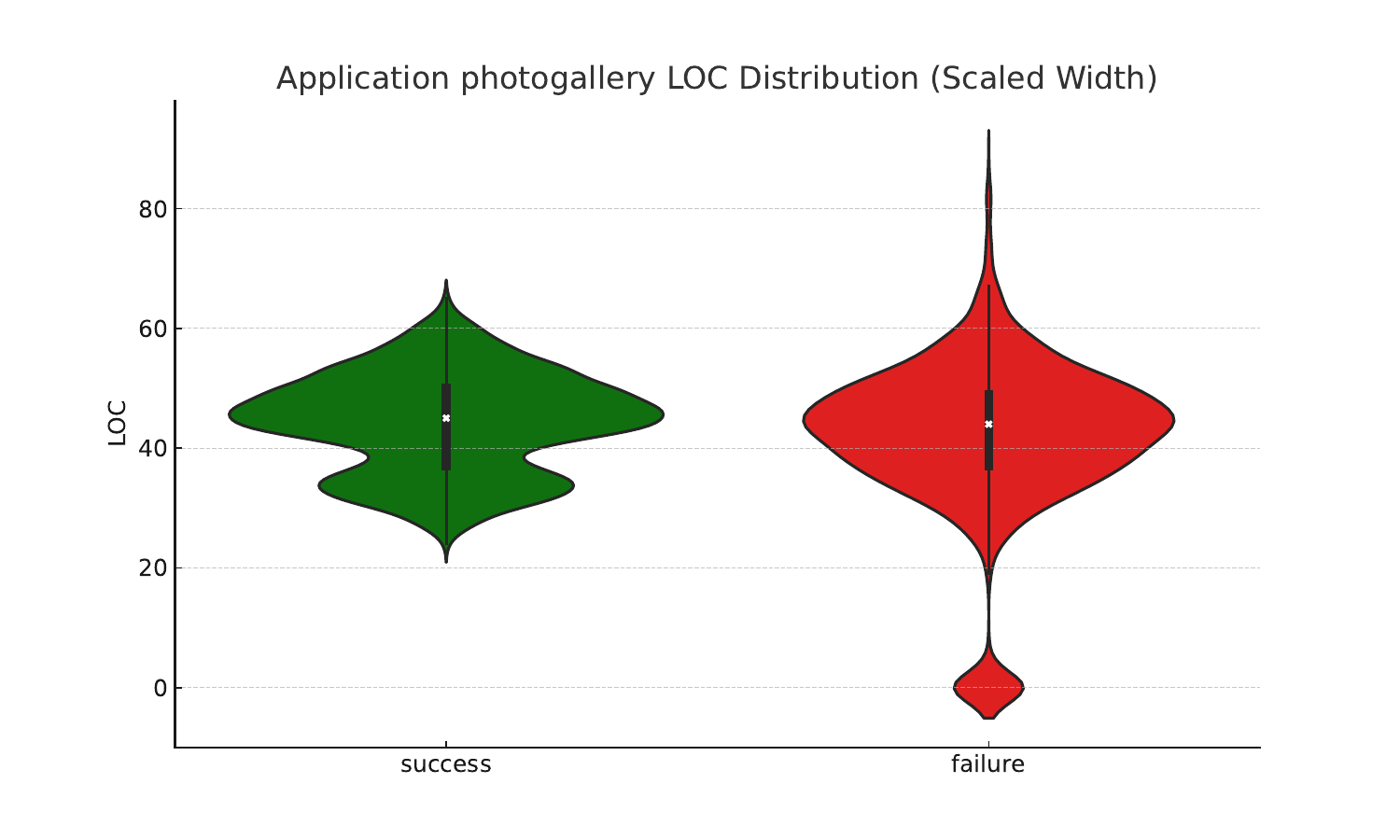}
        \caption{Photo Gallery (Mean LOC = 44)}
    \end{subfigure}

    \begin{subfigure}{0.49\textwidth}
        \centering
        \includegraphics[width=\linewidth]{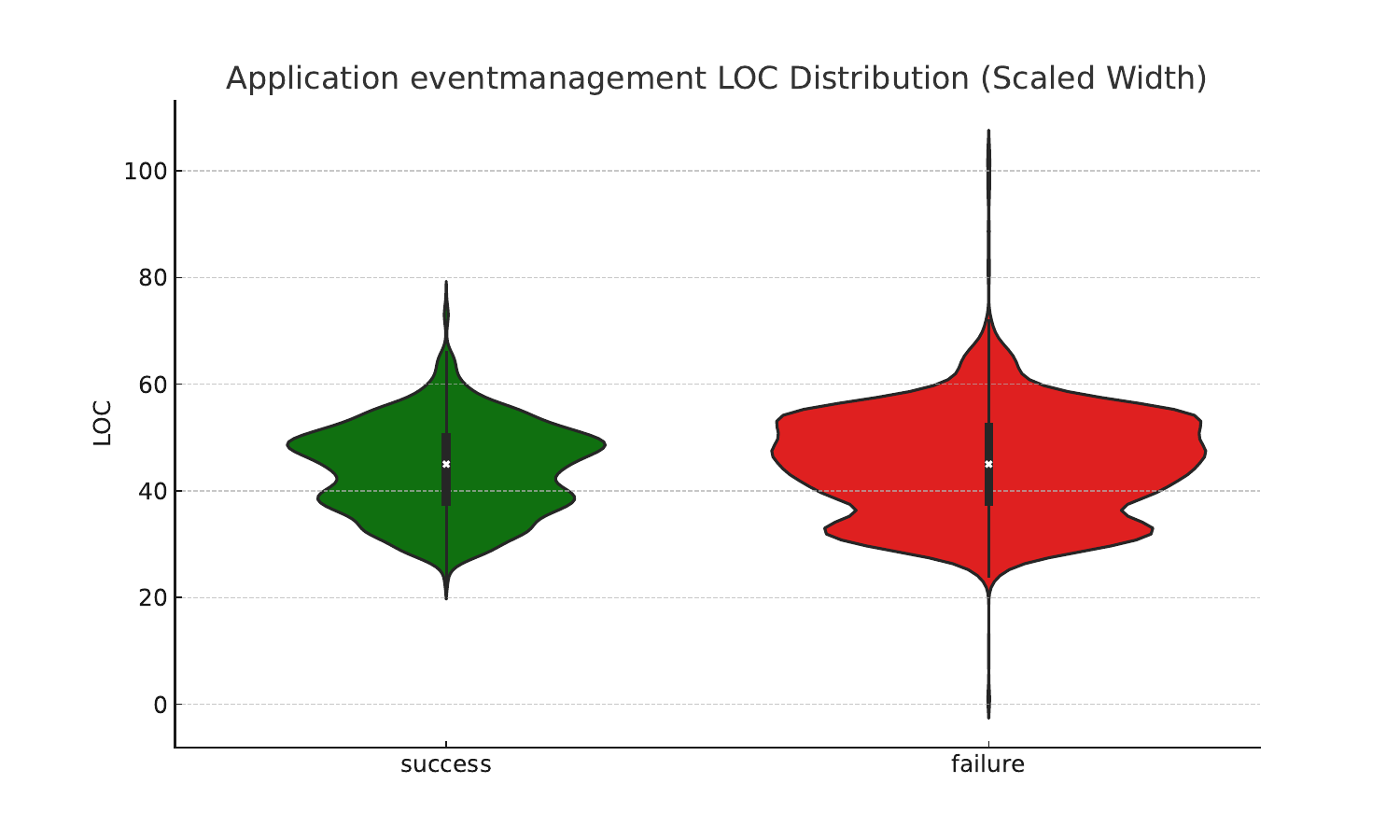}
        \caption{Event Management (Mean LOC = 45)}
    \end{subfigure}
    \begin{subfigure}{0.49\textwidth}
        \centering
        \includegraphics[width=\linewidth]{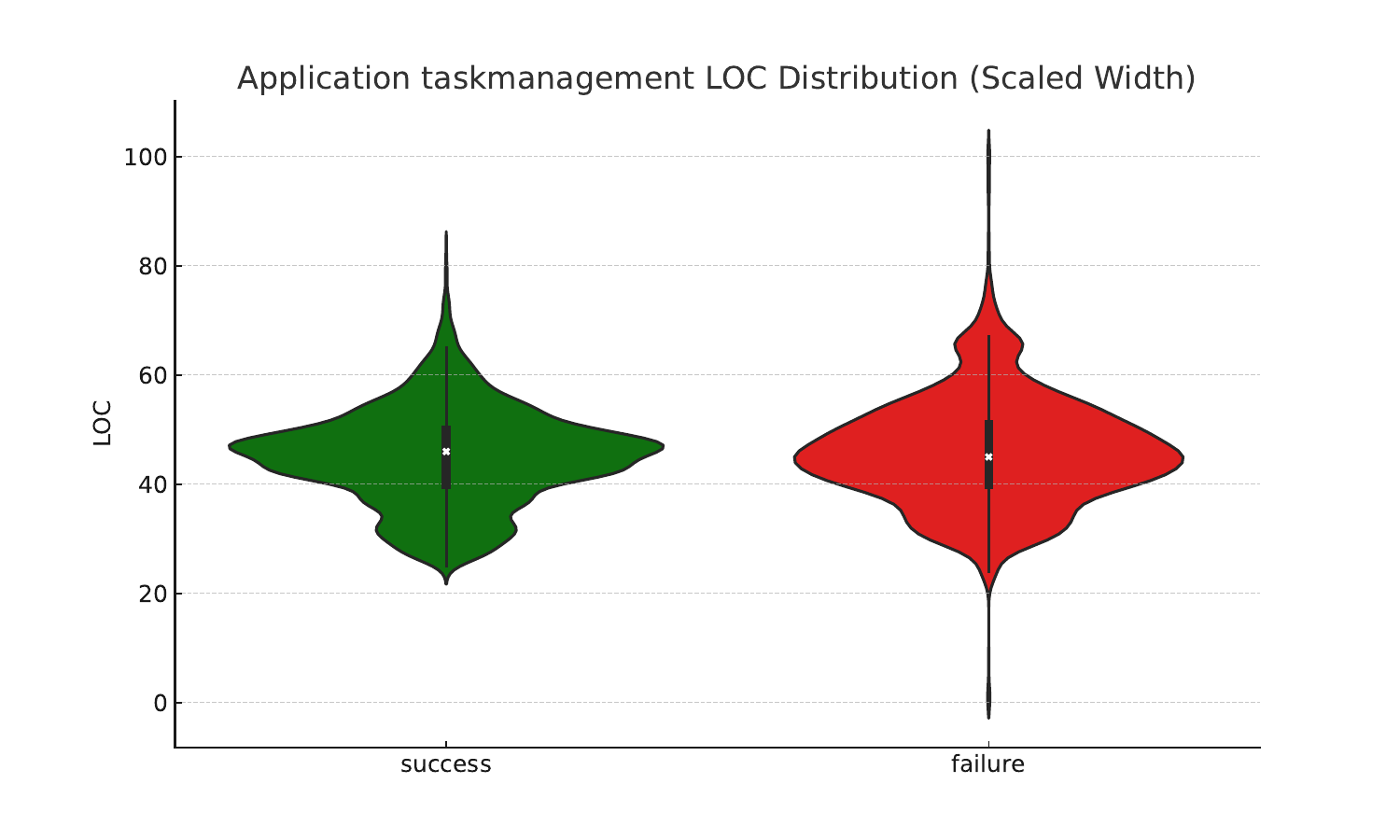}
        \caption{Task Management (Mean LOC = 46)}
    \end{subfigure}

    \caption{LOC Distribution by Application: Success vs Failure}
    \label{fig:loc_successfail_distribution_apps}
\end{figure}

Since each application assembles outputs from all models with full spectrum of performances, the success and failure data set are about the equal size. Similar to what we have observed in model-based sharding (Sec.~\ref{sec:loc_successfail}), the distribution pattern for success is equally or more complex than that for failure, summarized in Tab.~\ref{tab:loc_successfail_distribution_apps}.
\begin{table}[ht]
\centering
\begin{tabular}{|c|c|c|}
\hline
& \textbf{UniModal Success} & \textbf{MultiModal Success} \\
\hline
\textbf{UniModal Failure} & (b) (q) (t) & (c) (d) (f) (g) (h) (j) (k) (l) (m) (n) (o) (p) \\
\hline
\textbf{MultiModal Failure} &  & (a) (e) (i) (r) (s) \\
\hline
\end{tabular}
\caption{Summary of Fig.~\ref{fig:loc_successfail_distribution_apps}: Unimodal vs Multimodal}
\label{tab:loc_successfail_distribution_apps}
\end{table}
\section{Appendix: Error Distribution by Applications}\label{sec:errors_apps}
\begin{figure}[h!]
    \centering
    \includegraphics[width=0.9\textwidth]{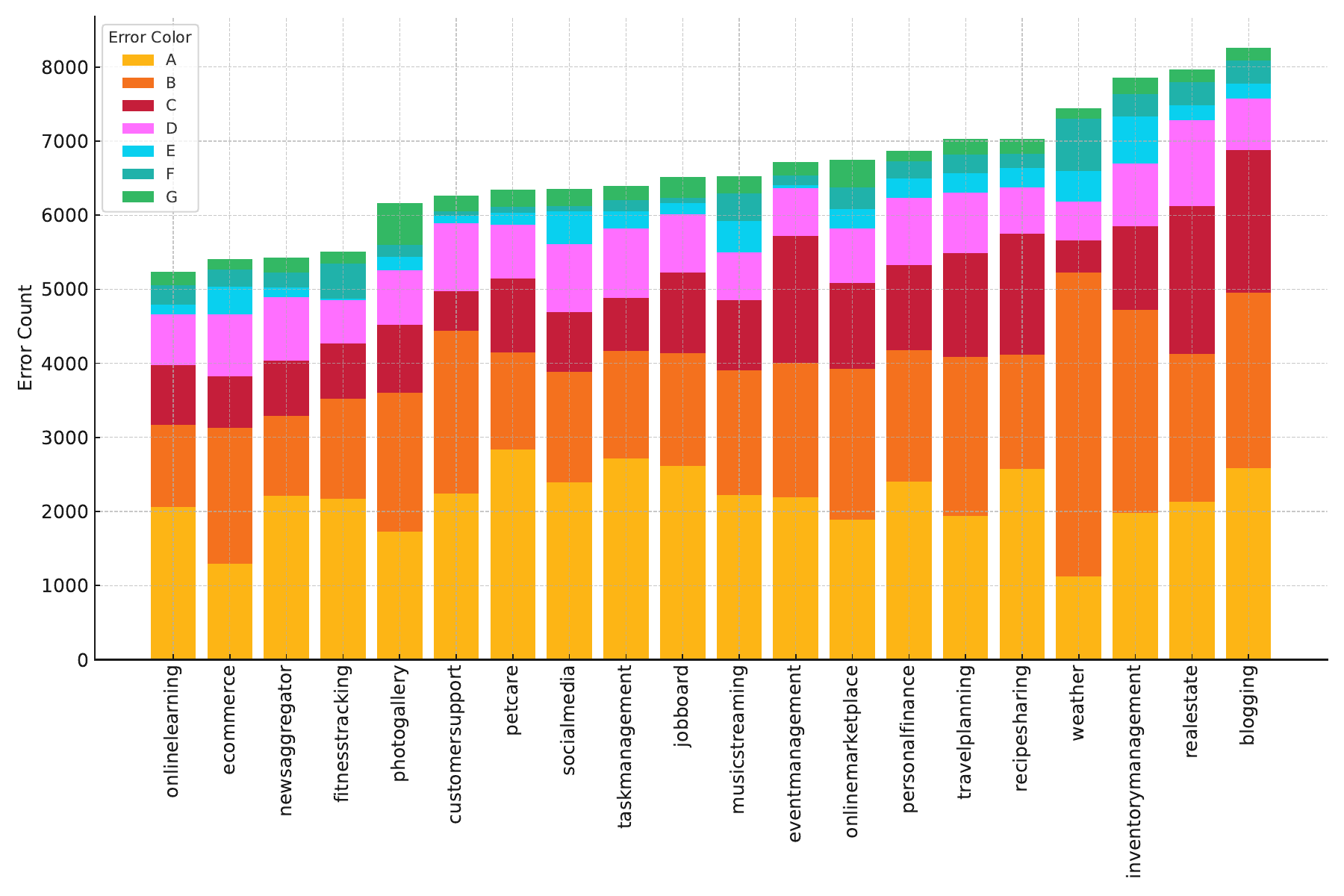}
    \caption{Errors by Applications}
    \label{fig:errors_apps}
\end{figure}

Fig.~\ref{fig:errors_apps} shows error distribution by applications. Since each application assembles outputs from all models, the raw error counts are at the same scale for all applications. We do not find any distinctive patterns. There is neither special error nor special application. 
\bibliographystyle{plainnat}
\bibliography{ref}
\end{document}